\documentclass[11pt,a4paper]{article}
\pdfoutput=1
\usepackage{jheppub}
\usepackage[utf8]{inputenc} 
\usepackage{multirow}
\usepackage{graphicx}
\usepackage{tabularx}
\usepackage{slashed}
\usepackage{url}
\usepackage{hyperref}
\usepackage[normalem]{ulem}

\pdfoutput = 1 
\graphicspath{{figs/}}

\newcommand{\arxiv}[1]{\href{http://arxiv.org/abs/#1}{arXiv:#1}}

\newcommand{\qqqquad}{\qquad \qquad \qquad}

\def\slashchar#1{\setbox0=\hbox{$#1$}           
   \dimen0=\wd0                                 
   \setbox1=\hbox{/} \dimen1=\wd1               
   \ifdim\dimen0>\dimen1                        
      \rlap{\hbox to \dimen0{\hfil/\hfil}}      
      #1                                        
   \else                                        
      \rlap{\hbox to \dimen1{\hfil$#1$\hfil}}   
      /                                         
   \fi}

\def\eg{{e.\,g.}\ }
\def\ie{{i.\,e.}\ }

\begin{document}


\title{Deep-learning Top Taggers or The End of QCD?} 

\author[a]{Gregor Kasieczka}
\affiliation[a]{Institute for Particle Physics, ETH Z\"urich, Switzerland}

\author[b]{Tilman Plehn}
\affiliation[b]{Institut f\"ur Theoretische Physik, Universit\"at Heidelberg, Germany}

\author[c]{Michael Russell}
\affiliation[c]{School of Physics and Astronomy, University of Glasgow, United Kingdom}

\author[b]{Torben Schell}

\emailAdd{gregor.kasieczka@cern.ch}
\emailAdd{plehn@uni-heidelberg.de}
\emailAdd{schell@thphys.uni-heidelberg.de}
\emailAdd{m.russell.2@research.gla.ac.uk}

\preprint{MCNET-17-07}

\abstract{ 
  Machine learning based on convolutional neural networks can be used
  to study jet images from the LHC.  Top tagging in fat jets offers a
  well-defined framework to establish our DeepTop approach and compare
  its performance to QCD-based top taggers.  We first optimize a
  network architecture to identify top quarks in Monte Carlo
  simulations of the Standard Model production channel. Using standard
  fat jets we then compare its performance to a multivariate
  QCD-based top tagger. We find that both approaches lead to
  comparable performance, establishing convolutional networks as a
  promising new approach for multivariate hypothesis-based top tagging.
}

\arxivnumber{1701.08784}

\maketitle

\clearpage

\section{Introduction}
\label{sec:intro}

Geometrically large, so-called fat jets have proven to be an exciting
as well as useful new analysis direction for many LHC Run~I and
Run~II analyses. Their jet substructure allows us to search for
hadronic decays for example of Higgs bosons~\cite{bdrs}, weak gauge
bosons~\cite{w_tag}, or top
quarks~\cite{early_top,hopkins,template,heptop1,heptop2,shower_deco} in a shower
evolution otherwise described by QCD
radiation~\cite{tagging_review}. Given this success, a straightforward
question to ask is whether we can analyze the same jet substructure patterns
without relying on advanced QCD algorithms. An example of such an
approach are wavelets, describing patterns of hadronic weak boson
decays~\cite{wavelet_tim,wavelet_monk}. Even more generally, we can
apply image recognition techniques to the two-dimensional azimuthal
angle vs rapidity plane, for example searching for hadronic decays of
weak bosons~\cite{slac1,slac2,irvine_w,gan} or top
quarks~\cite{ann_top}. The same techniques can be applied to separate
quark-like and gluon-like jets~\cite{quark_gluon}.

Many of the available machine learning applications in jet physics
have in common that we do not have an established, well-performing QCD
approach to compare to. Instead, machine learning techniques are
motivated by their potential to actually make such analyses
possible. Our study focuses on the question of how machine learning
compares to state-of-the-art top taggers in a well-defined fat jet
environment, \ie highly successful QCD-based tagging approaches
established at the LHC. Such a study allows us to answer the question
if QCD-based taggers have a future in hadron collider physics at all,
or if they should and will eventually be replaced with simple pattern
recognition.\medskip

On the machine learning side we will use algorithms known as
convolutional neural networks~\cite{convnet}. Such deep learning
techniques are routinely used in computer vision, targeting image or
face recognition, as well as in natural language processing. In jet
physics the basic idea is to view the azimuthal angle vs rapidity
plane with calorimeter entries as a sparsely filled image, where the
filled pixels correspond to the calorimeter cells with non-zero energy
deposits and the pixel intensities to the deposited energy.  After
some image pre-processing, a training sample of signal and background
images can be fed through a convolutional network, designed to learn
the signal-like and background-like features of these \textsl{jet
  images}~\cite{slac2,quark_gluon}.  The final layer of the network
converts the learned features of the image into a probability of it
being either signal or background. The performance can be expressed in
terms of its receiver operator characteristic (ROC), in complete
analogy to multivariate top tagger analyses~\cite{heptop3,heptop4}.

\subsection{Multivariate analysis tools}

Top tagging is a typical (binary) classification problem. Given a set
of variables $\{x_i\}$ we predict a signal or background label $y$.
In general, we train a classifier on a data set with known labels and
then test the performance of the classifier on a different data
set.\medskip

Rectangular cuts are sufficient if the variables $x_i$ contain orthogonal
information and the signal region variable space is simply
connected. A decision tree as a classifier is especially useful if
there are several disconnected signal regions, or if the shape of the
signal region is not a simple box. The classification is based on a
sequence of cuts to separate signal from background events, where each
criterion depends on the previous decision. Boosting sequentially
trains a collection of decision trees, where in each step the training
data is reweighted according to the result of the previous
classifier. The final classification of the boosted decision tree
(BDT) is based on the vote of all classifiers, and leads to an
increased performance and more stable results. BDTs are part of the
standard LHC toolbox, including modern top taggers~\cite{heptop4}. We
will use them for the QCD-based taggers in our comparison.\medskip

Artificial Neural Networks (ANN) mimic sets of connected neurons. Each
neuron (node) combines its inputs linearly, including biases, and
yields an output based on a non-linear activation function. The usual
implementation are feed-forward networks, where the input for a node
is given by a subset of the outputs of the nodes in the previous
layer. Here, nodes in the first layer work on the $\{x_i\}$, and the
last layer performs the classification. The internal layers are
referred to as hidden layers. The internal parameters of a network, \ie
the weights and biases of the nodes, are obtained by minimizing a
so-called cost or loss function.  Artificial networks with more than
one or two hidden layers are referred to as deep neural networks
(DNN). ANNs and DNNs are frequently used in LHC
analyses~\cite{dnn}.\medskip

In image and pattern recognition convolutional networks (ConvNets)
have shown impressive results. Their main feature is the structure of the
input, where for example in a two-dimensional image the information of
neighboring pixels should be correlated. If we attempt to extract
features in the image with standard DNN and fully connected
neurons in each layer to all pixels, the construction scales poorly
with the dimensionality of the image. Alternatively, we can first
convolute the pixels with linear kernels or filters. The convoluted
images are referred to as feature maps. On all pixels of the feature
map we can apply a non-linear activation function, such that the
feature maps serve as input for further convolution layers, where the
kernels mix information from all input feature maps. After the last
convolution step, the pixels of the feature maps are fed to a standard
DNN. While the convolution layers allow for the identification of
features in the image, the actual classification is performed by the
DNN. While an arbitrarily large non-convolutional DNN should be able to learn features
in the image directly, the convolution layers lead to much faster
convergence of the model.  Image recognition in terms of ConvNets has
only recently been tested for LHC applications~\cite{slac2,quark_gluon}. The machine learning
side of our comparison will be based on ConvNets.

\subsection{Image recognition}

Image recognition includes many operations, which
we will briefly review in this section. The convolutional neural
network starts from a two-dimensional input image and identifying
characteristic patterns using a stack of convolutional layers. We use
a set of standard operations, starting from the $n\times n$ image
input $I$:

\begin{itemize}
\item[--] ZeroPadding: $(n\times n) \to (n+2 \times n+2)$\\ We
  artificially increase the image by adding zeros at all boundaries in order to
  remove dependence on non-trivial boundary conditions,
  \begin{align} 
    I \to  
    \begin{pmatrix} 
      0 & \cdots &0 \\
      \vdots & I & \vdots\\
      0 & \cdots & 0
    \end{pmatrix} \; .
  \end{align}

\item[--] Convolution: $n'_\text{c-kernel} \times (n\times n) \to
  n_\text{c-kernel} \times((n-n_\text{c-size}+1) \times
  (n-n_\text{c-size}+1))$\\ To identify features in an $n \times n$
  image or feature map we linearly convolute the input with
  $n_\text{c-kernel}$ kernels of size $n_\text{c-size} \times
  n_\text{c-size}$. If in the previous step there are
  $n'_\text{c-kernel}>1$ layers, the kernels are moved over all input
  layers.  For each kernel this defines a feature map $\widetilde
  F^{k}$ which mixes information from all input layers
  \begin{align}
    \widetilde F^{k}_{ij} = \sum_{l=0}^{n'_\text{c-kernel}-1} \quad  \sum_{r,s=0}^{n_\text{c-size}-1} 
    \widetilde{W}^{kl}_{rs}  \;
    I^{l}_{i+r,j+s} + b_k
    \qquad \text{for} \quad
    k = 0,...,n_\text{c-kernel}-1 \; .
  \label{eq:def_conv}
  \end{align}

\item[--] Activation: $(n\times n) \to (n \times n)$\\ This non-linear
  element allows us to create more complex features. A common choice is
  the rectified linear activation function (ReLU) which sets pixel with
  negative values to zero, $f_\text{act}(x) = \max (0, x)$. In this
  case we define for example
  \begin{align} 
    F^{k}_{ij} = f_\text{act}(\widetilde F^{k}_{ij})
              = \max \left( 0, \widetilde F^{k}_{ij} \right) \; .
  \end{align}
  Instead of introducing an additional unit performing the activation,
  it can also be considered as part of the previous layer.

\item[--] Pooling: $(n\times n) \to (n/p \times n/p)$\\ We can reduce
  the size of the feature map by dividing the input into patches of
  fixed size $p \times p$ (sub-sampling) and assign a single value to
  each patch
  \begin{align} 
    F'_{ij} = f_\text{pool}(F_{(ip\dots (i+1)p-1,jp\dots (j+1)p-1})
    \; .
  \end{align}
  MaxPooling returns the maximum value of the subsample
  $f_\text{pool}(F)=\max_\text{patch}(F_{ij})$.

\end{itemize}
\medskip

A convolutional layer consists of a ZeroPadding, Convolution, and
Activation step each. We then combine $n_\text{c-layer}$ of these
layers, followed by a pooling step, into a block. Each of our
$n_\text{c-block}$ blocks therefore works with essentially the same
size of the feature maps, while the pooling step between the blocks
strongly reduces the size of the feature maps. This ConvNet setup
efficiently identifies structures in two-dimensional jet images,
encoded in a set of kernels $W$ transforming the original picture into
a feature map. In a second step of our analysis the ConvNet output
constitutes the input of a fully connected DNN, which translates the
feature map into an output label $y$:

\begin{itemize}
\item[--] Flattening: $(n\times n) \to (n^2 \times 1)$\\ While the
  ConvNet uses two-dimensional inputs and produces a set of
  corresponding feature maps, the actual classification is done by a
  DNN in one dimension. The transition between the formats reads
  \begin{align} 
    x = \left( F_{11},\dots,F_{1n},\dots,F_{n1},\dots,F_{nn} 
        \right) \; .
  \end{align}

\item[--] Fully connected (dense) layers: $n^2 \to
  n_\text{d-node}$\\ The output of a standard DNN is the weighted sum
  of all inputs, including a bias, passed through an activation
  function. Using rectified linear activation it reads
  \begin{align}
    y_i 
        = \max \left( 0, \sum_{j=0}^{n^2-1} W_{ij} x_j + b_i \right) \; .
  \label{eq:def_dnn}
  \end{align}
  For the last layer we apply a specific SoftMax activation function 
  \begin{align}
    y_i = \frac{\exp \left( W_{ij} x_j + b_i \right)}
               { \sum_i \exp \left( W_{ij} x_j + b_i \right)} \; .
  \end{align}
   It ensures $y_i \in [0,1]$, so the label can be interpreted as a
   signal or background probability.

\end{itemize}
\medskip

In a third step we define a cost or loss function, which we use to
train our network to a training data set. For a fixed architecture a
parameter point $\theta$ is given by the ConvNet weights
$\widetilde{W}_{rs}^{kl}$ defined in Eq.\eqref{eq:def_conv} combined
with the DNN weights $W_{ij}$ and biases $b_i$ defined in
Eq.\eqref{eq:def_dnn}.  We minimize the the mean squared error
\begin{align}
  L(\theta) = \frac 1 N \sum_{i=0}^{N} \left( y(\theta; x_i) - y_i \right)^2
  \;, 
\end{align}
where $y(\theta;x_i)$ is the predicted binary label of the input $x_i$
and $y_i$ is its true value.
This choice of loss function does not optimize the learning performance
or the probabalistic information, but it will work fine for our purpose. 
Eventually, it could for example be replaced by the cross  entropy.
For a given parameter point $\theta$ we compute the gradient of the
loss function $L(\theta)$ and first shift the parameter point from
$\theta_n$ to $\theta_{n+1}$ in the direction of the gradient $\nabla
L(\theta_{n})$.  In addition, we can include the direction of the
previous change such that the combined shift in parameter space is
\begin{align}
  \theta_{n+1} = \theta_{n} - \eta_L \nabla L(\theta_{n}) + \alpha
  (\theta_{n} - \theta_{n-1}) \;.
\end{align}
The learning rate $\eta_L$ determines the step size and can be chosen to
decay with each step (decay rate).  The parameter $\alpha$, referred
to as momentum, dampens the effect of rapidly changing gradients and
improves convergence. The Nesterov algorithm changes the point of
evaluation of the gradient to
\begin{align}
  \theta_{n+1} = \theta_{n} - \eta_L \nabla L(\theta_{n} + \alpha
  (\theta_{n} - \theta_{n-1})) + \alpha
  (\theta_{n} - \theta_{n-1}) \;.
\end{align}
Each training step (epoch) uses the full set of training events.

\section{Machine learning setup}
\label{sec:machine}

The goal of our analysis is to determine if a machine learning
approach to top tagging at the LHC offers a significant advantage over
the established QCD-based taggers, and to understand the learning
pattern of such a convolutional network. To reliably answer this
question we build a flexible neural network setup, define an
appropriate interface with LHC data through the jet images, and
optimize the ConvNet/DNN architecture and parameters for top tagging
in fat jets. To build our neural network we use the Python package
\textsc{Theano}~\cite{theano}, with a \textsc{Keras}
front-end~\cite{keras}. An optimized speed or CPU usage is not
part of our performance study.

\subsection{Jet images and pre-processing}
\label{sec:machine_image}

The basis of our study are calorimeter images, which we produce using
standard Monte Carlo simulations --- obviously, in an actual
application they should come from data.  In recent years, many
strategies have been developed to define appropriate signal samples
which allow us to benchmark top taggers. Typically, they rely on top
pair production with an identified leptonic top decay recoiling
against a top jet. The lepton kinematics can then be used to estimate
the transverse momentum of the hadronically decaying top quark.  We
simulate a 14 TeV LHC hadronic $t\bar{t}$ sample and a QCD dijet
sample with \textsc{Pythia8}~\cite{pythia}, ignoring multiple
interactions. While one could clearly include pile-up in the simulations,
understanding and removing it requires information beyond the calorimeter
images, for examples from tracks. For our early study we do not include 
track information in our jet image, some ideas in this direction are
pointed out in Ref.~\cite{quark_gluon}

All events are passed through a fast detector simulation with
\textsc{Delphes3}~\cite{delphes} with calorimeter towers of size
$\Delta\eta \times \Delta\phi = 0.1 \times 5^\circ$ and a threshold of
1~GeV.  We cluster these towers with \textsc{FastJet3}~\cite{fastjet}
to $R=1.5$ anti-$k_T$~\cite{anti_kt} jets with $|\eta |< 1.0$. These
anti-$k_T$ jets give us a smooth outer shape of the fat jet and a
well-defined jet area for our jet image.  To ensure that the jet
substructure in the jet image is consistent with QCD we re-cluster the
anti-$k_T$ jet constituents with an $R=1.5$ C/A
jet~\cite{ca_algo}. Its substructures define the actual jet image.
When we identify calorimeter towers with pixels of the jet image, it
is not clear whether the information should be the energy $E$ or only
its transverse component $E_T$.

Our fat jets have to fulfill $|\eta_\text{fat}| < 1.0$, to guarantee
that they are entirely in the central part of the detector and to
justify our calorimeter tower size. For this paper we focus on the
range $p_{T,\text{fat}} = 350~...~450$~GeV, such that all top decay
products can be easily captured in the fat jet.  For signal events, we
require that the fat jet can be associated with a Monte-Carlo truth
top quark within $\Delta R < 1.2$.\medskip

\begin{figure}[t]
  \includegraphics[width=0.44 \textwidth]{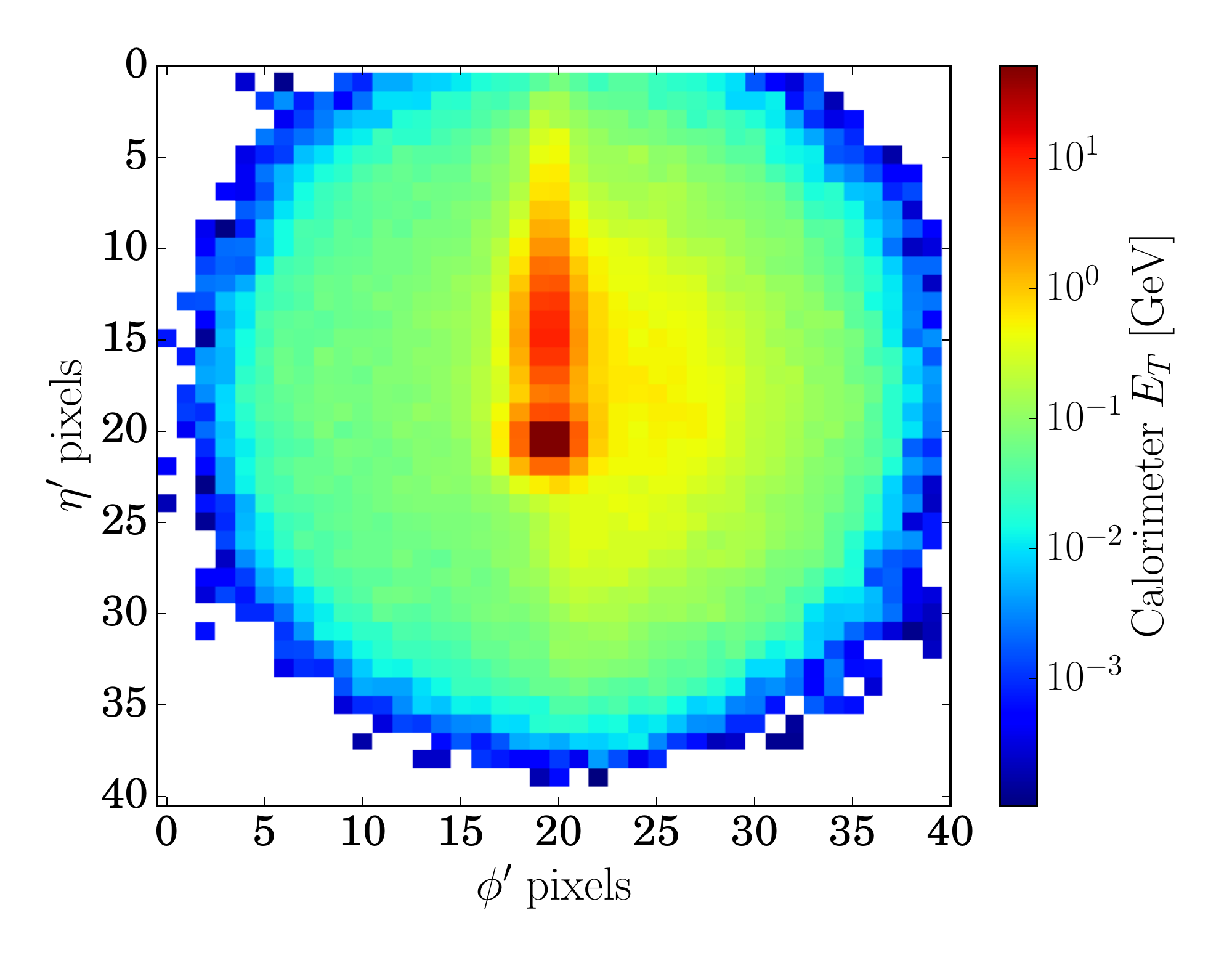}
   \hspace*{0.05\textwidth}
   \includegraphics[width=0.44 \textwidth]{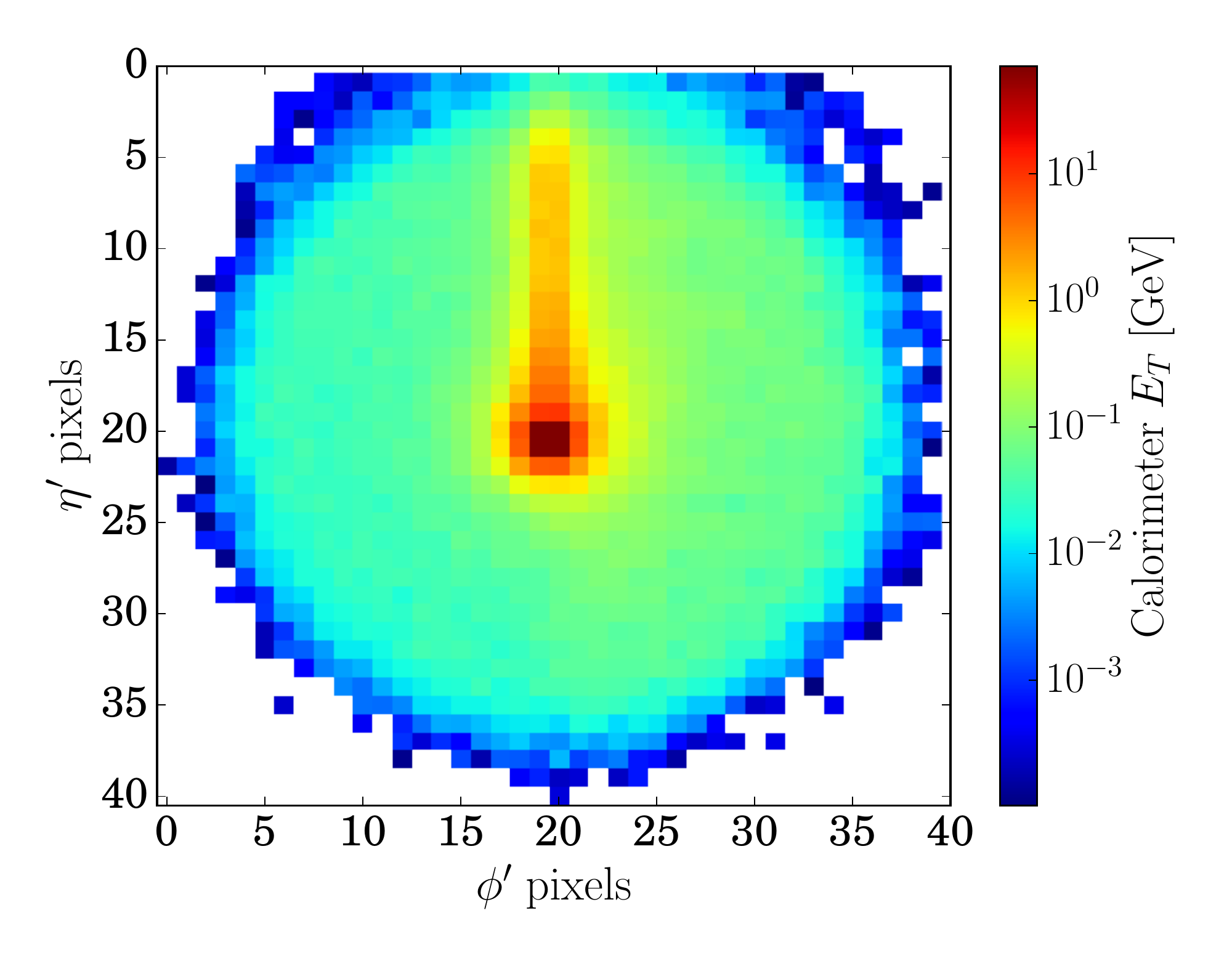}
  \caption{Jet image after pre-processing for the signal (left) and
    background (right). Each picture is averaged over 10,000 actual
    images.}
  \label{fig:averaged_images}
\end{figure} 

We can speed up the learning process or illustrate the ConvNet
performance by applying a set of pre-processing steps:
\begin{enumerate}
\item Find maxima: before we can align any image we have to identify
  characteristic points. Using a filter of size $3\times3$ pixels, we
  localize the three leading maxima in the image;
\item Shift: we then shift the image to center the global maximum
  taking into account the periodicity in the azimuthal angle direction;
\item Rotation: next, we rotate the image such that the second maximum
  is in the 12 o'clock position. The interpolation is done linearly;
\item Flip: next we flip the image to ensure the third maximum
  is in the right half-plane;
\item Crop: finally, we crop the image to $40 \times 40$
  pixels.
\end{enumerate}
Throughout the paper we will apply two pre-processing setups: for
minimal pre-processing we apply steps~1, 2 and~5 to define a centered
jet image of given size. Alternatively, for full pre-processing we
apply all five steps.  In figure~\ref{fig:averaged_images} we show
averaged signal and background images based on the transverse energy
from 10,000 individual images after full pre-processing.  The leading
subjet is in the center of the image, the second subjet is in the 12
o'clock position, and a third subjet from the top decay is smeared
over the right half of the signal images. These images indicate that
fully pre-processed images might lose a small amount of information
at the end of the 12 o'clock axis.\medskip

\begin{figure}[t]
  \includegraphics[width=0.47\textwidth]{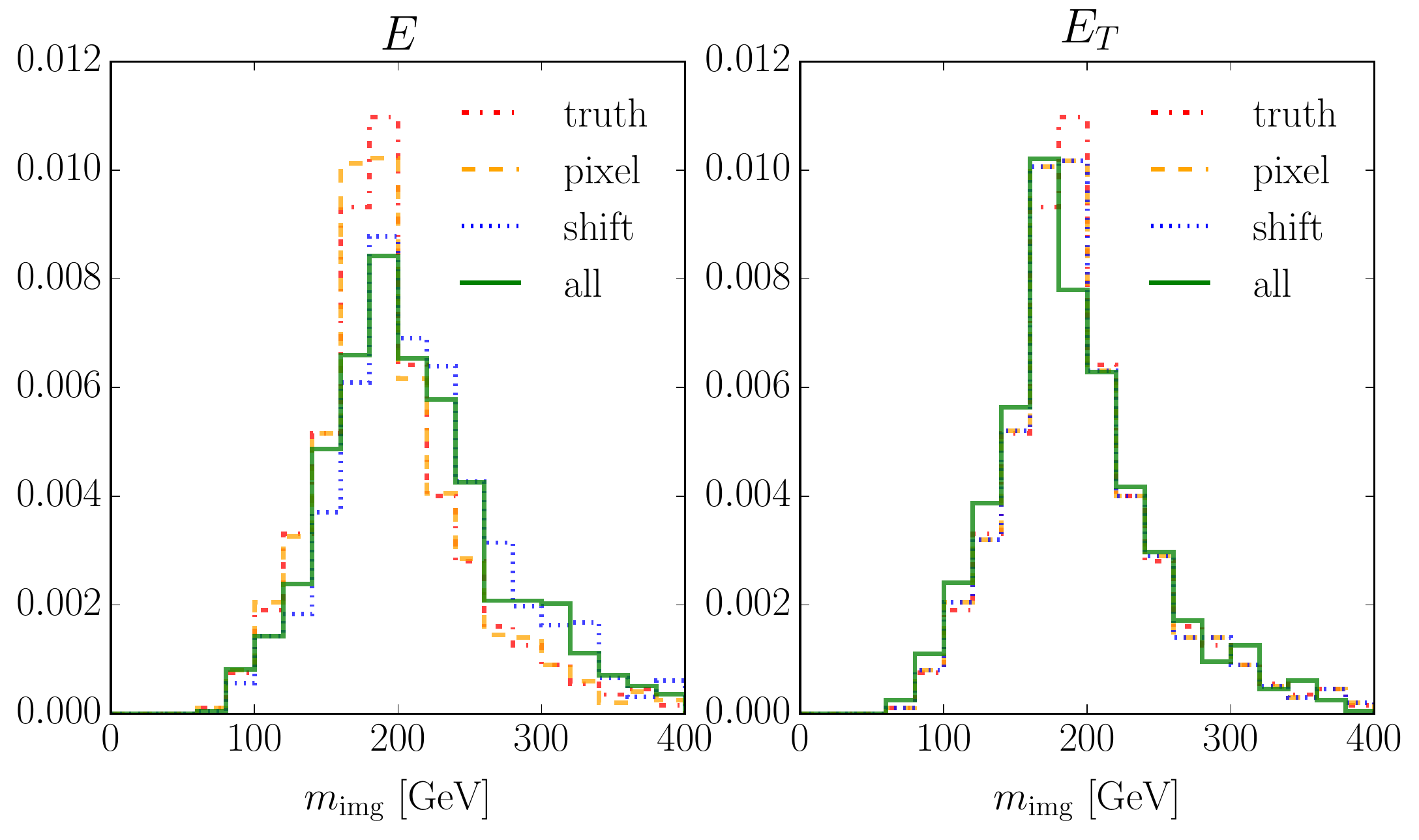}
  \quad
  \includegraphics[width=0.47\textwidth]{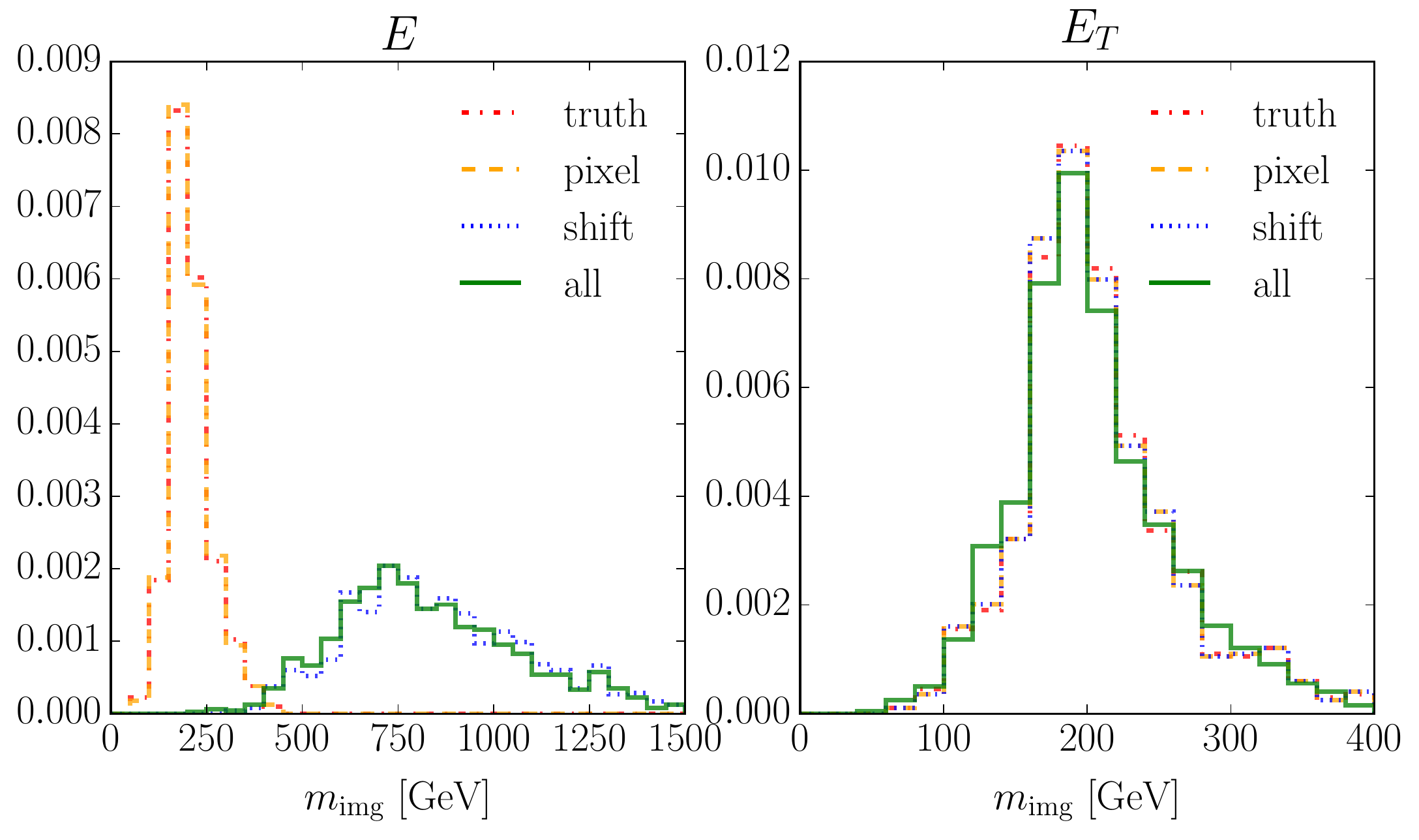}\\
  \includegraphics[width=0.47\textwidth]{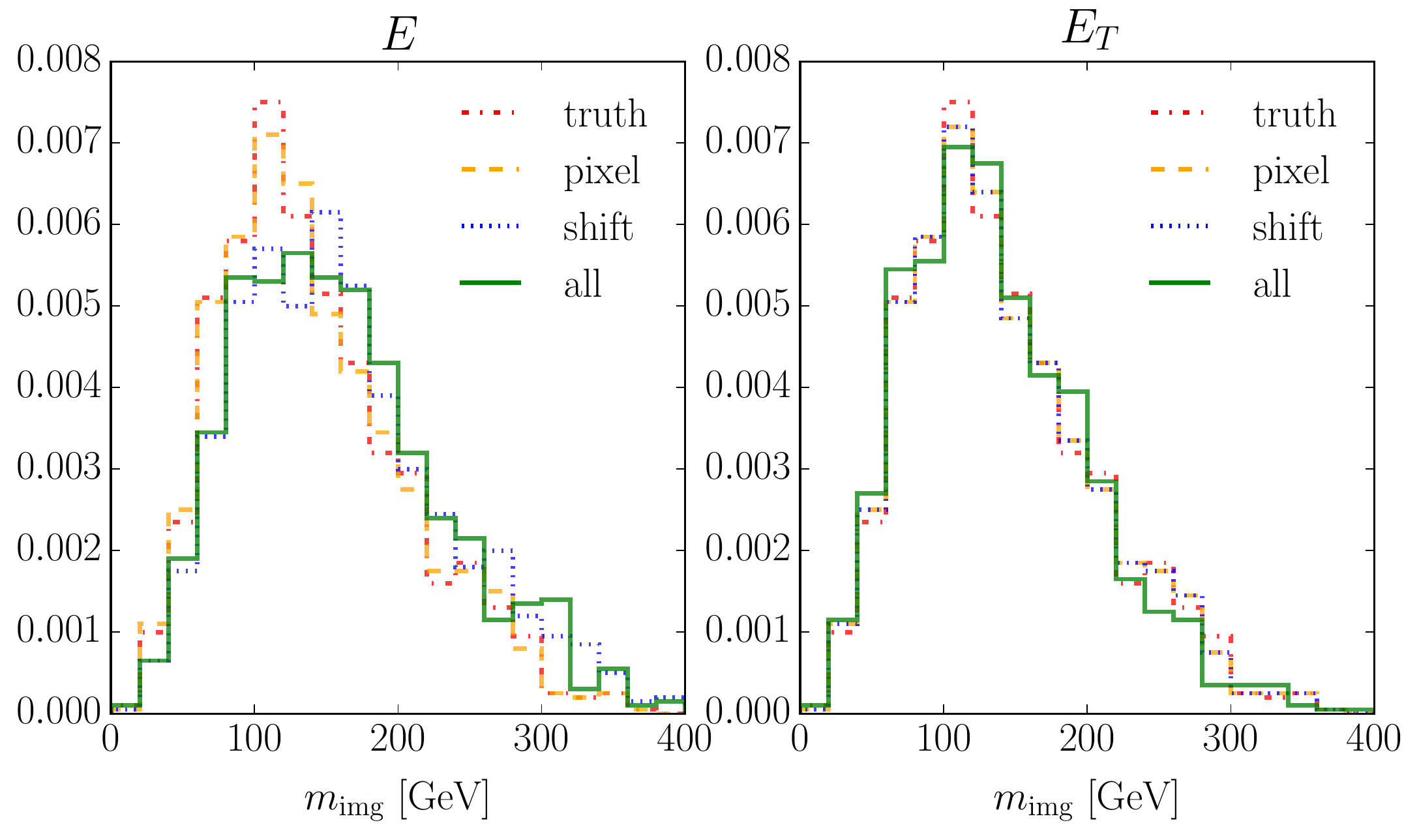}
  \quad
  \includegraphics[width=0.47\textwidth]{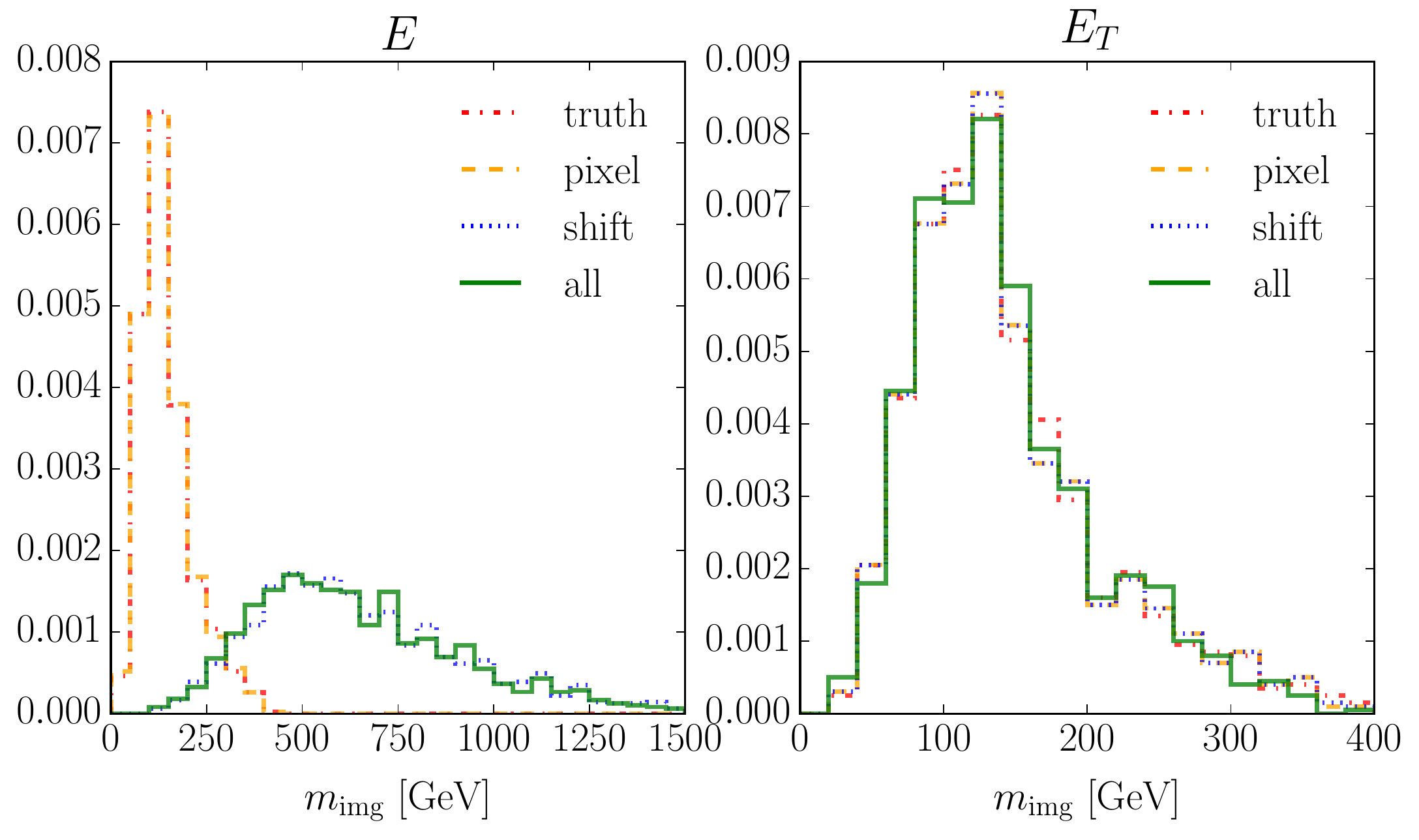}
  \caption{Effect of the preprocessing on the image mass calculated
    from $E$-(left) and $E_T$-images (right) of signal (top) and
    background(bottom). The right set of plots illustrates the
    situation for forward jets with $|\eta| > 2$.}
  \label{fig:img_mass_step}
\end{figure}

A non-trivial pre-processing step is the shift in the $\eta$ direction,
since the jet energy $E$ is not invariant under a longitudinal
boost. Following Ref.~\cite{slac2} we investigate the effect on the
mass information contained in the images,
\begin{align}
  m_\text{img}^2 = \left[ \sum_i E_i \left(1,\ \frac{\cos\phi'_i}{\cosh\eta'_i},\
      \frac{\sin\phi'_i}{\cosh\eta'_i},
      \frac{\sinh\eta'_i}{\cosh\eta'_i}\right)\right]^2 \qquad E_i =
  E_{T,i} \cosh\eta'_i \;,
\end{align}
where $\eta'_i$ and $\phi'_i$ are the center of the $i$th pixel after
pre-processing.  The study of all pre-processing steps and their
effect on the image mass in figure~\ref{fig:img_mass_step} illustrates
that indeed the rapidity shift has the largest effect on the $E$
images, but this effect is not large.  For the $E_T$ images the jet
mass distribution is unaffected by the shift pre-processing step. The
reason why our effect on the $E$ images is much milder than the one
observed in Ref.~\cite{slac2} is our condition $|\eta_\text{fat}| <
1$. In the the lower panels of figure~\ref{fig:img_mass_step} we
illustrate the effect of pre-processing on fat jets with $|\eta| > 2$,
where the image masses changes dramatically. Independent of these
details we use pre-processed $E_T$ images as our machine learning
input~\cite{convnet,sklearn,keras,theano}. Since networks prefer small
numbers, we scale the images to keep most pixel entries between 0 and 1.

\subsection{Network architecture}
\label{sec:machine_net}

\begin{figure}[t]
  \includegraphics[width=0.46\textwidth]{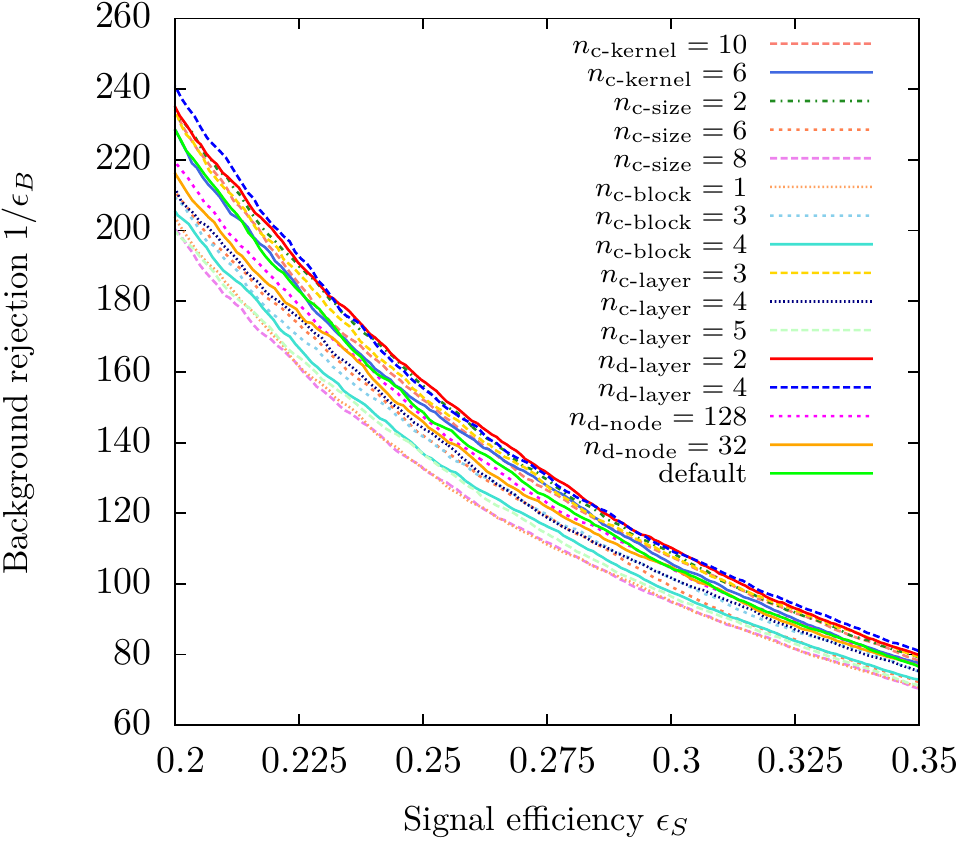}
  \hspace*{0.05\textwidth}
  \includegraphics[width=0.44\textwidth]{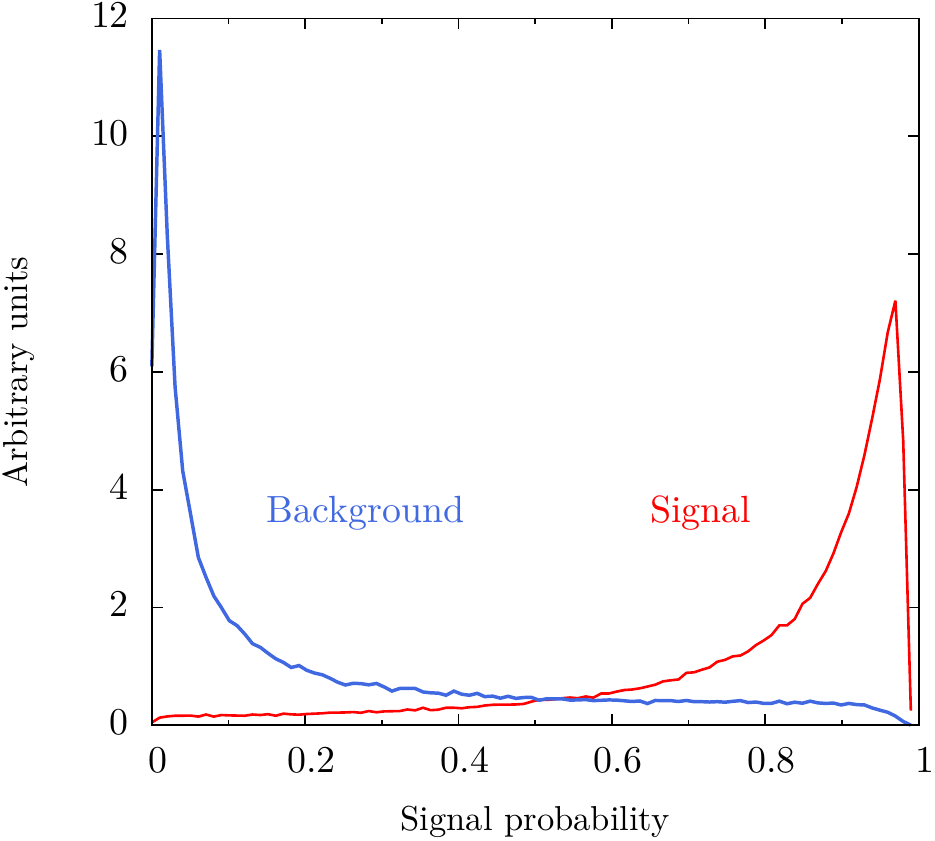}
  \caption{Left: performance of some of our tested architectures for
    full pre-processing in terms of an ROC curve, including the
    default \textsc{DeepTop} network. Right: discrimination power or
    predicted signal probability for signal events and background probability for
    background events. We use the default network.}
  \label{fig:arc_scan}
\end{figure}

To identify a suitable \textsc{DeepTop} network architecture, we scan
over several possible realizations or hyper-parameters.  As discussed
in the last section, we start with jet images of size $40 \times 40$.
For architecture testing we split our total signal and background
samples of 600,000 images each into three sub-samples: 
training (150,000 signal and background events each),
validation/optimization (150,000 signal and background events each),
and final test (300,000 signal and background events each).
Networks are trained on the training sample. No early stopping is
performed, but the set of weights minimizing loss on the
validation/optimization sample is used to avoid overfitting.


In a first step we need to optimize our network architecture.  The
ConvNet side is organized in $n_\text{c-block}$ blocks, each
containing $n_\text{c-layer}$ sequences of ZeroPadding, Convolution
and Activation steps.  For activation we choose the ReLU step function 
while weights are initialized by drawing from a Glorot uniform 
distribution~\cite{glorot}.
Inside each block the size of the feature maps can be slightly reduced
due to boundary effects. For each convolution we globally set a filter
size or convolutional size $n_\text{c-size} \times n_\text{c-size}$.
The global number of kernels of corresponding feature maps is given by
$n_\text{c-kernel}$. Two blocks are separated by a pooling step, in
our case using MaxPooling, which significantly reduces the size of the
feature maps. For a quadratic pool size of $p \times p$ fitting into
the $n \times n$ size of each feature map, the initial size of the new
block's input feature maps is $n/p \times n/p$.  The final output
feature maps are used as input to a DNN with $n_\text{d-layer}$ fully
connected layers and $n_\text{d-node}$ nodes per layer.

\begin{table}[b!]
\begin{center}
\begin{tabular}{ l | c | c}
\hline
hyper-parameter & scan range & default \\
\hline
$n_\text{c-block}$  & 1,2,3,4  & 2 \\
$n_\text{c-layer}$  & 2,3,4,5  & 2 \\
$n_\text{c-kernel}$ & 6,8,10   & 8 \\
$n_\text{c-size}$   & 2,4,6,8   & 4 \\
$n_\text{d-layer}$  & 2,3,4     & 3 \\
$n_\text{d-nodes}$ & 32,64,128  & 64 \\
$p$                & 0,2,4     & 2  \\
\hline
\end{tabular}
\caption{Range of parameters defining the combined ConvNet and DNN
  architecture, leading to the range of efficiencies shown in the left
  panel of figure~\ref{fig:arc_scan} for fully pre-processed images.}
\label{tab:arc_best}
\end{center}
\end{table}

\begin{figure}[t]
  \includegraphics[width=\textwidth]{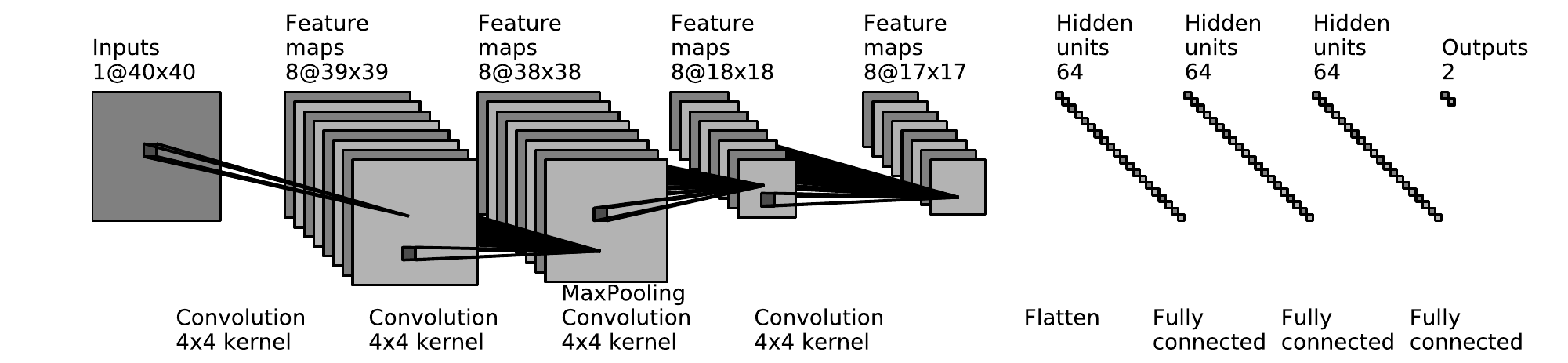} 
  \caption{Architecture~\cite{draw_convnet} of our default networks for fully
    pre-processed images, defined in Tab.~\ref{tab:arc_best}.}
  \label{fig:arc_best}
\end{figure}

In the left panel of figure~\ref{fig:arc_scan} we show the performance
of some test architectures. We give the complete list of tested
hyper-parameters in Tab.~\ref{tab:arc_best}. As our default we choose
one of the best-performing networks on the validation/optimization sample after explicitly ensuring its
stability with respect to changing its hyper-parameters. The
hyper-parameters of the default network we use for fully as well as
minimally pre-processed images are given in Tab.~\ref{tab:arc_best}.
In figure~\ref{fig:arc_best} we illustrate this default
architecture.\medskip

\begin{figure}[b!]
  \includegraphics[width=0.119\textwidth]{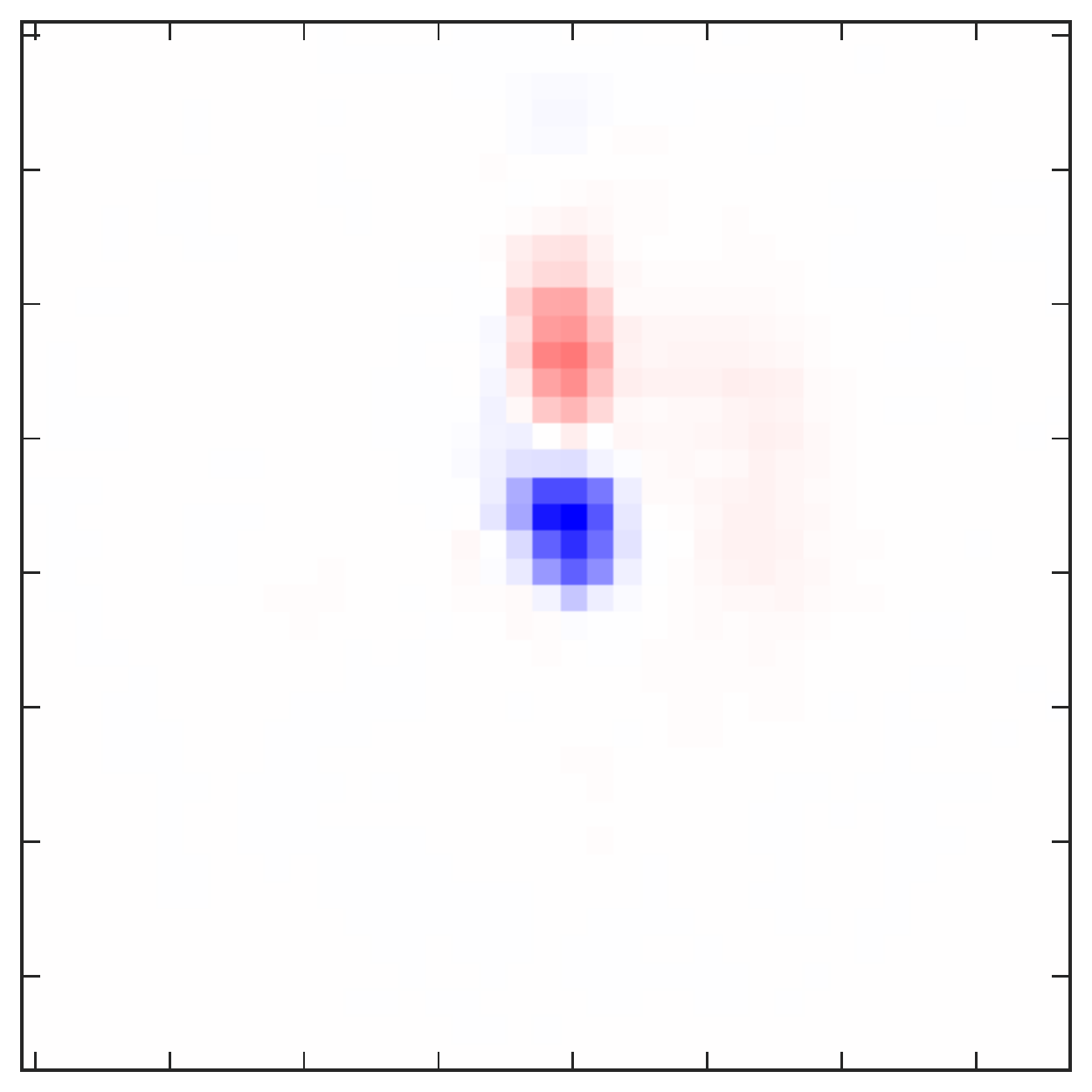}
  \includegraphics[width=0.119\textwidth]{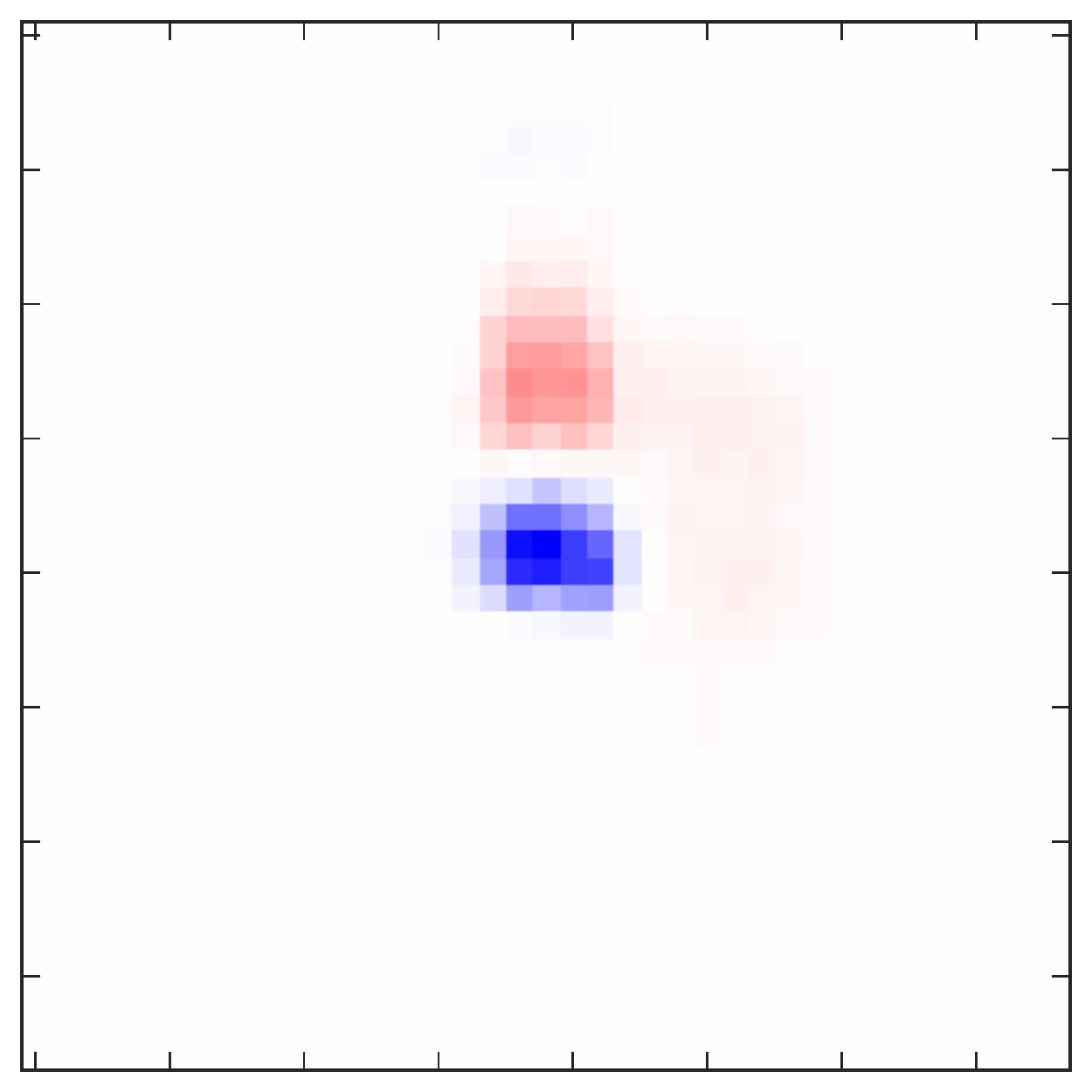}
  \includegraphics[width=0.119\textwidth]{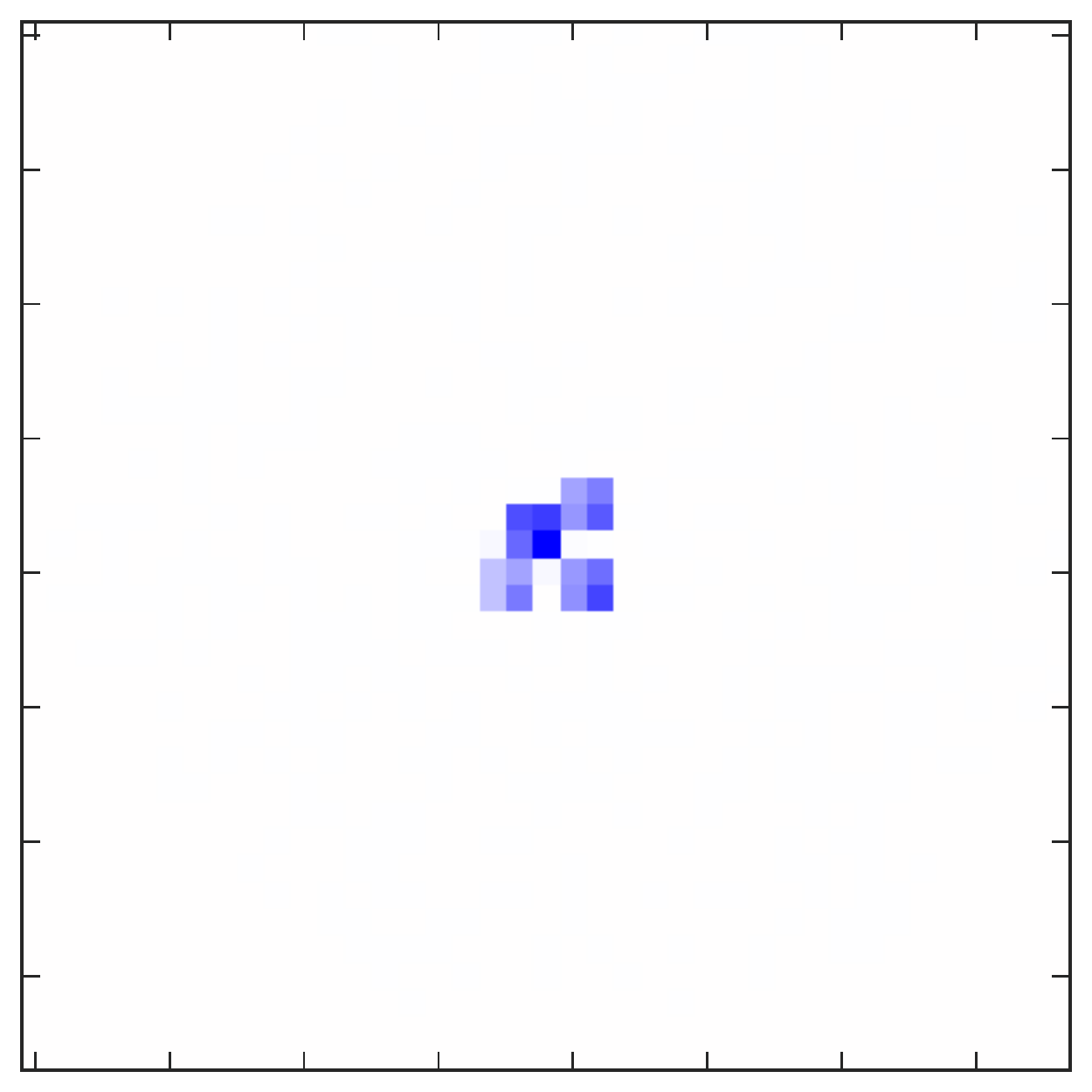}
  \includegraphics[width=0.119\textwidth]{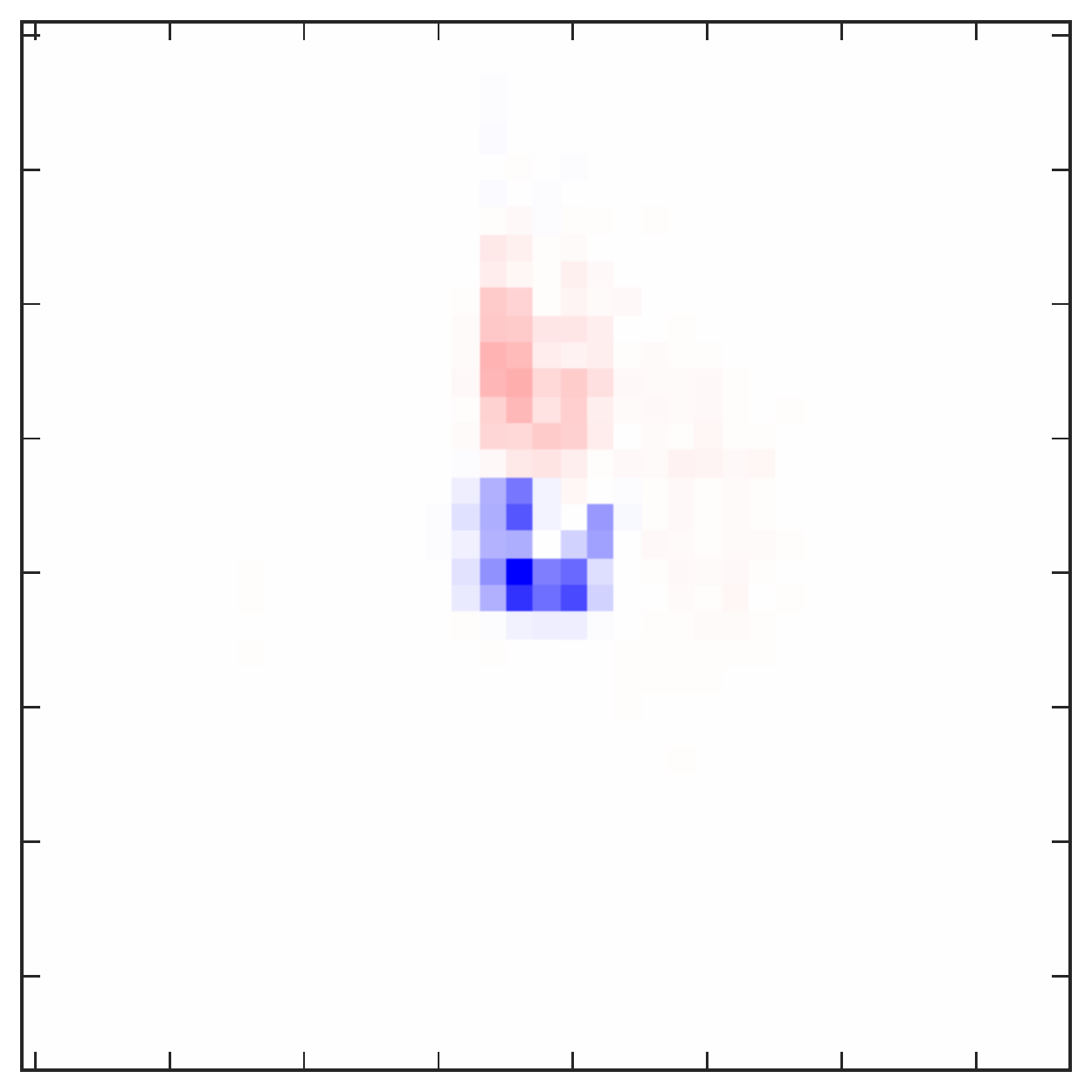}
  \includegraphics[width=0.119\textwidth]{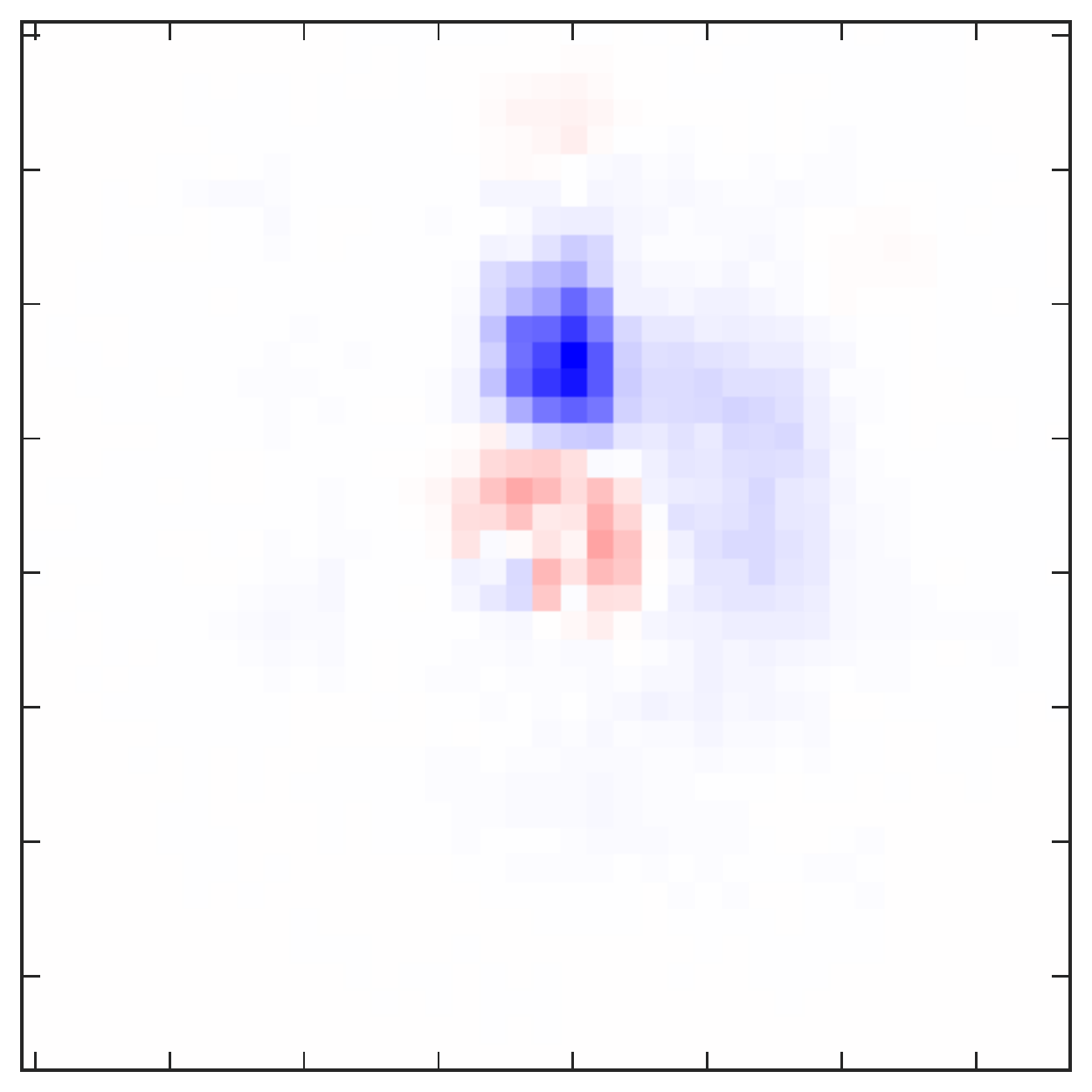}
  \includegraphics[width=0.119\textwidth]{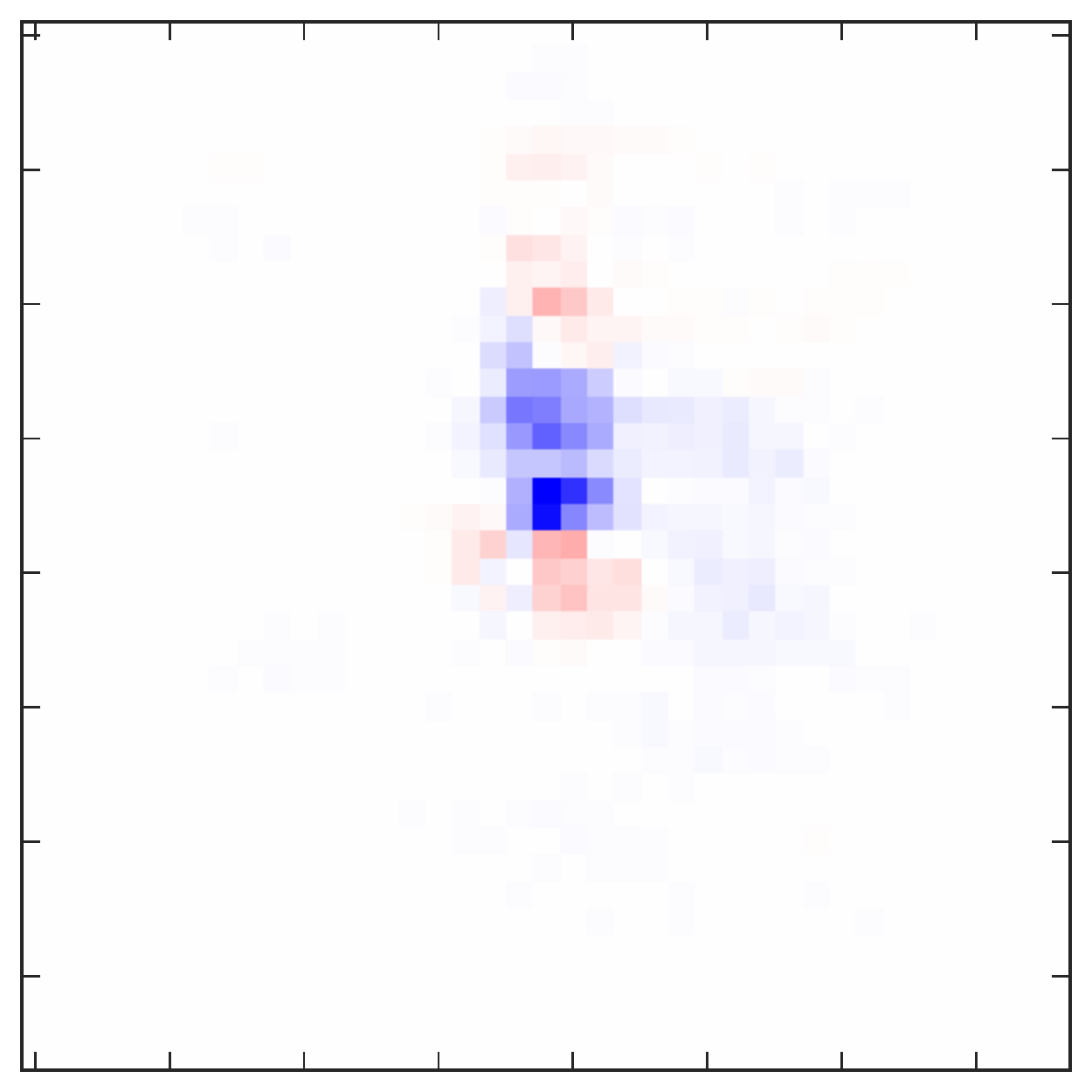} 
  \includegraphics[width=0.119\textwidth]{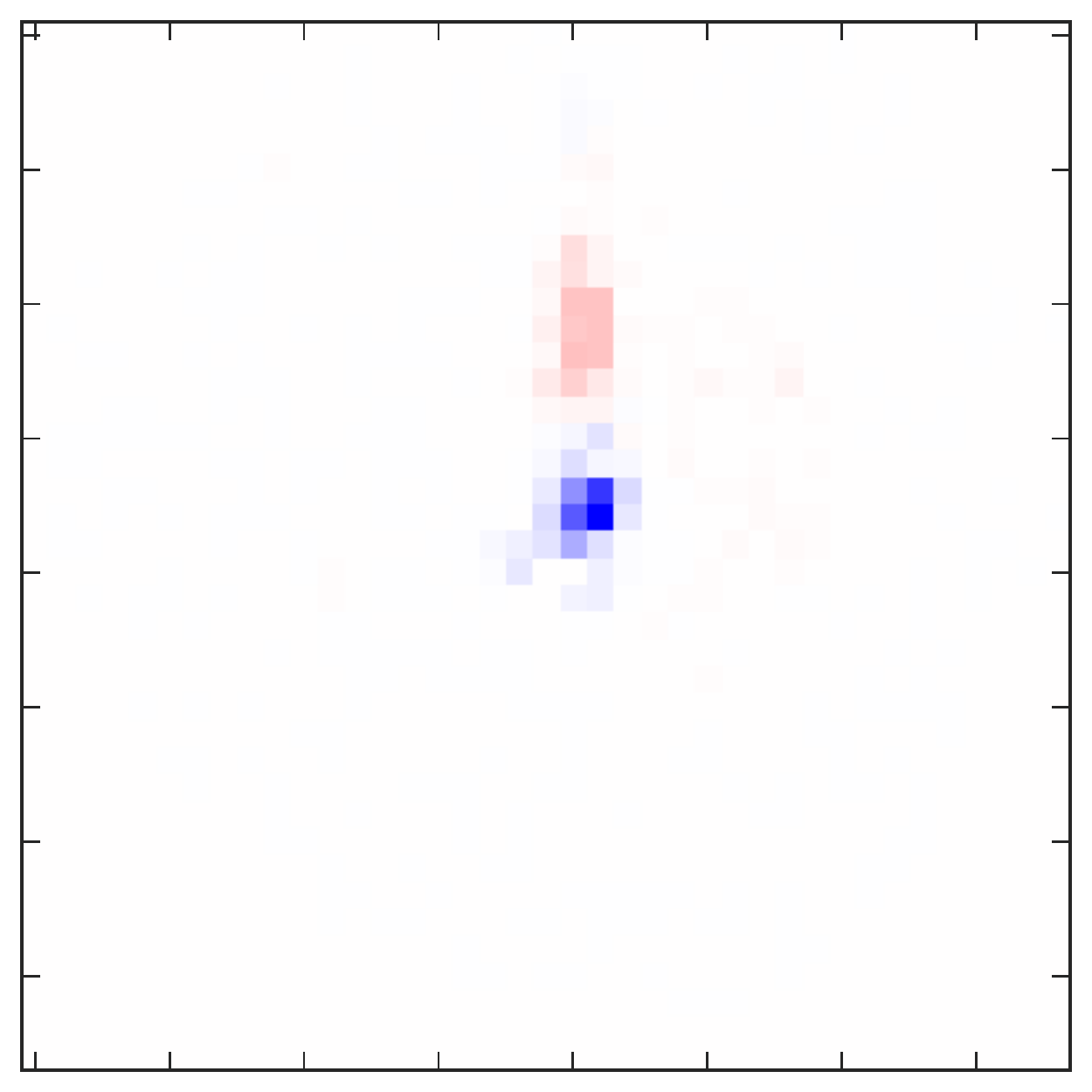} 
  \includegraphics[width=0.119\textwidth]{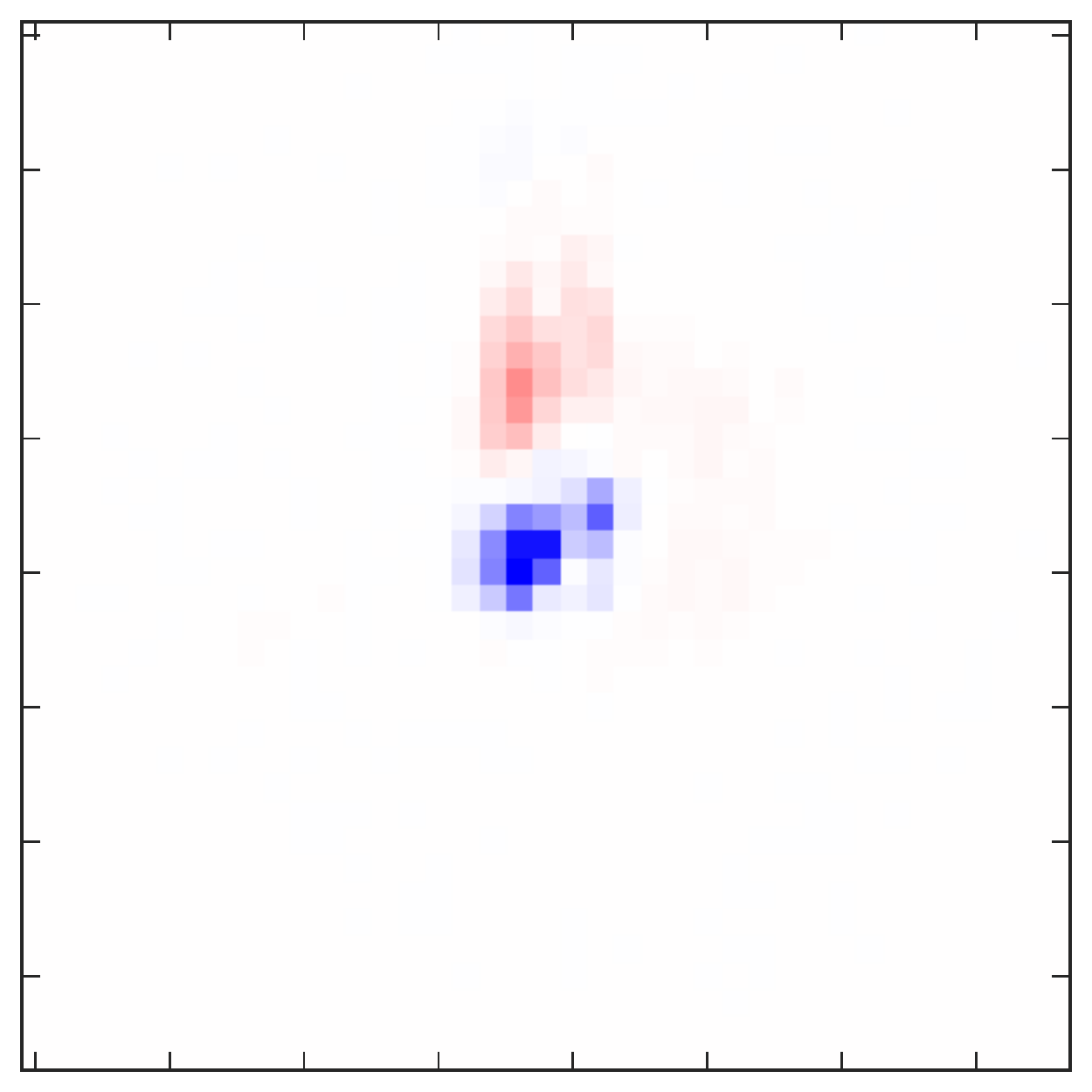} \\
  \includegraphics[width=0.119\textwidth]{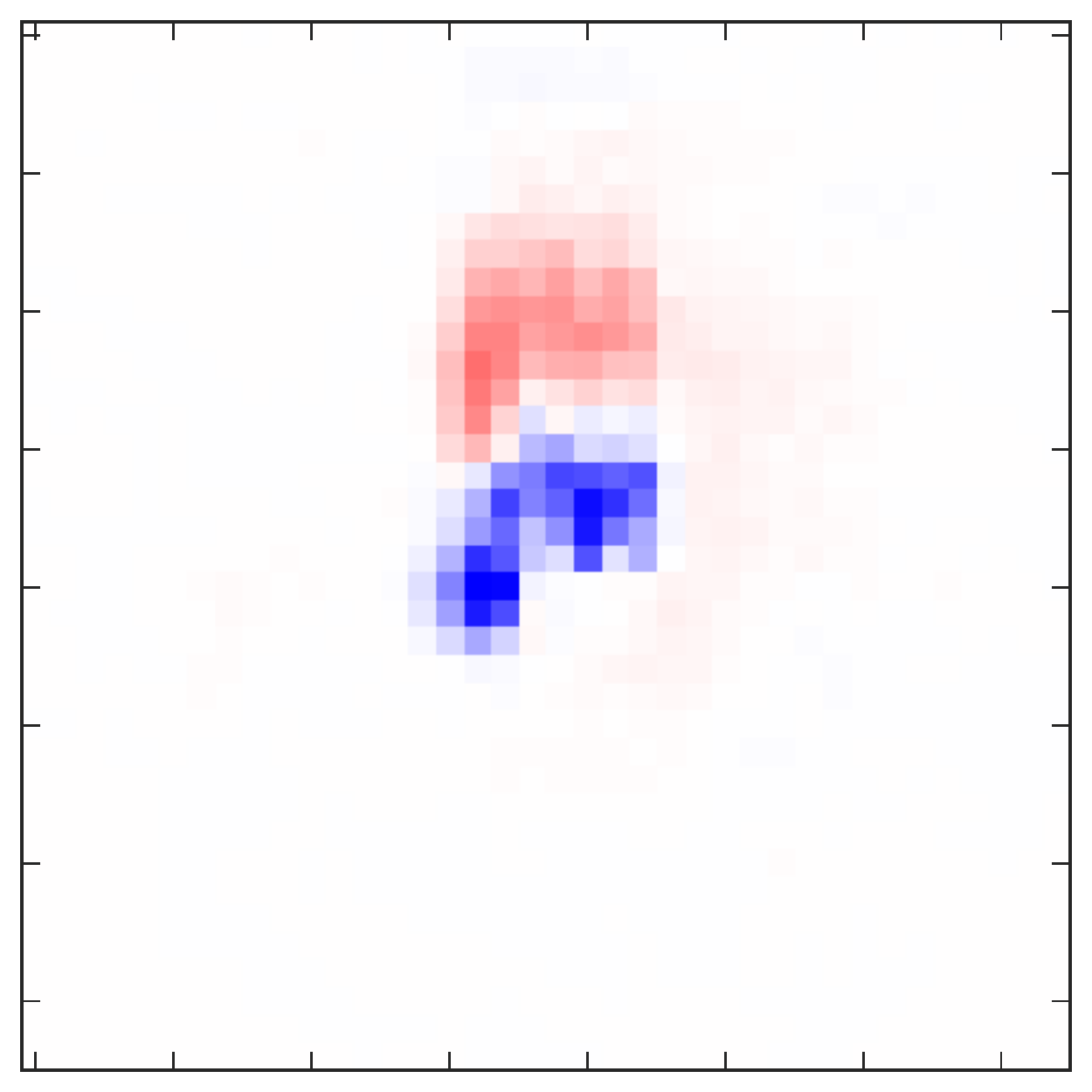}
  \includegraphics[width=0.119\textwidth]{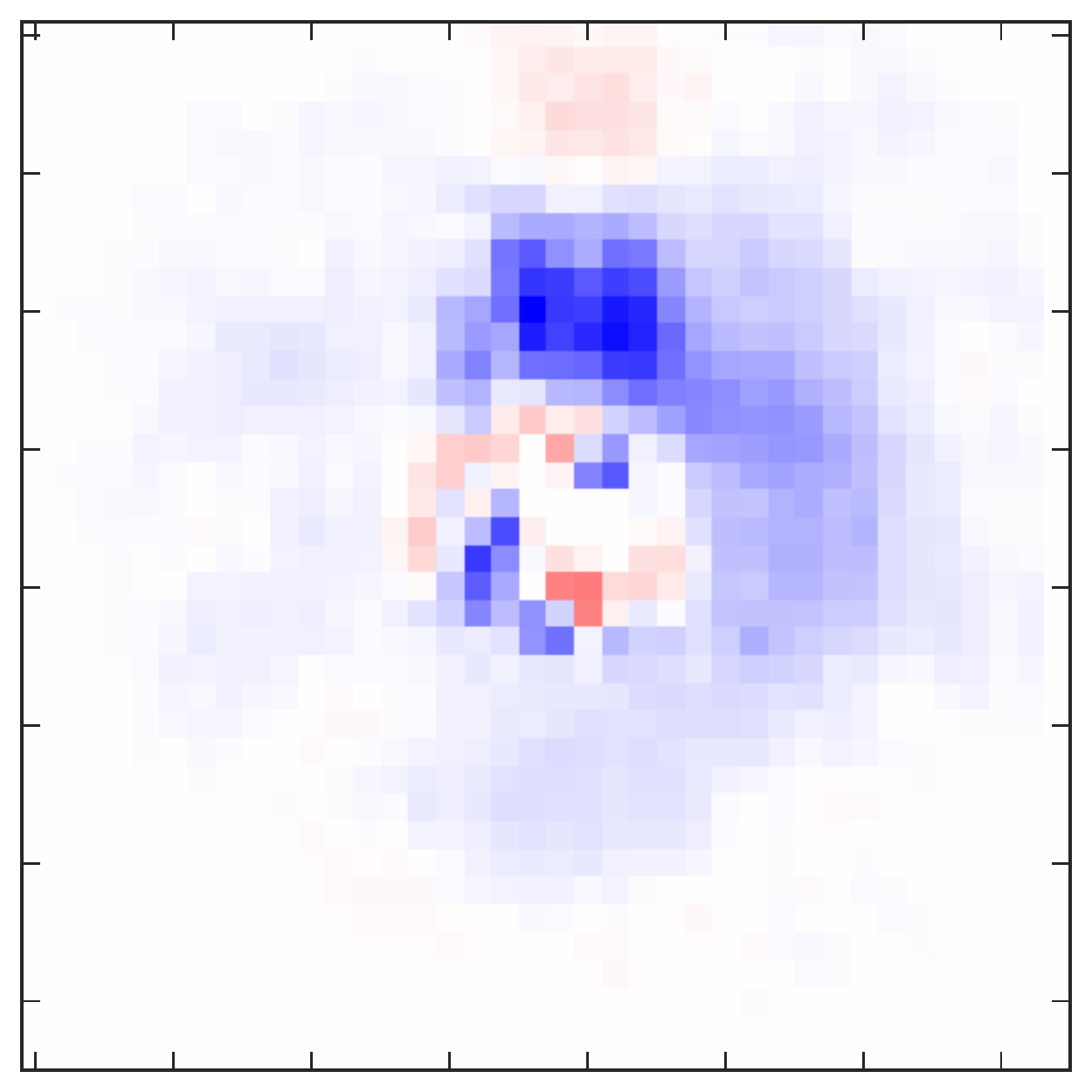}
  \includegraphics[width=0.119\textwidth]{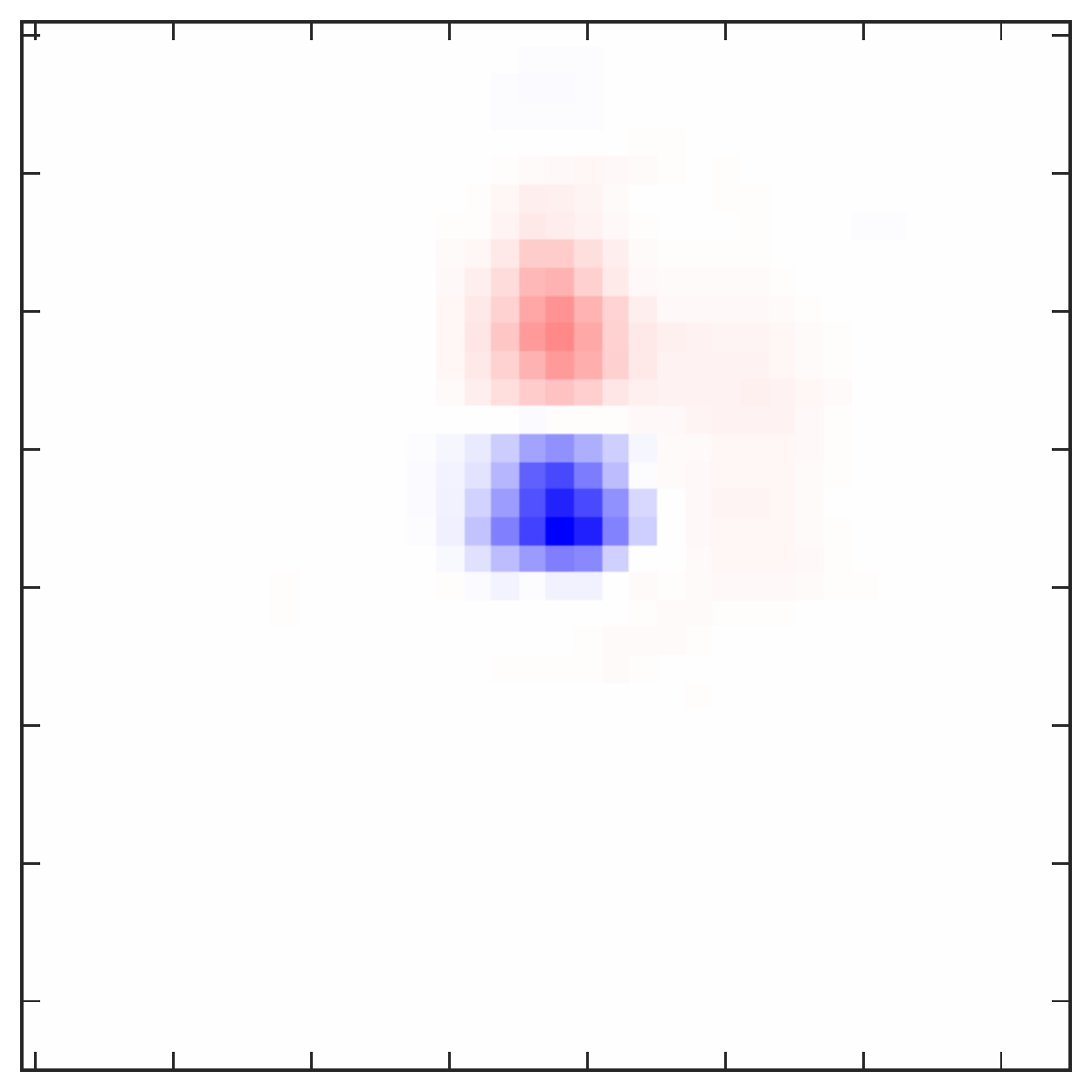}
  \includegraphics[width=0.119\textwidth]{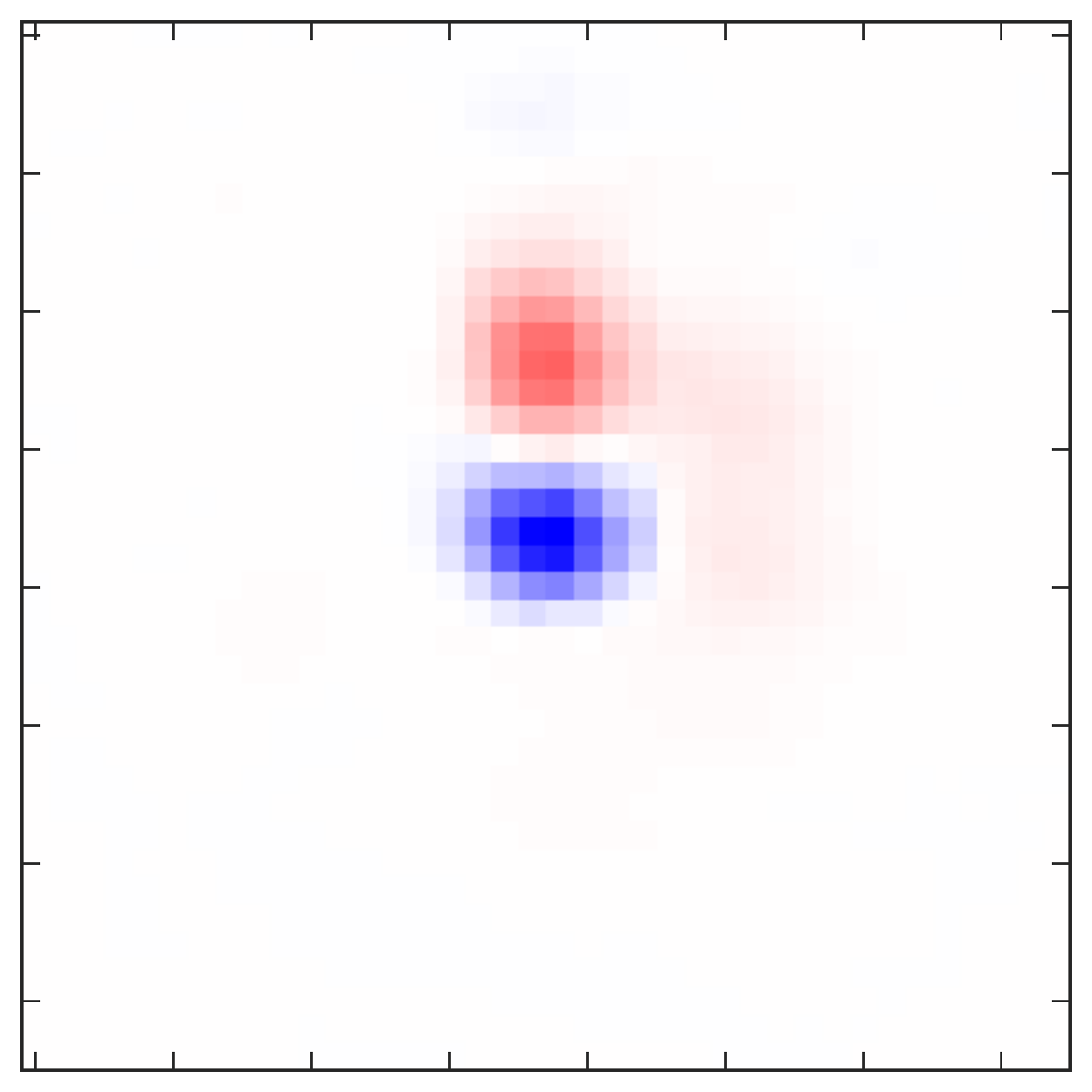}
  \includegraphics[width=0.119\textwidth]{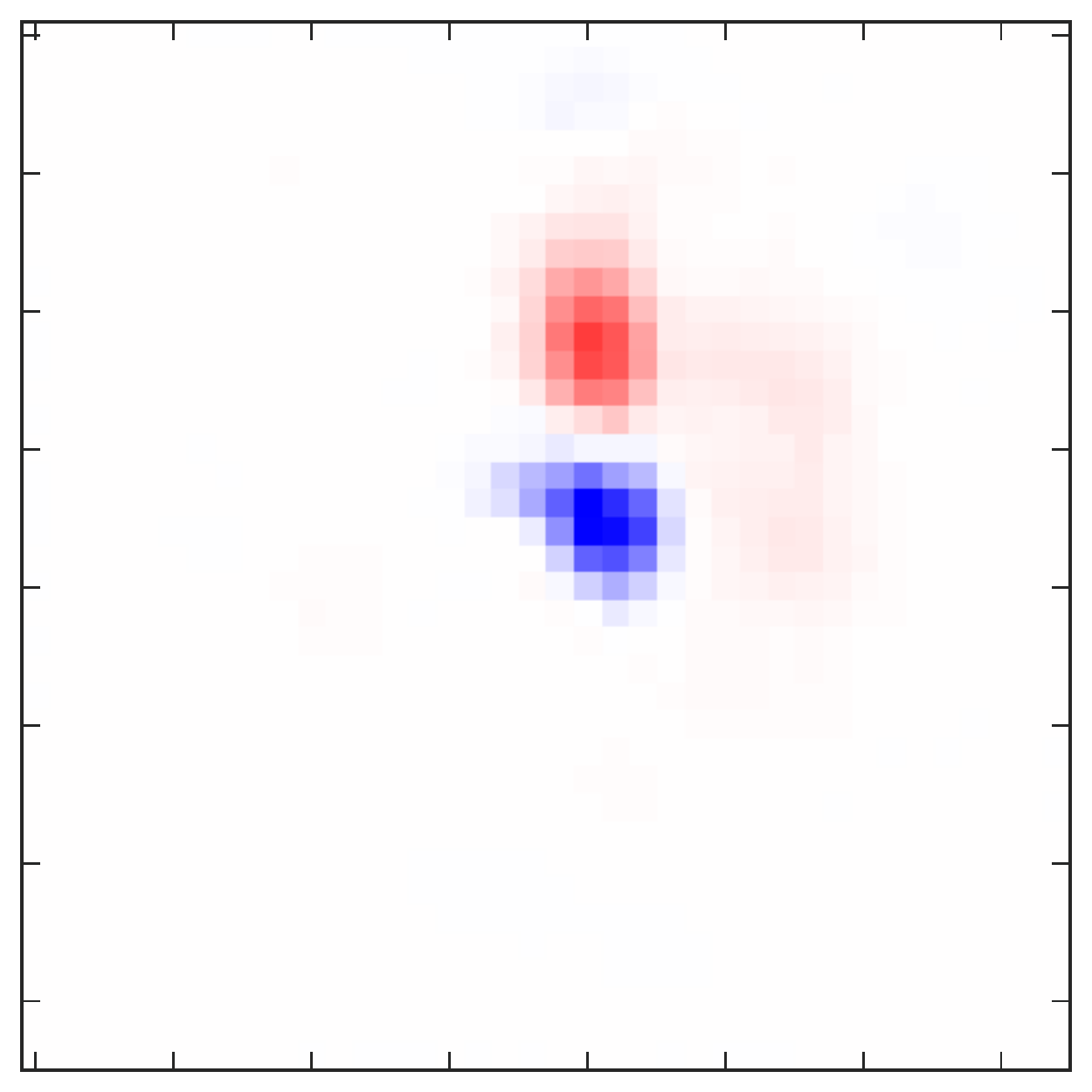}
  \includegraphics[width=0.119\textwidth]{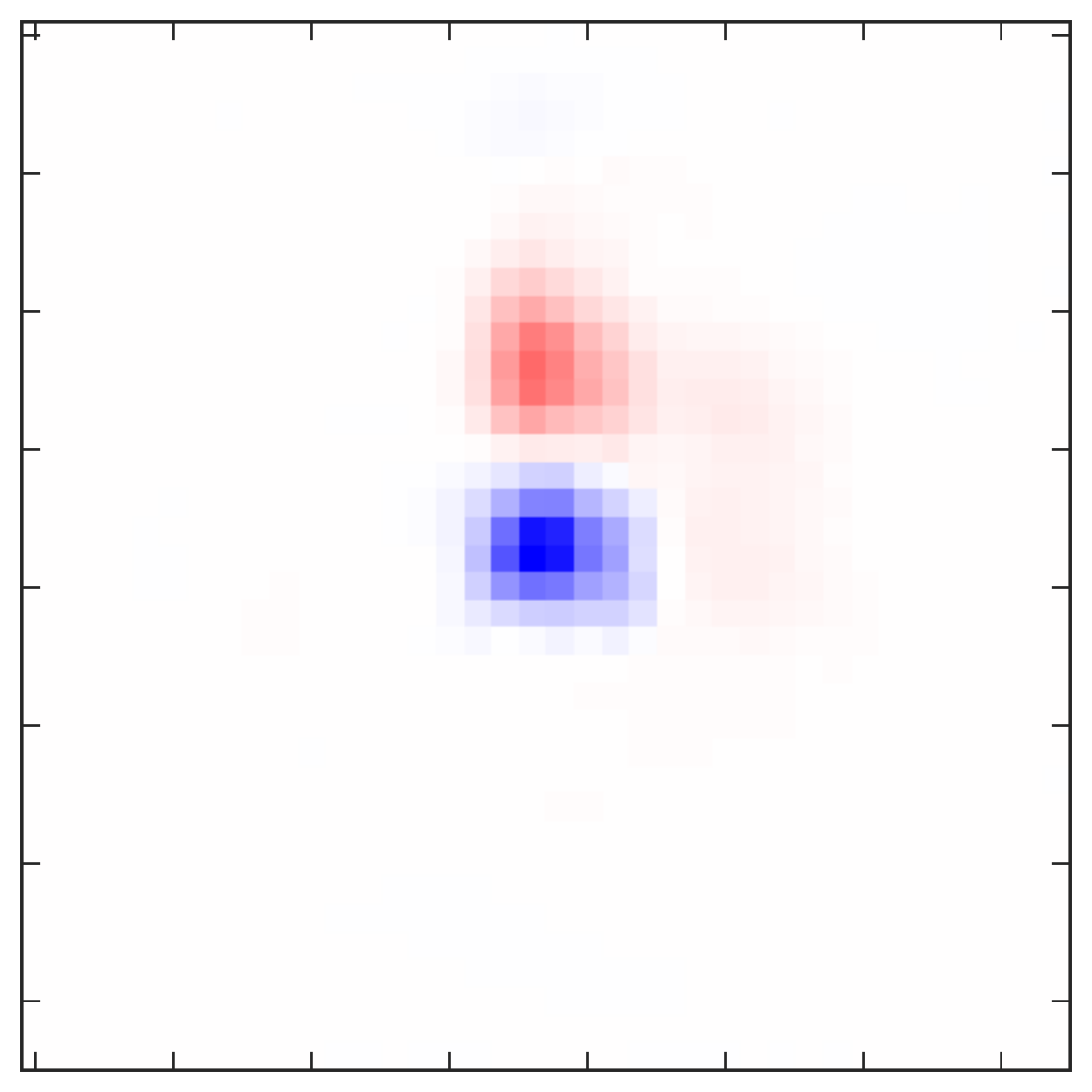} 
  \includegraphics[width=0.119\textwidth]{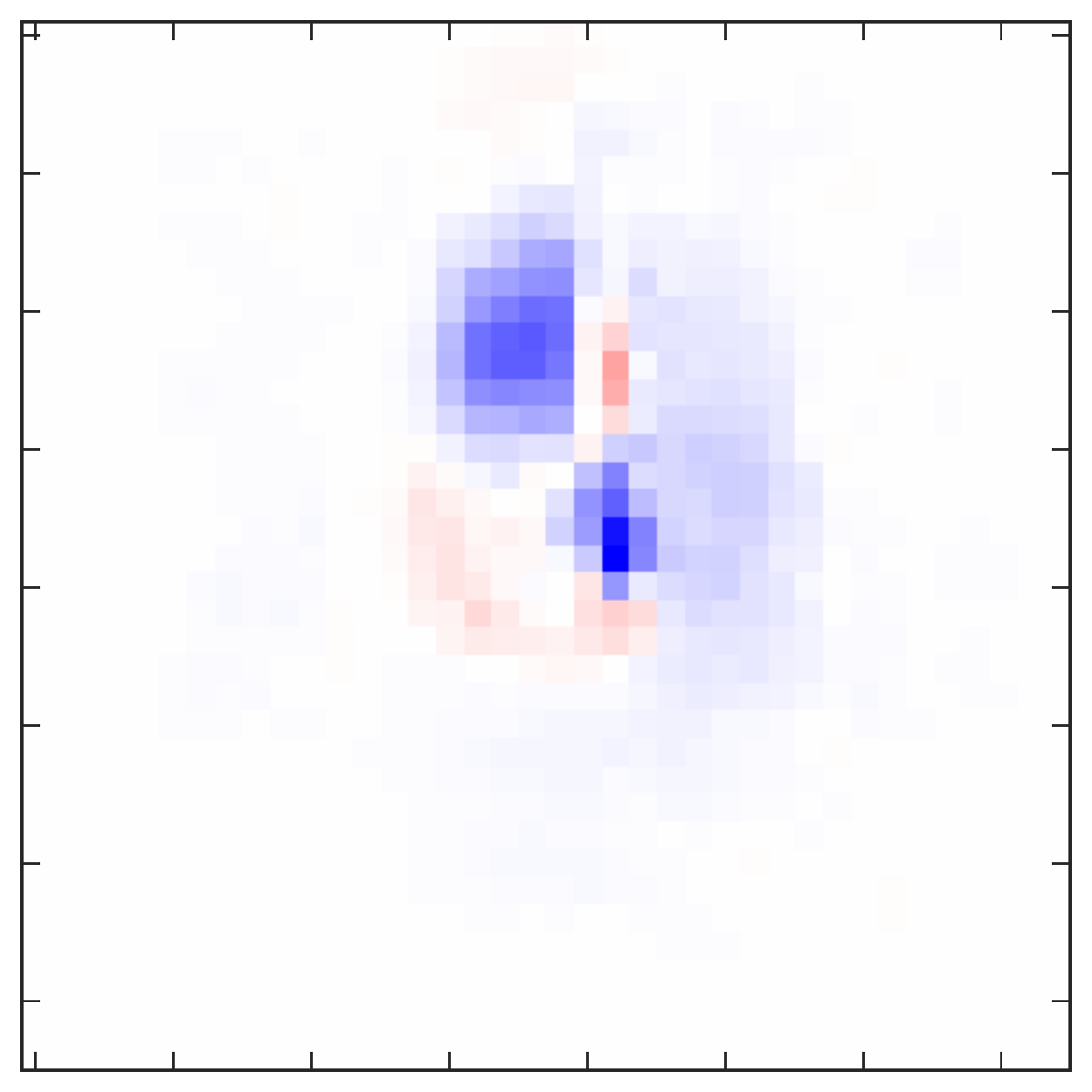} 
  \includegraphics[width=0.119\textwidth]{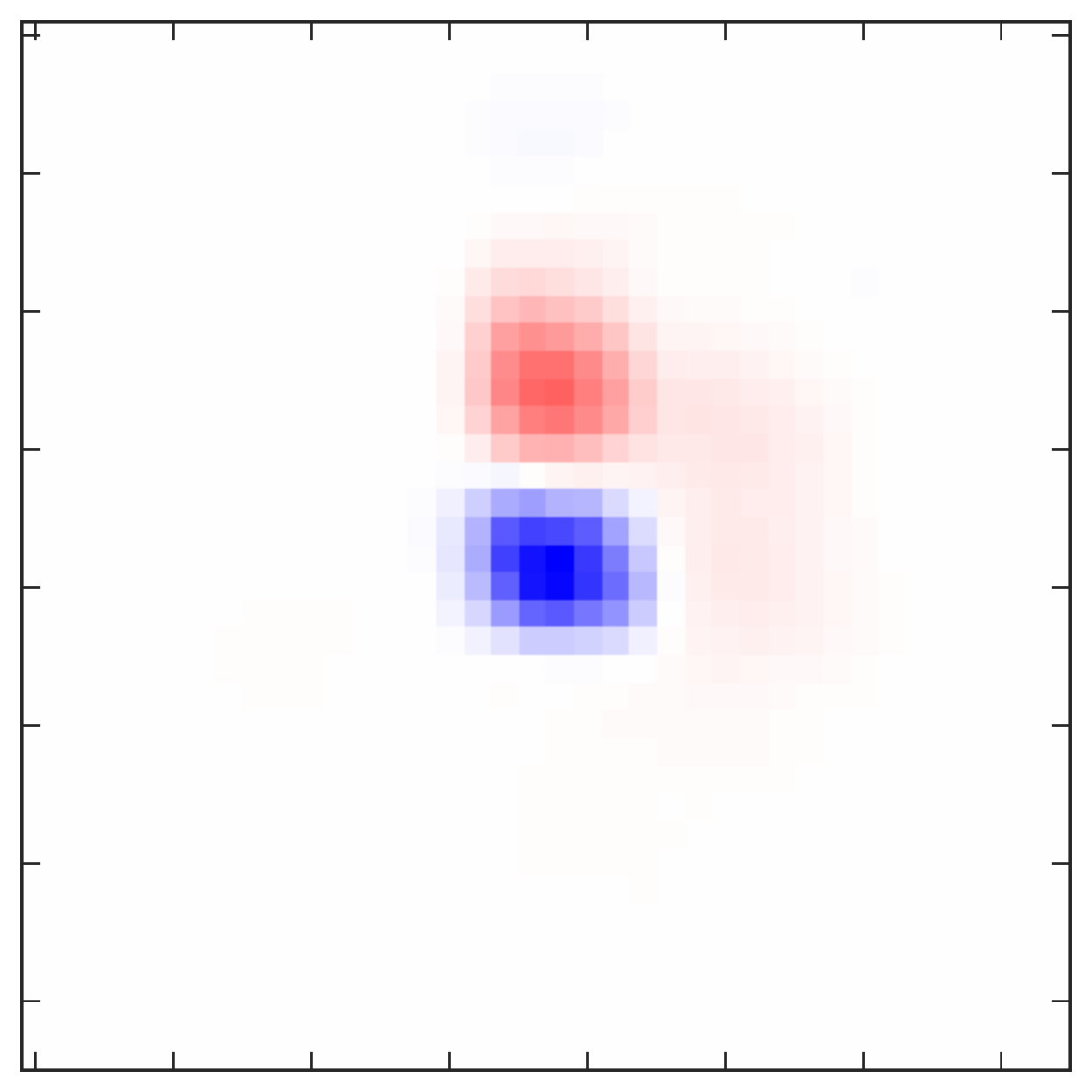} \\
  \includegraphics[width=0.119\textwidth]{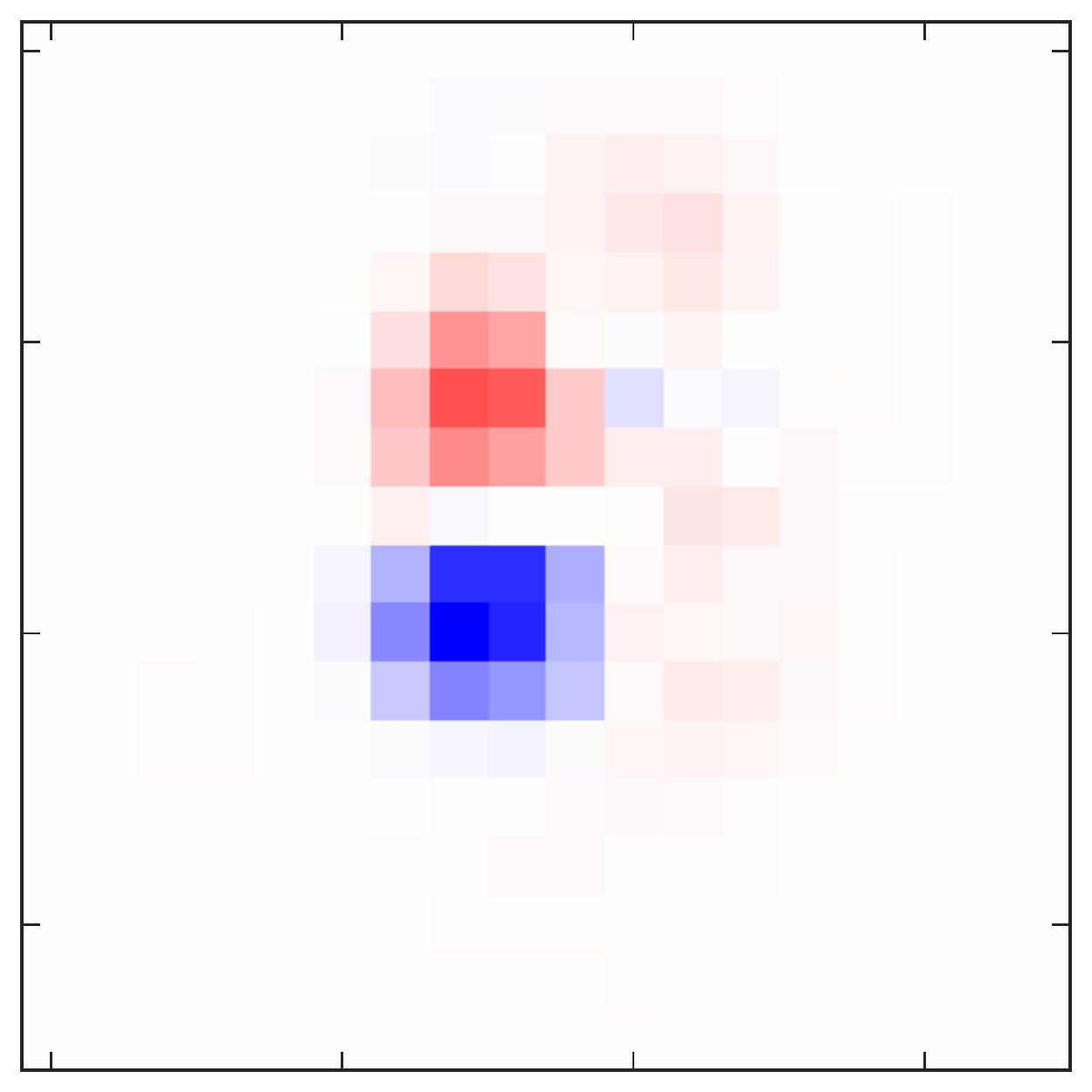}
  \includegraphics[width=0.119\textwidth]{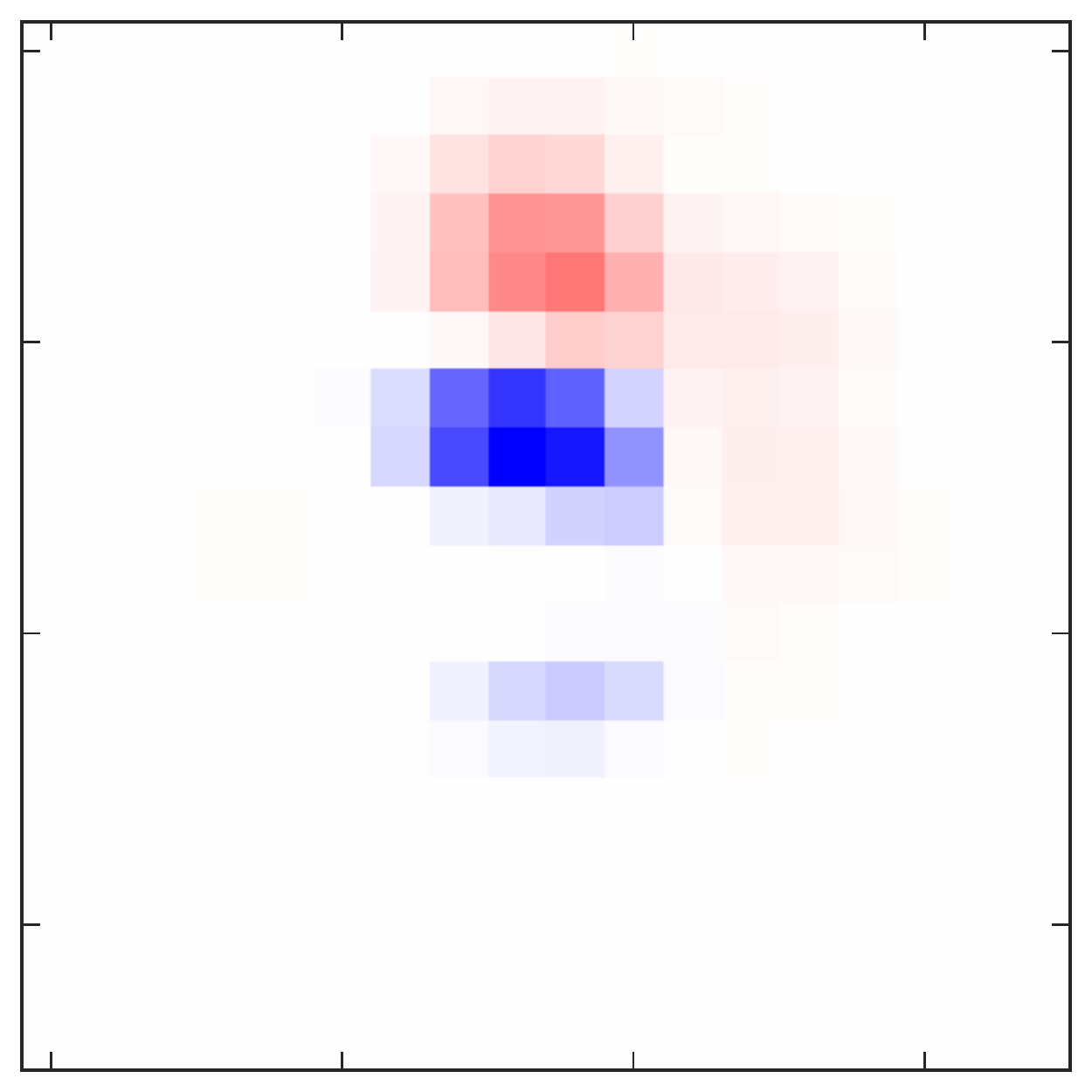}
  \includegraphics[width=0.119\textwidth]{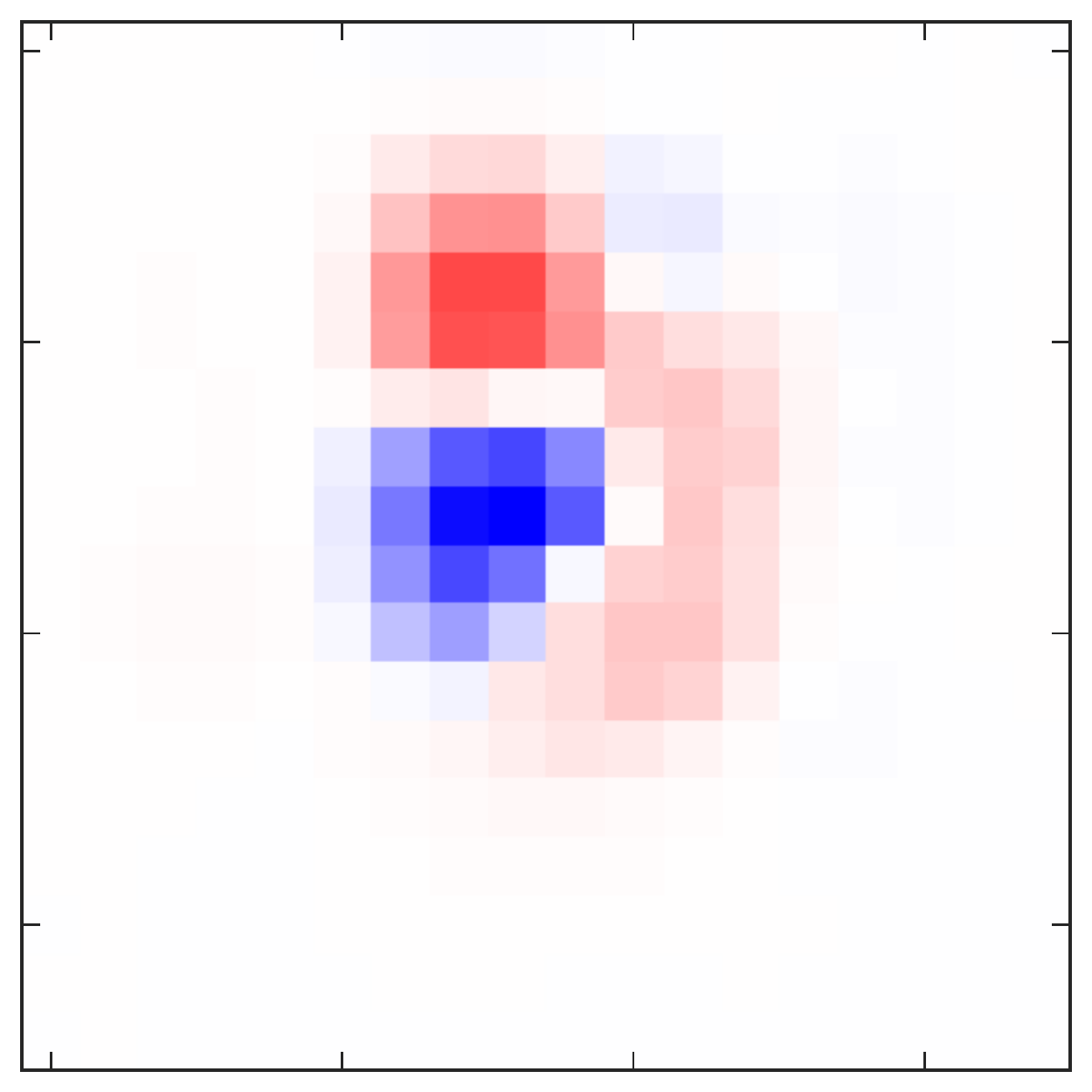}
  \includegraphics[width=0.119\textwidth]{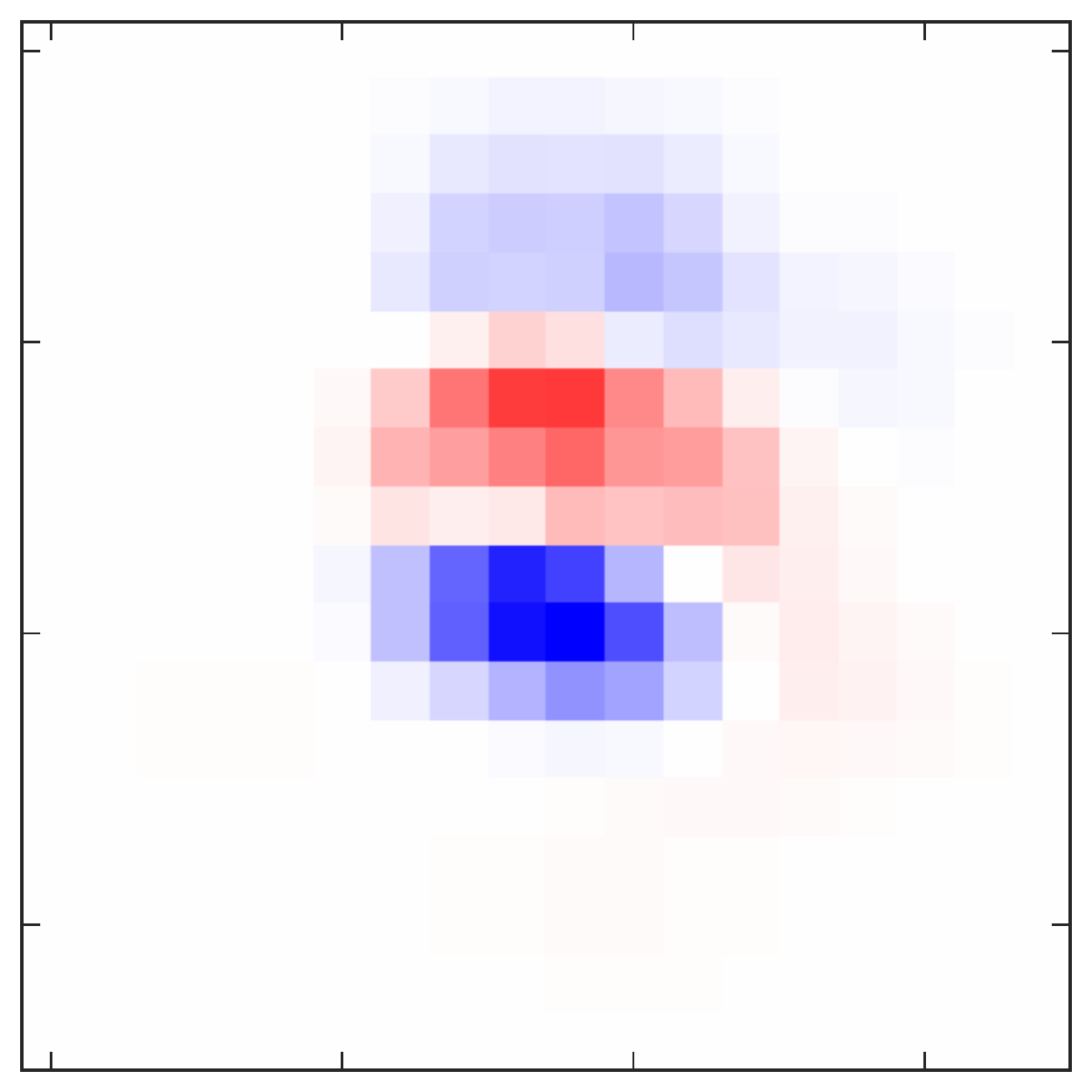}
  \includegraphics[width=0.119\textwidth]{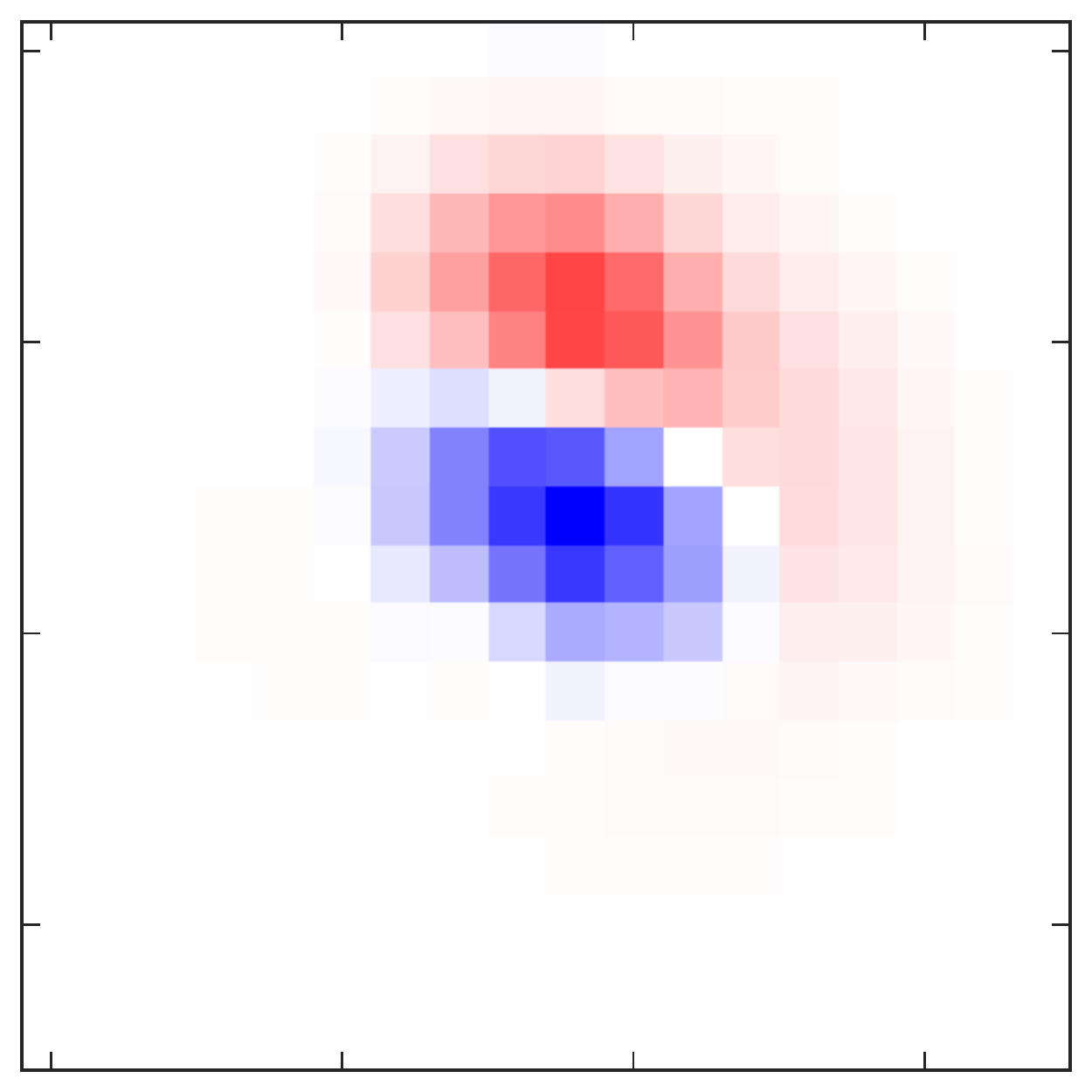}
  \includegraphics[width=0.119\textwidth]{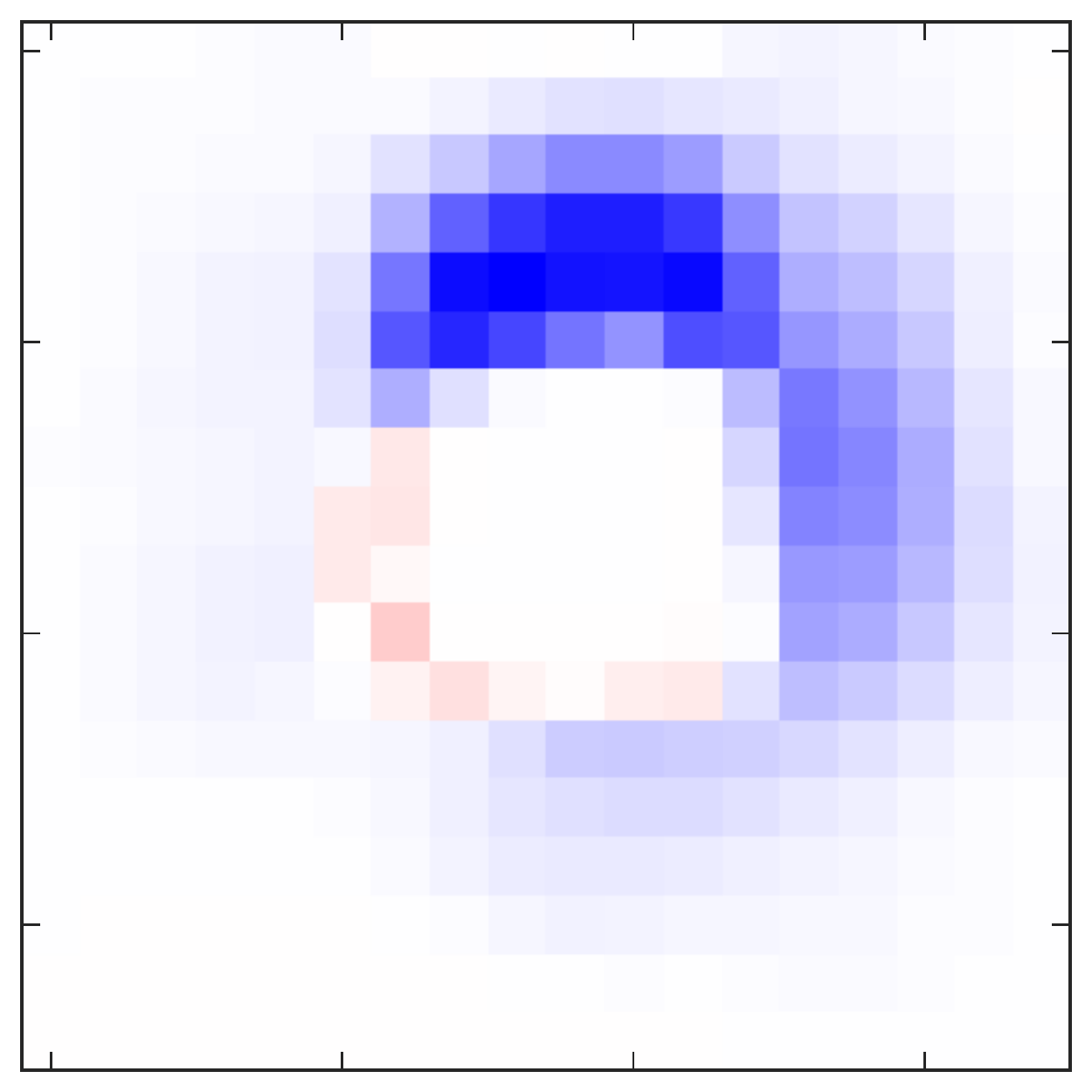} 
  \includegraphics[width=0.119\textwidth]{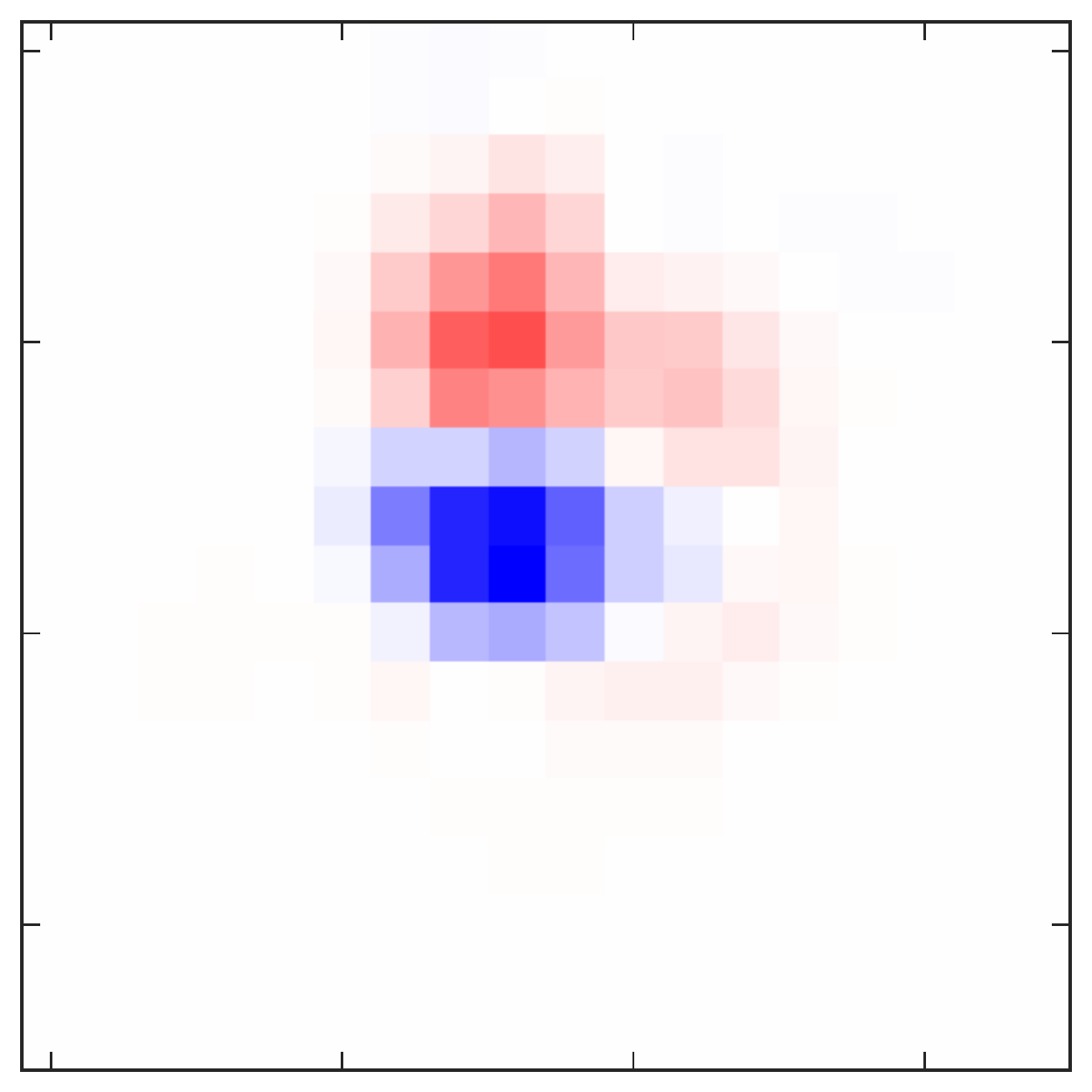} 
  \includegraphics[width=0.119\textwidth]{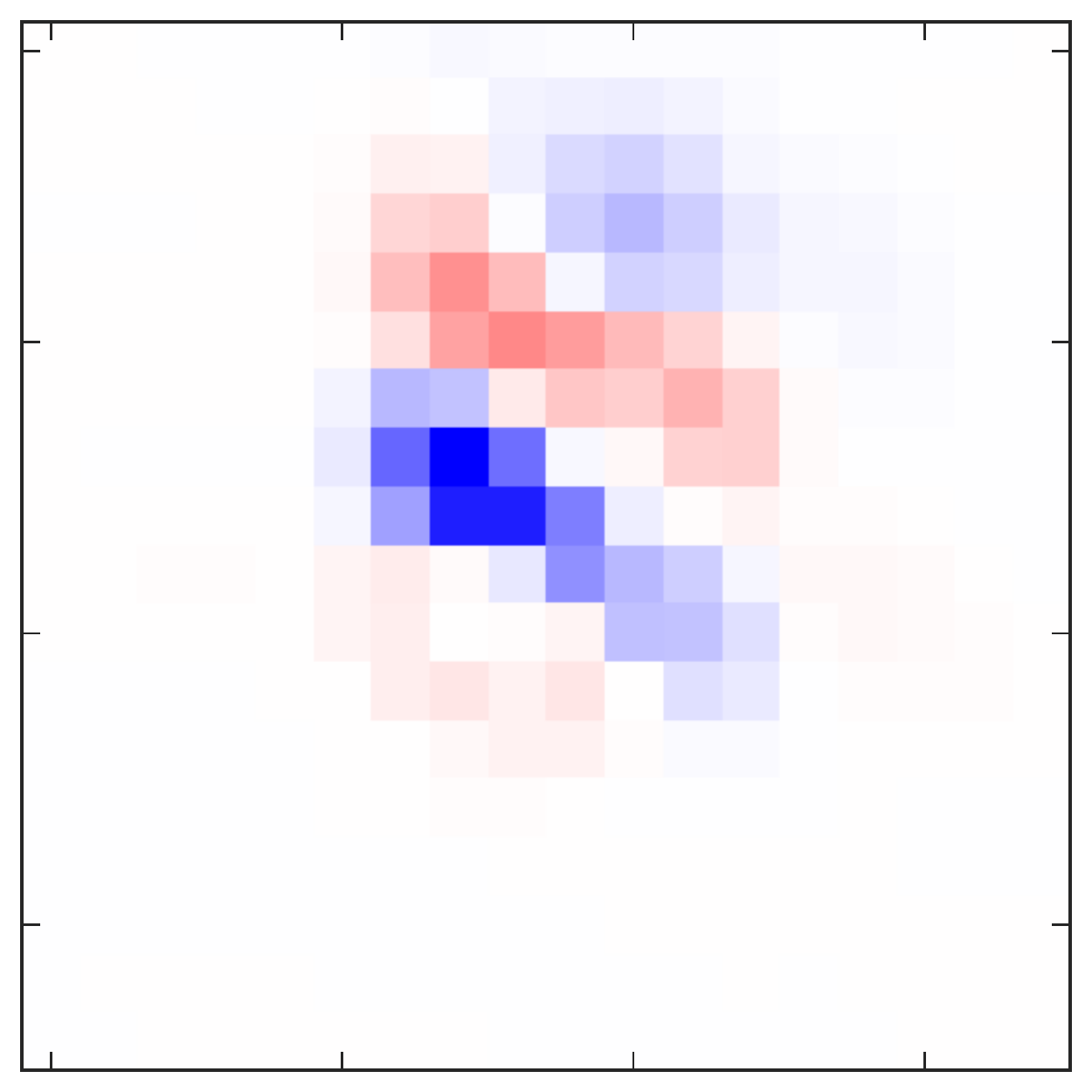} \\
  \includegraphics[width=0.119\textwidth]{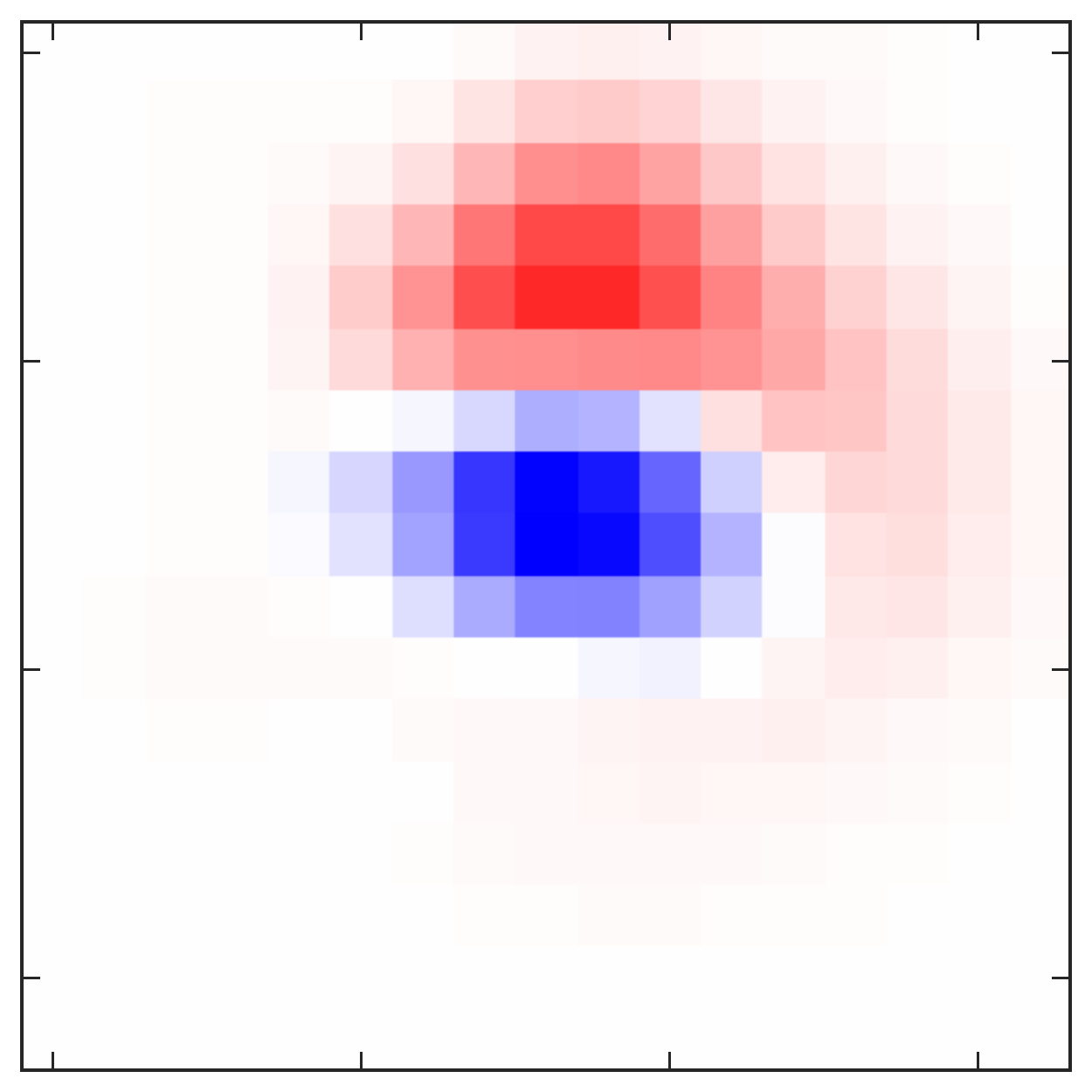}
  \includegraphics[width=0.119\textwidth]{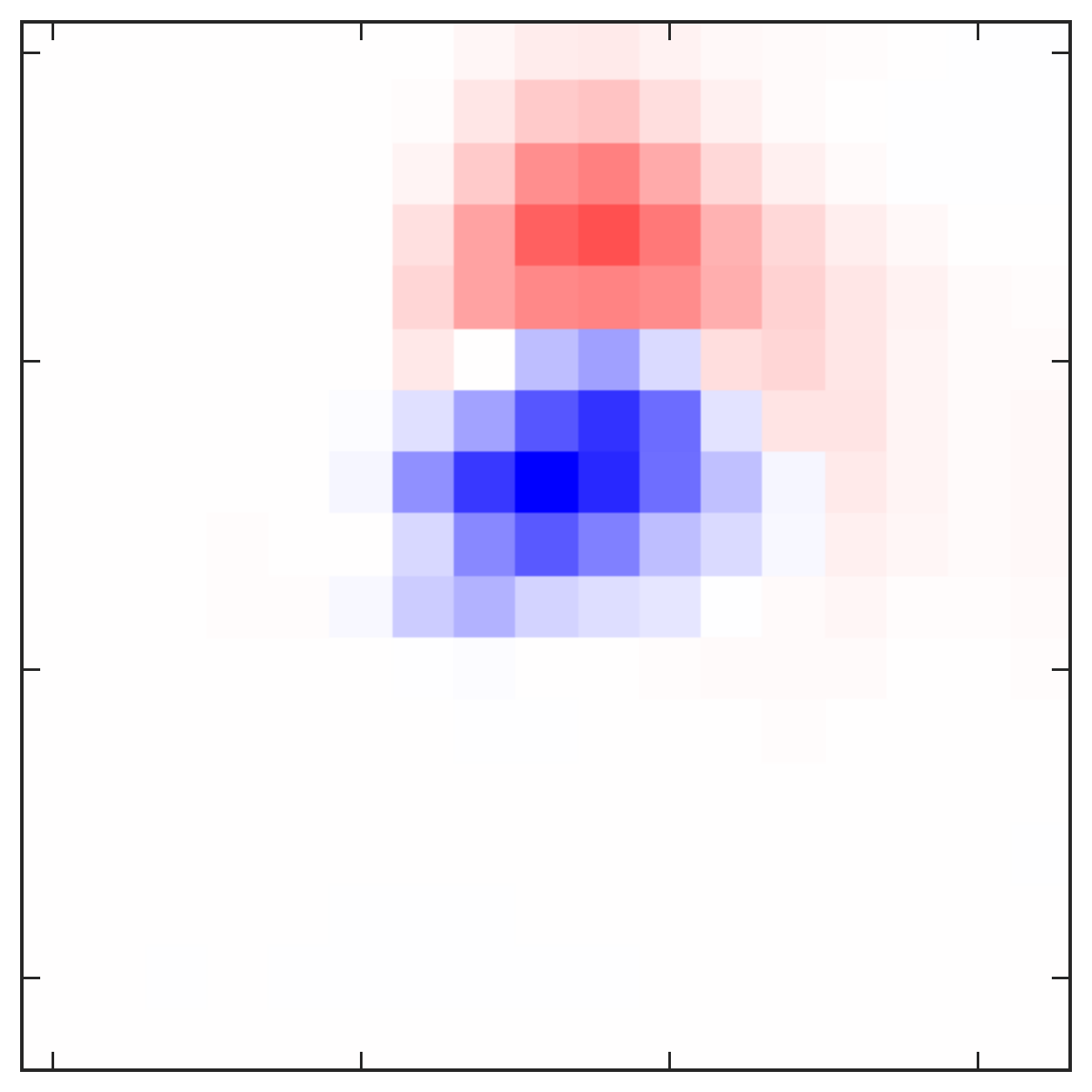}
  \includegraphics[width=0.119\textwidth]{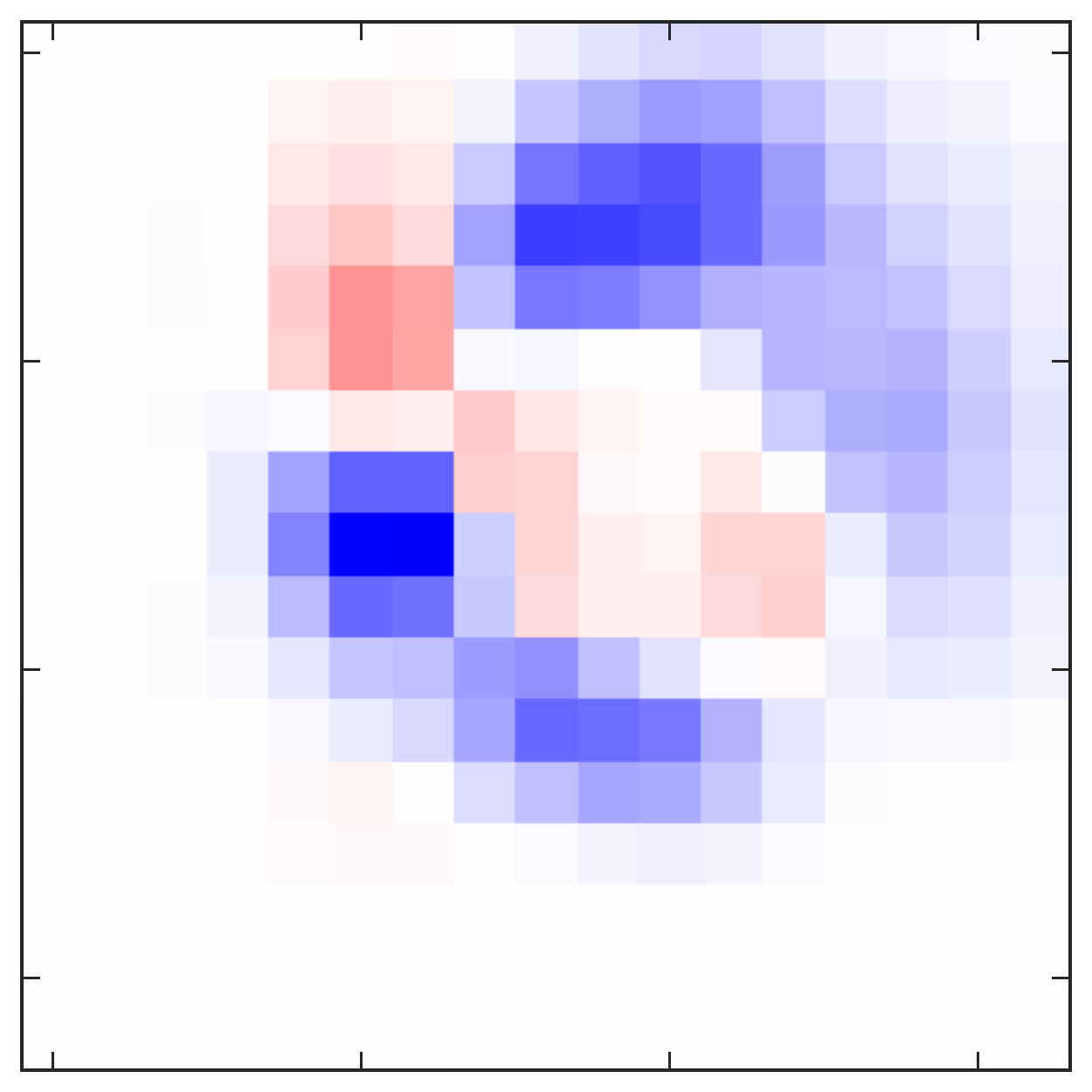}
  \includegraphics[width=0.119\textwidth]{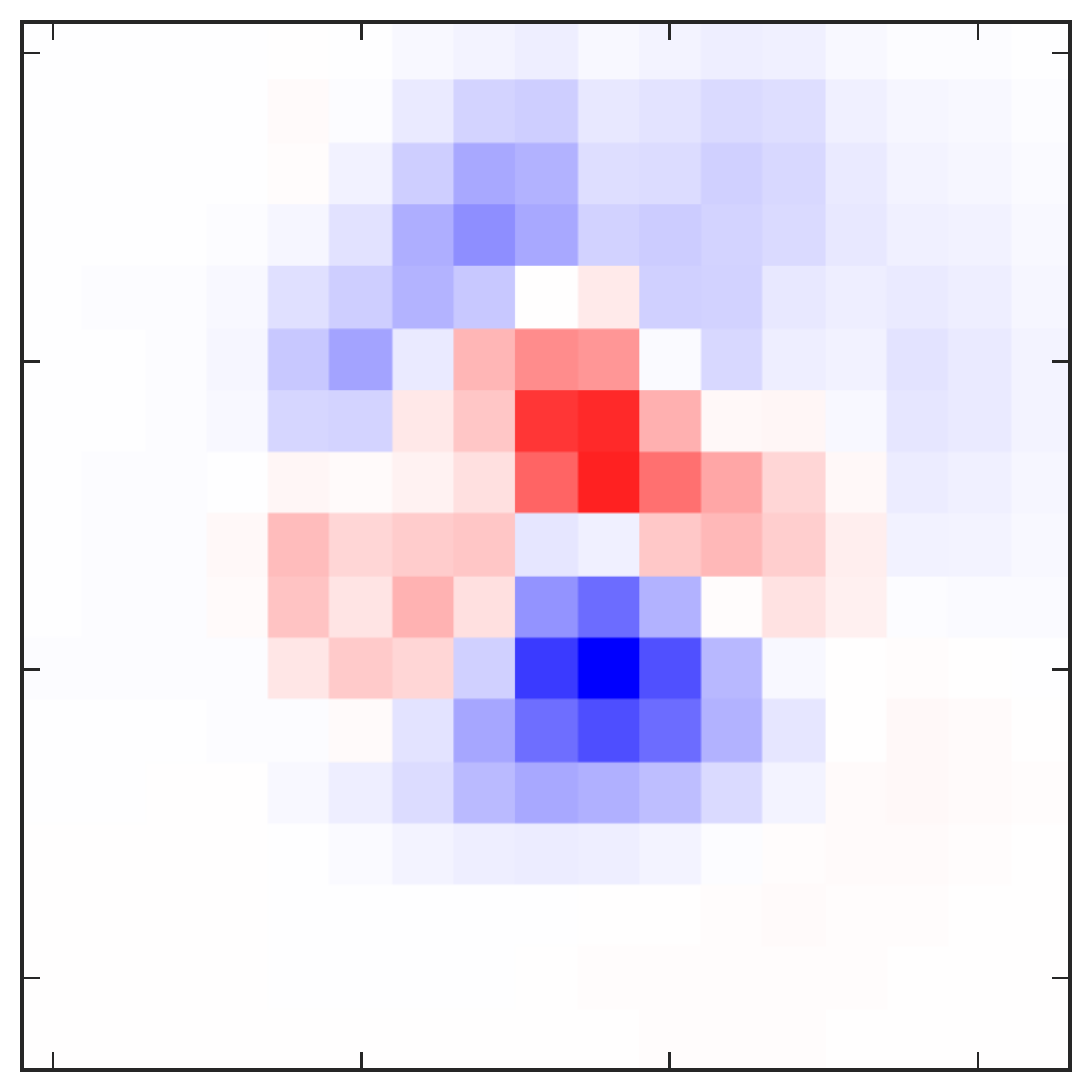}
  \includegraphics[width=0.119\textwidth]{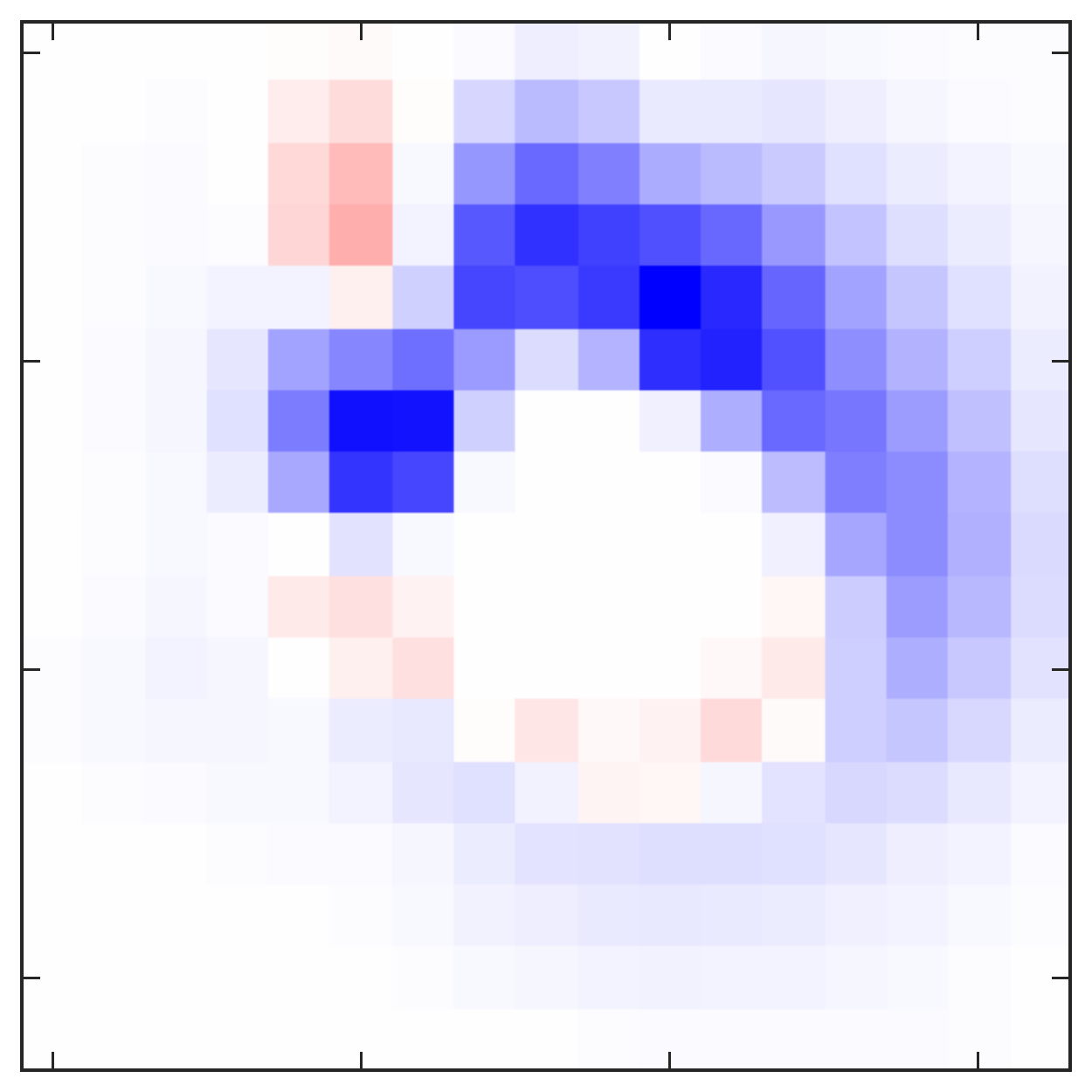}
  \includegraphics[width=0.119\textwidth]{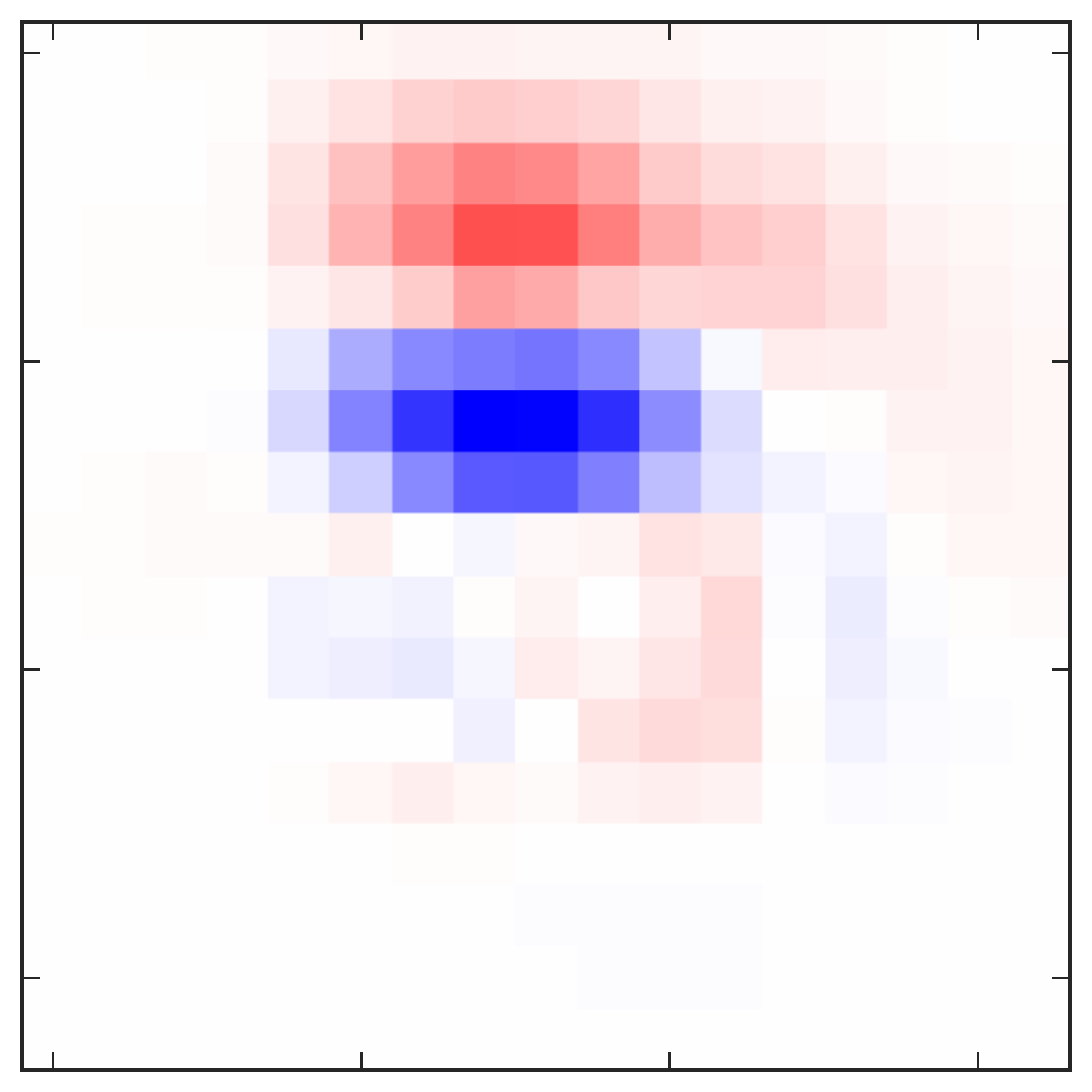}
  \includegraphics[width=0.119\textwidth]{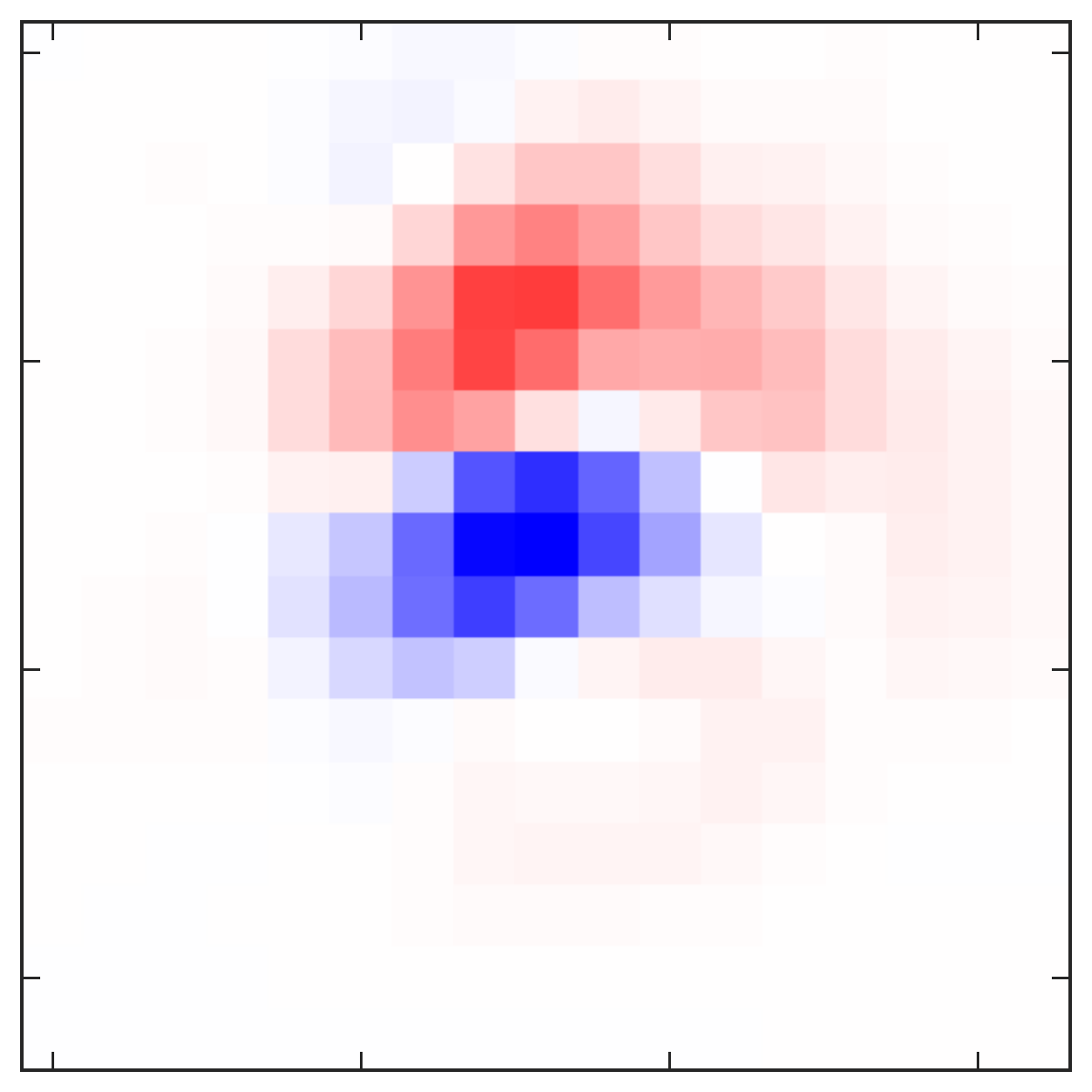}
  \includegraphics[width=0.119\textwidth]{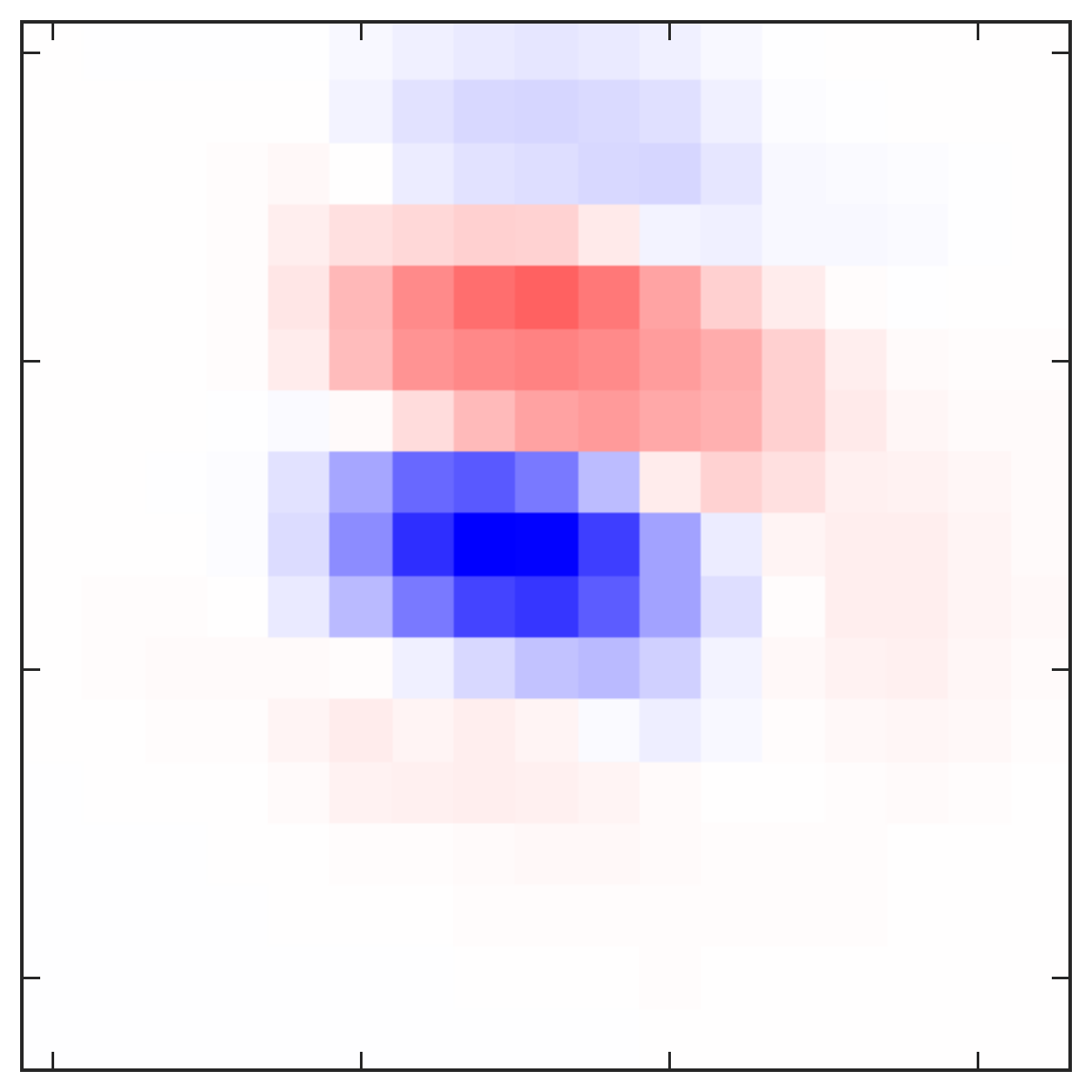}
  \caption{Averaged signal minus background for our default network
    and full pre-processing. The rows correspond to ConvNet layers one
    to four. After two rows MaxPooling reduces the number of pixels by
    roughly a factor of four. The columns indicate the feature maps
    one to eight. Red areas indicate signal-like regions, blue areas
    indicate background-like regions.}
  \label{fig:conv_layers}
\end{figure}

\begin{figure}[t]
  \centering{
    \includegraphics[width=0.65\textwidth]{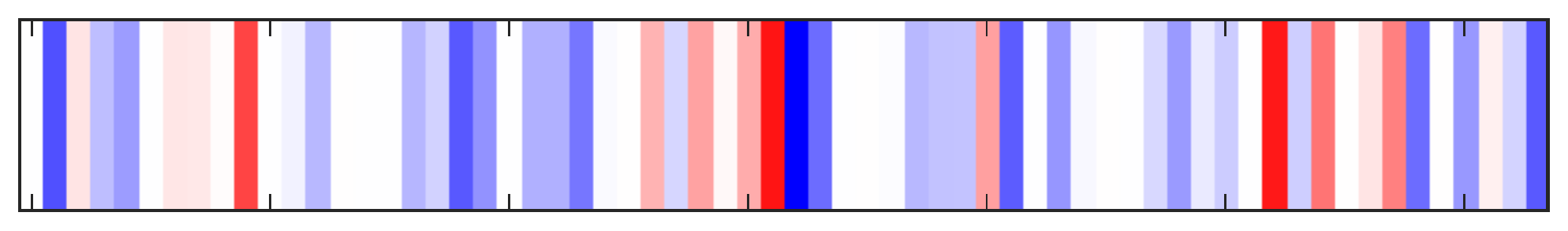}
    \includegraphics[width=0.65\textwidth]{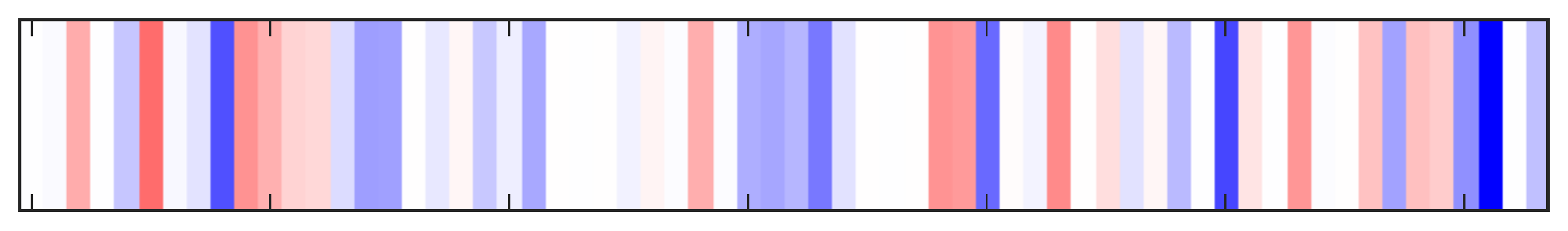}
    \includegraphics[width=0.65\textwidth]{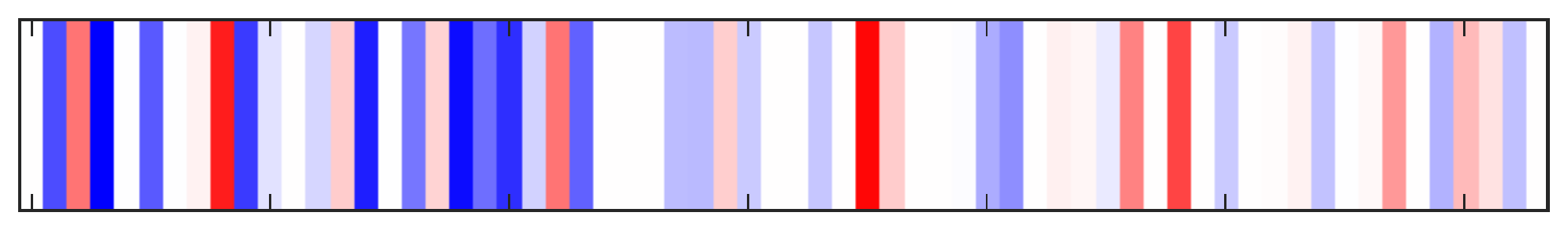}
  }
  \caption{Averaged signal minus background for our default network
    and full pre-processing. The rows show the three dense DNN
    layers. Red areas indicate signal-like regions, blue areas
    indicate background-like regions.}
  \label{fig:deep_layers}
\end{figure}

In the second step we train each network architecture using the mean
squared error as loss function and a the Nesterov algorithm with an
initial learning rate $\eta_L = 0.003$ and no momentum. Training is
performed on mini-batches with a size of 1000 images per batch. We
train our default setup over up to 1000 epochs and use the network
configuration minimizing the loss function calculated on the validation/optimization
sample. Different learning parameters were used to ensure convergence
when training on the minimally pre-processed and the scale-smeared
samples. Because the DNN output is a signal and background
probability, the minimum signal probability required for signal
classification is a parameter that allows to link the signal
efficiency $\epsilon_S$ with the mis-tagging rate of background events
$\epsilon_B$.

In section~\ref{sec:test} we will use this trained network to test the
performance in terms of ROC curves, correlating the signal efficiency
and the mis-tagging rate.\medskip

Before we move to the performance study, we can get a feeling for what
is happening inside the trained ConvNet by looking at the output of
the different layers in the case of fully pre-processed images. In
figure~\ref{fig:conv_layers} we show the difference of the averaged
output for 100 signal and 100 background images. For each of those two
categories, we require a classifier output of at least 0.8. Each
row illustrates the output of a convolutional layer. Signal-like red
areas are typical for jet images originating from top decays; blue
areas are typical for backgrounds. The first layer seems to
consistently capture a well-separated second subjet, and some kernels
of the later layers seem to capture the third signal subjet in the
right half-plane. While one should keep in mind that there is no
one-to-one correspondence between the location in feature maps of
later layers and the pixels in the input image, we will discuss these
kinds of structures in the jet image below.

In figure~\ref{fig:deep_layers} we show the same kind of intermediate
result for the two fully connected DNN layers. Each of the 64 linear
bars represents a node of the layer. We see that individual nodes are
quite distinctive for signal and background images, but they cannot be
linked to any pattern in the jet image. This illustrates how the
two-dimensional ConvNet approach is more promising that a regular
neural net. The fact that some nodes are not discriminative indicates
that in the interest of speed the number of nodes could be reduced
slightly. The output of the DNN is essentially the same as the
probabilities shown in the right panel of figure~\ref{fig:arc_scan},
ignoring the central probability range between 20\% and 80\%.\medskip

\begin{figure}[b!]
\begin{center}
  \includegraphics[width=0.4\textwidth]{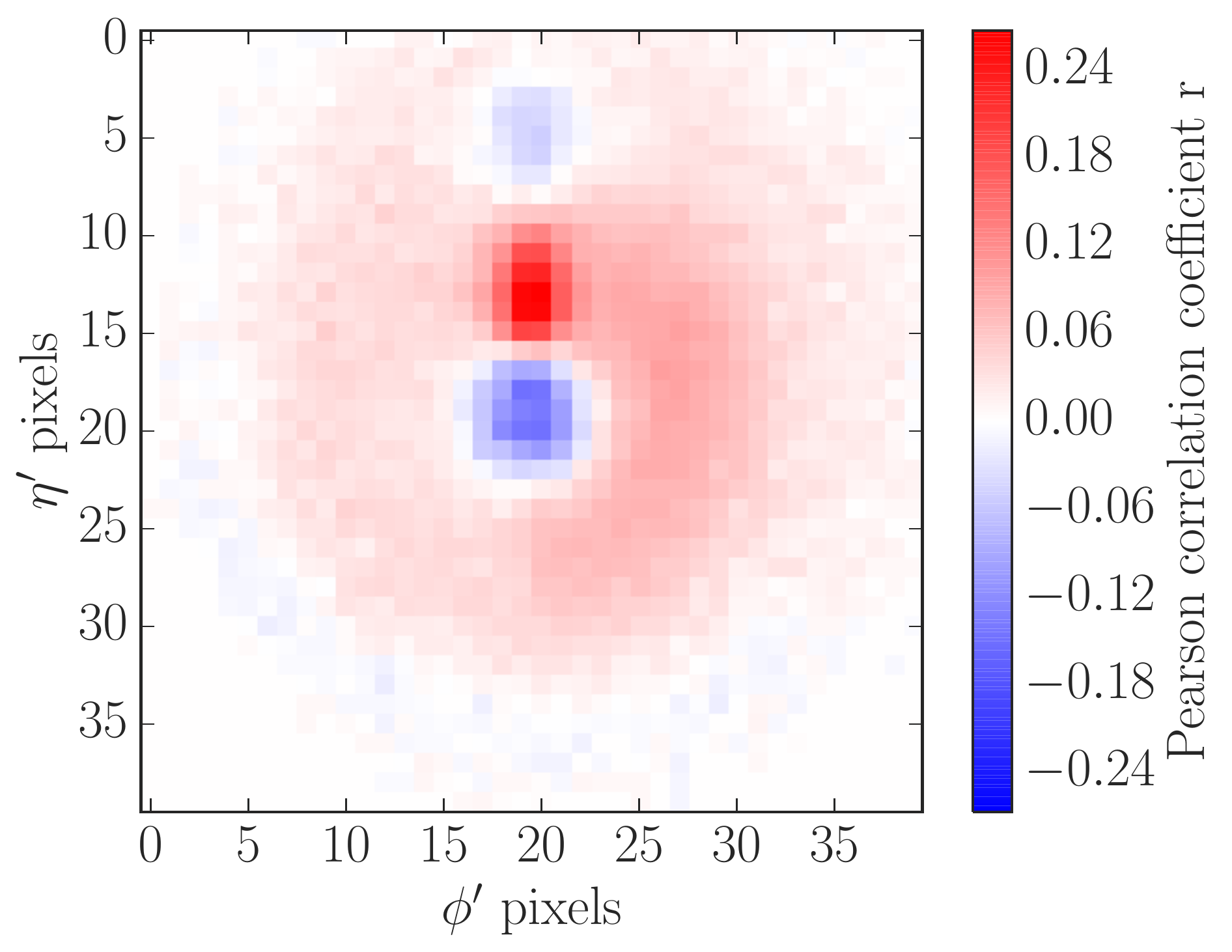}
  \caption{Pearson correlation coefficient for 10,000 signal and
    background images each. The corresponding jet image is illustrated
    in figure~\ref{fig:averaged_images}.  Red areas indicate signal-like
    regions, blue areas indicate background-like regions.}
  \label{fig:pcc}
\end{center}
\end{figure}

To see which pixels of the fully pre-processed $40 \times 40$ jet image have
an impact on the signal vs background label, we can correlate the
deviation of a pixel $x_{ij}$ from its mean value $\bar{x}_{ij}$ with
the deviation of the label $y$ from its mean value $\bar{y}$. A
properly normalized correlation function for a given set of combined
signal and background images can be defined as
\begin{align}
  r_{ij} = \frac{\sum_\text{images}\left(x_{ij} - \bar{x}_{ij}\right)
    \left(y - \bar{y}\right)}
  { \sqrt{ \sum_\text{images}\left(x_{ij} -
        \bar{x}_{ij}\right)^2} \sqrt{ \sum_\text{images}\left(y - \bar{y}\right)^2 }
  } \;.
\end{align}
It is usually referred to as the Pearson correlation coefficient. From
the definition we see that for a signal probability $y$ positive
values of $r_{ij}$ indicate signal-like patterns.  In
figure~\ref{fig:pcc} we show this correlation for our network
architecture.  A large energy deposition in the center leads to
classification as background. A secondary energy deposition in the 12
o'clock position combined with additional energy deposits in the right
half-plane lead to a classification as signal. This is consistent with
our expectations after full pre-processing, shown in
figure~\ref{fig:averaged_images}.

\section{Performance test}
\label{sec:test}
 
Given our optimized machine learning setup introduced in
section~\ref{sec:machine} and the fact that we can understand its
workings and trust its outcome, we can now compare its performance
with state-of-the-art top taggers. The details of the signal and background
samples and jet images are discussed in section~\ref{sec:machine_image};
essentially, we attempt to separate a top decay inside a fat jet from
a QCD fat jet including fast detector simulation and for the
transverse momentum range $p_{T,\text{fat}} = 350~...~450$~GeV. Other
transverse momentum ranges for the fat jet can be targeted using the
same DNN method. 

Because we focus on a comparing the performance of the DNN approach
with the performance of standard multivariate top taggers we take our
Monte Carlo training and testing sample as a replacement of actual
data. This distinguishes our approach from tagging methods which use
Monte Carlo simulations for training, like the \textsc{Template
  Tagger}~\cite{template}.  This means that for our performance test
we do not have to include uncertainties in our \textsc{Pythia}
simulations compared to other Monte Carlo simulations and
data~\cite{aussies}.

\subsection{QCD-based taggers}
\label{sec:test_qcd}

Acting on the same calorimeter entries in the rapidity vs azimuthal
angle plane which define the jet image, we can employ QCD-based
algorithms to determine the origin of a given configuration.  Based on
QCD jet algorithms, for example the multivariate
\textsc{HEPTopTagger2}~\cite{heptop1,heptop2,heptop3,heptop4} extracts
hard substructures using a mass drop condition~\cite{bdrs}
\begin{align}
  \max(m_1,m_2) > f_\text{drop} \; m_{1+2}
  \label{eq:mod_mass_drop}
\end{align}
with a given $f_\text{drop} = 0.8$. Provided that at least three hard
substructures exist, different constraints on the invariant masses of
combinations of filtered substructures~\cite{bdrs} define a top
tag. One of the features of the \textsc{HEPTopTagger} is that even in
the multivariate analysis it will always identify a top candidate with
a three-prong decay and the correct reconstructed top mass.  An
alternative approach is to groom the fat jet using the
\textsc{SoftDrop} criterion~\cite{softdrop}
\begin{align}
  \frac{\min(p_{T1},p_{T2})}{p_{T1}+p_{T2}} > z_\text{cut} \left(
    \frac{\Delta R_{12}}{R_0} \right)^\beta
  \label{eq:soft_drop}
\end{align}
and employ the groomed jet mass. It can be thought of a combination of
$p_T$-drop criterion~\cite{hopkins} with a soft-collinear extension of
pruning~\cite{pruning}. We use the \textsc{SoftDrop} parameters
$\beta=1$ and $z_\text{cut}=0.2$. The main difference between the
\textsc{HEPTopTagger} and \textsc{SoftDrop} is that the latter does
not explicitly target the top and $W$ decays, needs an additional
condition on a mass scale to work as a tagger, and will not
reconstruct the top 4-momentum. Because of these much weaker
constraints on the top candidate kinematics a \textsc{SoftDrop}
construction is ideally suited for a hypothesis test differentiating
between fat QCD jets and top decay jets.\medskip

The QCD shower-based taggers alone are known to not fully use the
available calorimeter information. However, they can be complemented
by a simple observable quantifying the number of constituents inside
the fat jet or the number of prongs in the top decay. Adding 
the $N$-subjettiness~\cite{nsubjettiness} variables
\begin{align}
  \tau_N = \frac{1}{R_0 \sum_k p_{T,k}} \sum_k p_{T,k} \; \min \left(
    \Delta R _{1,k}, \Delta R _{2,k}, \cdots, \Delta R _{N,k} \right)^\beta \;.
  \label{eq:tau_N}
\end{align}
to the \textsc{HEPTopTagger} or \textsc{SoftDrop} picks up this
additional information and also induces the three-prong top decay
structure into \textsc{SoftDrop}.  We use $N$ $k_T$-axes, $\beta=1$
and the reference distance $R_0$. A small value $\tau_N$ indicates
consistency with $N$ or less substructure axes, so an $N$-prong decays
give rise to a small ratio $\tau_N/\tau_{N-1}$. For top tagging
$\tau_{3}/\tau_{2}$ is particularly useful in combination with QCD
taggers in a multivariate setup~\cite{heptop4}. The $N$-subjettiness
variables $\tau_j$ can be defined based on the complete fat jet or
based on the fat jet after applying the \textsc{SoftDrop}
criterion. Using $\tau_j$ and $\tau^\text{sd}_j$ in a multivariate
analysis usually leads to optimal result.

\subsection{Comparison}
\label{sec:comp}

\begin{figure}[t]
  \begin{center}
  \includegraphics[width=0.46\textwidth]{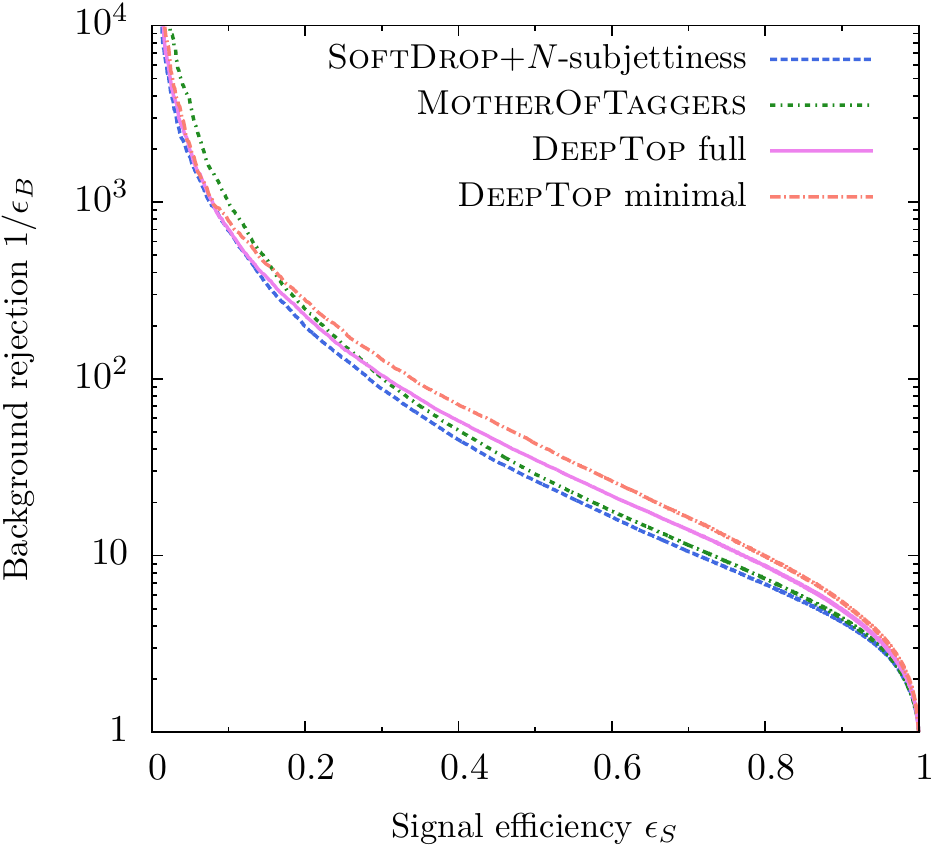}
  \end{center}
  \caption{Performance of the neural network tagger compared to the QCD-based
    approaches \textsc{SoftDrop} plus $N$-subjettiness and including
    the \textsc{HEPTopTagger} variables.}
\label{fig:performance}
\end{figure}

To benchmark the performance of our \textsc{DeepTop} DNN, we compare
its ROC curve with standard Boosted Decision Trees based on the C/A
jets using \textsc{SoftDrop} combined with $N$-subjettiness.  From
figure~\ref{fig:arc_scan} we know the spread of performance for the
different network architectures for fully pre-processed images.  In
figure~\ref{fig:performance} we see that minimal pre-processing actually
leads to slightly better results, because the combination or rotation
and cropping described in section~\ref{sec:machine_image} leads to a
small loss in information. Altogether, the band of different machine
learning results indicates how large the spread of performance will be
whenever for example binning issues in $p_{T,\text{fat}}$ are taken
into account, in which case we we would no longer be using the perfect
network for each fat jet.\medskip

For our BDT we use \textsc{GradientBoost} in the Python package
\textsc{sklearn}~\cite{sklearn} with 200 trees, a maximum depth of 2,
a learning rate of~0.1, and a sub-sampling fraction of $90\%$ for the
kinematic variables
\begin{align}
  \{\ m_\text{sd}, m_\text{fat}, \tau_2, \tau_3, \tau_2^\text{sd}, \tau_3^\text{sd}  \ \}
  \qqqquad \text{(\textsc{SoftDrop} + $N$-subjettiness)} \; ,
\end{align}
where $m_\text{fat}$ is the un-groomed mass of the fat jet.  This is
similar to standard experimental approaches for our transverse momentum range
$p_{T,\text{fat}} = 350~...~400$~GeV. In addition, we include the
\textsc{HEPTopTagger2} information from filtering combined with a mass
drop criterion,
\begin{align}
  \{\ m_\text{sd}, m_\text{fat}, m_\text{rec}, f_\text{rec}, \Delta R_\text{opt}, \tau_2, \tau_3, \tau_2^\text{sd}, \tau_3^\text{sd} \ \} 
  \qqqquad \text{(\textsc{MotherOfTaggers})} \; .
\label{eq:def_mother}
\end{align}
\medskip

In figure~\ref{fig:performance} we compare these two
QCD-based approaches with our best neural networks.  Firstly, we see
that both QCD-based BDT analyses and the two neural network setups are
close in performance. Indeed, adding \textsc{HEPTopTagger} information
slightly improves the \textsc{SoftDrop}+$N$-subjettiness setup,
reflecting the fact that our transverse momentum range is close to the
low-boost scenario where one should rely on the better-performing
\textsc{HEPTopTagger}. Second, we see that the difference between the
two pre-processing scenarios is in the same range as the difference
between the different approaches. Running the \textsc{DeepTop}
framework over signal samples with a 2-prong $W'$ decay to two jets
with $m_{W'} = m_t$ and over signal samples with a shifted value of
$m_t$ we have confirmed that the neural network setup learns both, the
number of decay subjets and the mass scale.\medskip

\begin{figure}[t]
  \includegraphics[width=0.49\textwidth]{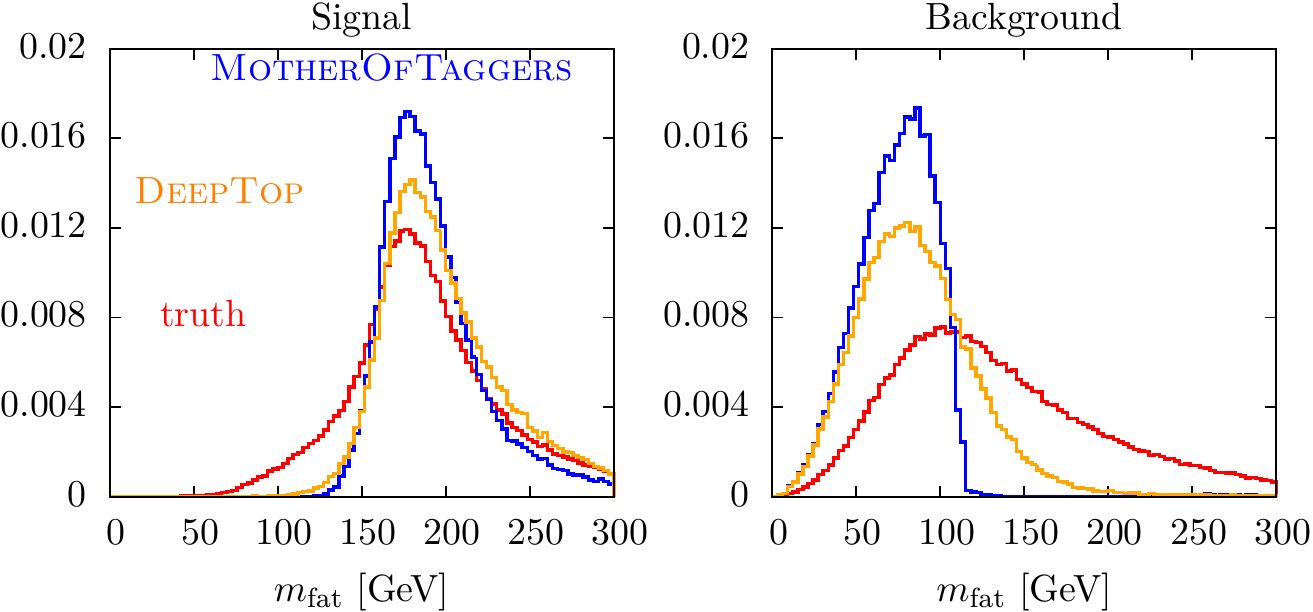}
  \includegraphics[width=0.49\textwidth]{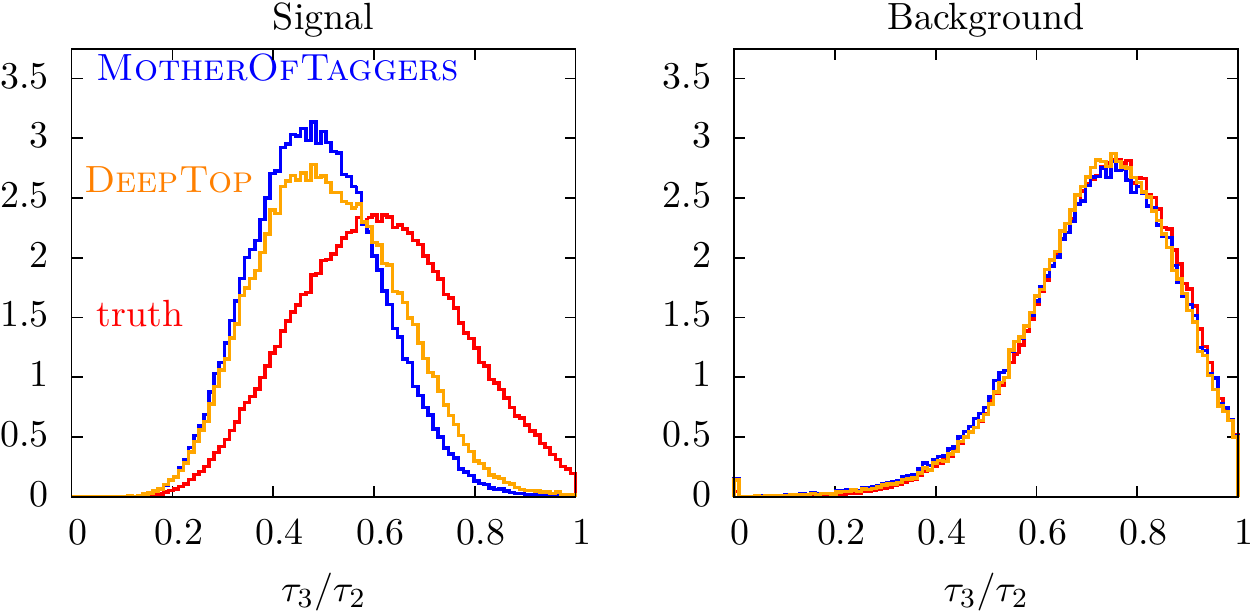}
  \caption{Kinematics observables $m_\text{fat}$ and $\tau_3/\tau_2$
    for events correctly determined to be signal or background by the
    \textsc{DeepTop} neural network and by the
    \textsc{MotherOfTaggers} BDT, as well as Monte Carlo truth.}
  \label{fig:inputs}
\end{figure}

Following up on on the observation that the neural network and the
QCD-based taggers show similar performance in tagging a boosted top
decay inside a fat jet, we can check what kind of information
is used in this distinction.

Both for the DNN and for the
\textsc{MotherOfTaggers} BDT output we can study signal-like learned
patterns in actual signal events by cutting on the output label $y$
corresponding to the 30\% most signal like events
shown on the right of figure~\ref{fig:arc_scan}.  Similarly, we
can require the 30\% most background like events to test if the background
patterns are learned correctly. In addition, we can compare the
kinematic distributions in both cases to the Monte Carlo truth. In
figure~\ref{fig:inputs} we show the distributions for $m_\text{fat}$ and
$\tau_3/\tau_2$, both part the set of observables defined in
Eq.\eqref{eq:def_mother}. We see that the DNN and BDT tagger indeed
learn essentially the same structures. The fact that their results are
more pronounced signal-like than the Monte Carlo truth is linked to
our stiff cut on $y$, which for the DNN and BDT tagger cases removes
events where the signal kinematic features is less pronounced. The
\textsc{MotherOfTaggers} curves for the signal are more peaked than
the \textsc{DeepTop} curves is due to the fact that the observables
are exactly the basis choice of the BDT, while for the neural network
they are derived quantities. In App.~\ref{app:what} we extend this
comparison to more kinematic observables.

Finally, a relevant question is to what degree the information used by
the neural network is dominated by low-$p_T$ effects. We can apply a
cutoff, for example including only pixels with a transverse energy
deposition $E_T > 5$~GeV. This is the typical energy scale where the
DNN performance starts to degrade, as we discuss in more detail in
App.~\ref{app:detector}.

\section{Conclusions}

Fat jets which can include the decay products of a boosted,
hadronically decaying top quark are an excellent basis to establish
machine learning based on fat jet images and compare their performance
to QCD-based top taggers. Here, machine learning is the logical next
step after developing multivariate top taggers which test QCD vs top
decay hypotheses rather than identifying and reconstructing an actual
top decay. This includes the assumption that our ConvNet
\textsc{DeepTop} approach will be trained purely on data.\medskip

We have constructed a ConvNet setup, inspired by standard image
recognition techniques~\cite{slac2,quark_gluon}. To optimize the
network architecture, train the network, and test the performance we
have used independent event samples. First, we have found that changes
in the network architecture only have a small impact on the top
tagging performance. Pre-processing the fat jet images is useful to
visualize, understand, and follow the network learning procedure for
example using the Pearson correlation coefficient, but has little
influence on the network performance.\medskip

As a base line we have constructed a \textsc{MotherOfTaggers}
QCD-based top tagger, implemented as a multivariate BDT. This allowed
us to quantify the performance of the \textsc{DeepTop} network and to
test which kinematic observables in the fat jet have been learned by
the neural network. We have also confirmed that the neural network is
not dominated by low-$p_T$ calorimeter entries and extraordinarily
stable with respect to changing the jet energy scale. In
figure~\ref{fig:performance} we found that the performance of the two
approaches is comparable, giving us all the freedom to define future
experimental strategies for top tagging, ranging from proper top
reconstruction to multivariate hypothesis testing and, finally, data-based
machine learning.

\bigskip
\acknowledgments

T.P. would like to thank the \textsl{Jets\@@LHC} workshop at the ICTS
in Bangalore, where the \textsc{DeepTop} results were discussed for
the first time, and where the paper was finalized.  T.S. acknowledges
support form the \textsl{International Max Planck Research School for
  Precision Test of Fundamental Symmetries} and from the DFG Research
Training Group \textsl{Particle Physics Beyond the Standard Model}.
M.R. was supported by the European Union Marie Curie Research Training
Network MCnetITN, under contract PITN-GA-2012-315877.  Our work was
supported by a grant from the Swiss National Supercomputing Centre
(CSCS) under project D61. There, we trained our networks on the Piz
Daint system using nodes equipped with Nvidia Tesla P100 GPUs.

\newpage
\appendix
\section{What the machine learns}
\label{app:what}

For our performance comparison of the QCD-based tagger approach and
the neural network it is crucial that we understand what the
\textsc{DeepTop} network learns in terms of physics variables. The
relevant jet substructure observables differentiating between QCD jets
and top jets are those which which we evaluate in the
\textsc{MotherOfTaggers} BDT, Eq.\eqref{eq:def_mother}.

To quantify which signal features the DNN and the BDT tagger have
correctly extracted we show observables for signal event correctly
identified as such, \ie requiring events with a classifier response $y$ corresponding  to the 30\% most signal like events.
 Following figure~\ref{fig:arc_scan} this cut value captures a
large fraction of correctly identified events. The same we also do for
the $30\%$ most background like events identified by each classifier.\medskip

\begin{figure}[h!]
  \includegraphics[width=0.49\textwidth]{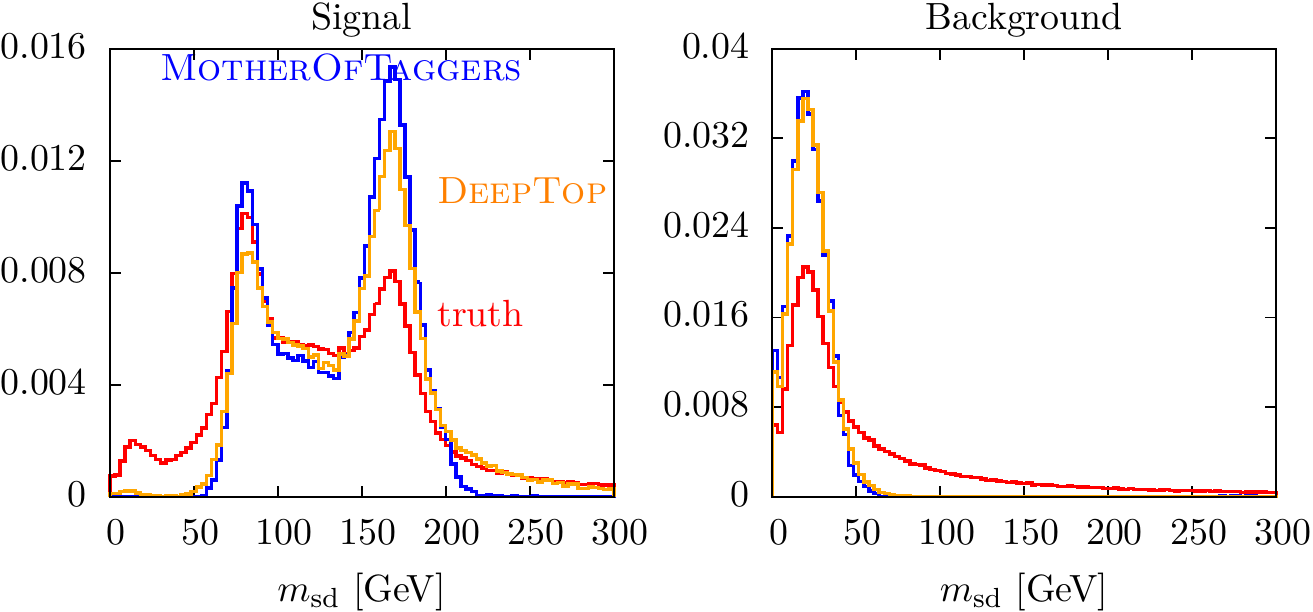}
  \includegraphics[width=0.49\textwidth]{mass_ungroomed_cut} \\
  \includegraphics[width=0.49\textwidth]{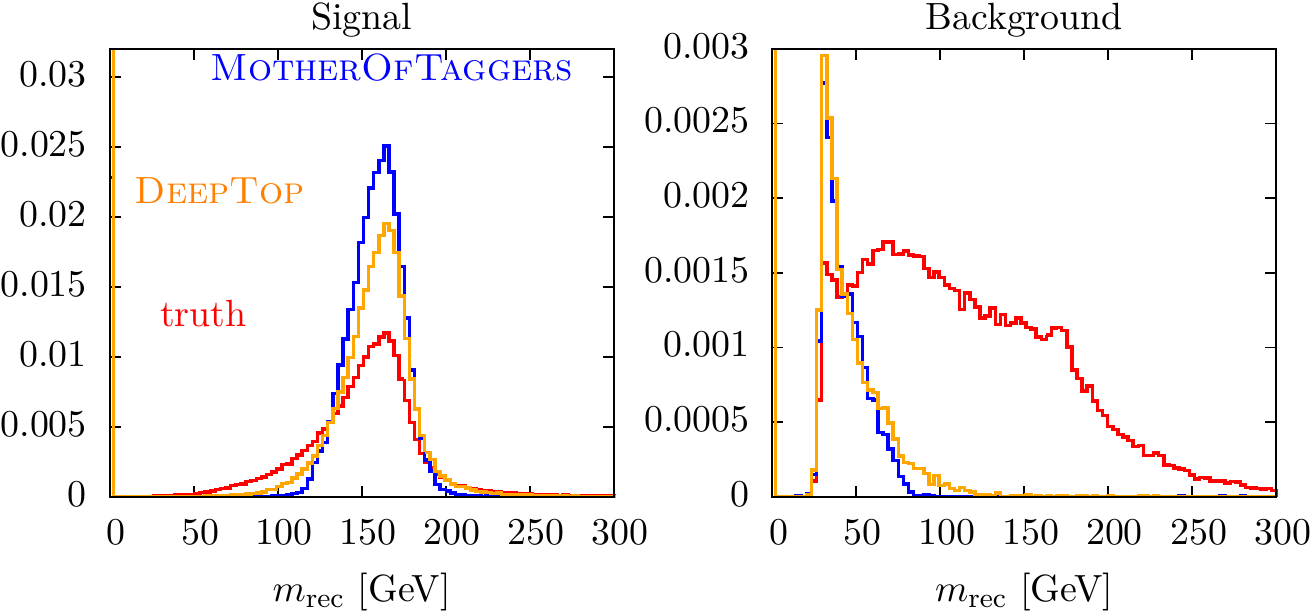} 
  \phantom{\includegraphics[width=0.49\textwidth]{HTT_mass_cut}} \\
  \includegraphics[width=0.49\textwidth]{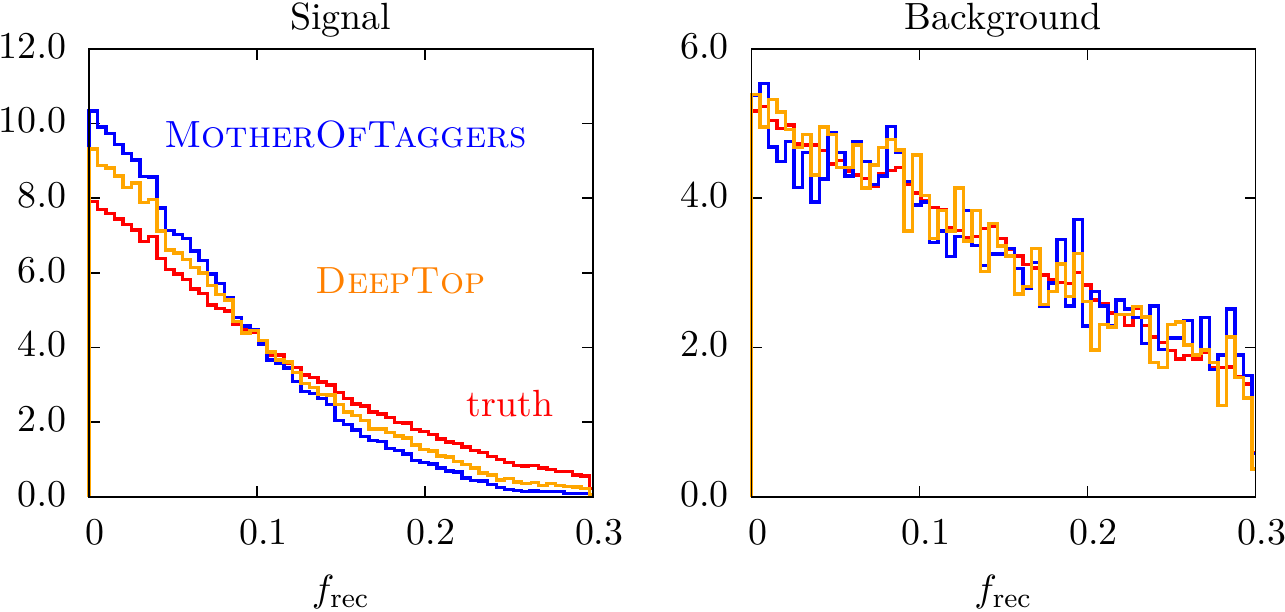}
  \includegraphics[width=0.49\textwidth]{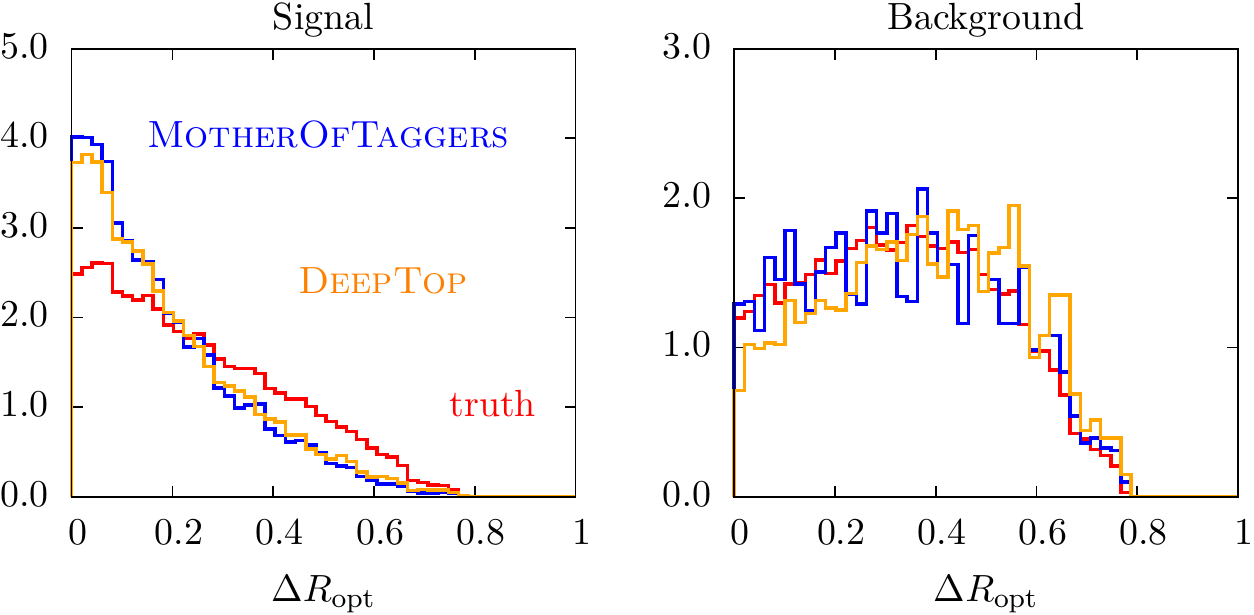} \\
  \includegraphics[width=0.49\textwidth]{tau32_cut}
  \includegraphics[width=0.49\textwidth]{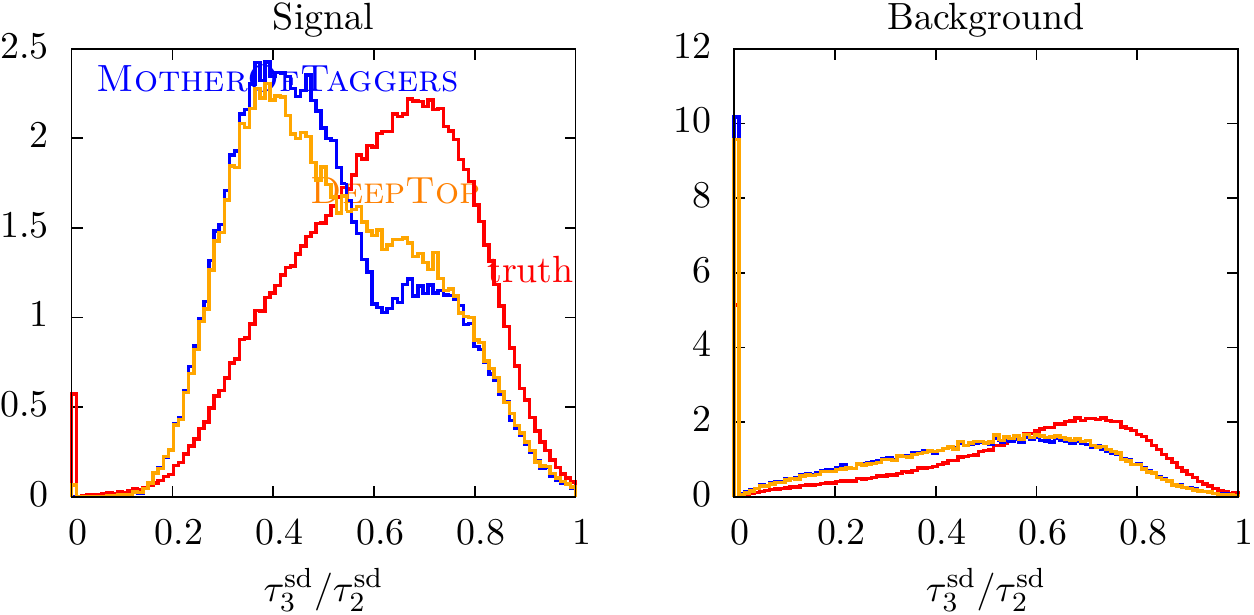}
  \caption{Kinematics observables defined in Eq.\eqref{eq:def_mother}
    for events correctly determined to be signal or background by the
    \textsc{DeepTop} neural network and by the
    \textsc{MotherOfTaggers} BDT, as well as Monte Carlo
    truth. Extended version of figure~\ref{fig:inputs}.}
\label{fig:physics}
\end{figure}

The upper two rows in figure~\ref{fig:physics} show the different mass
variables describing the fat jet. We see that the DNN and the BDT
tagger results are consistent, with a slightly better performance of
the BDT tagger for clear signal events. For the background the BDT output
is more pronounced as well. The deviation from
the true mass for the \textsc{HEPTopTagger} background performance is
explained by the fact that many events with no valid top candidate
return $m_\text{rec} = 0$.  Aside from generally comforting results we
observe a peculiarity: the \textsc{SoftDrop} mass identifies the
correct top mass in fewer that half of the correctly identified signal
events, while the fat jet mass $m_\text{fat}$ does correctly reproduce
the top mass. The reason why the \textsc{SoftDrop} mass is
nevertheless an excellent tool to identify top decays is that its
background distribution peaks at very low values, around $m_\text{sd}
\approx 20$~GeV. Even for $m_\text{sd} \approx m_W$ the hypothesis
test between top signal and QCD background can clearly identify a
massive particle decay.

In the third row we see that the \textsc{HEPTopTagger} $W$-to-top mass
ratio $f_\text{rec}$ only has little significance for the transverse
momentum range studied. For the optimalR variable $\Delta
R_\text{opt}$~\cite{heptop4} the DNN and the BDT tagger again give consistent
results. Finally, for the $N$-subjettiness ratio $\tau_3/\tau_2$
before and after applying the \textsc{SoftDrop} criterion the results
are again consistent for the two tagging approaches.\medskip

\begin{figure}[t]
  \includegraphics[width=0.49\textwidth]{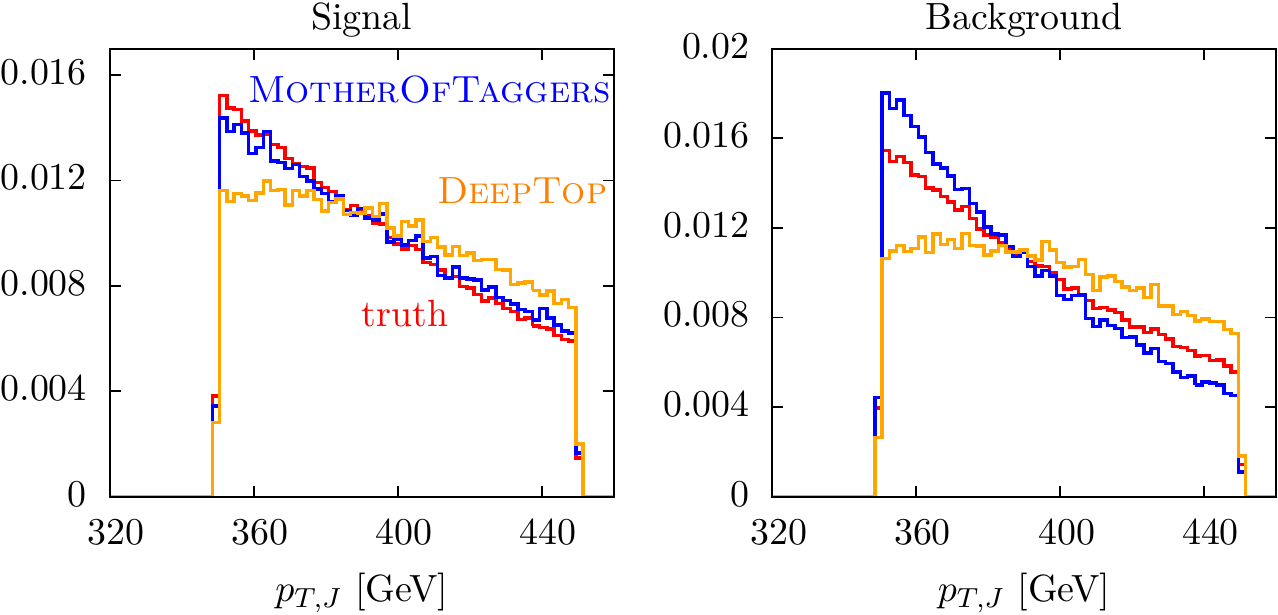}
  \includegraphics[width=0.49\textwidth]{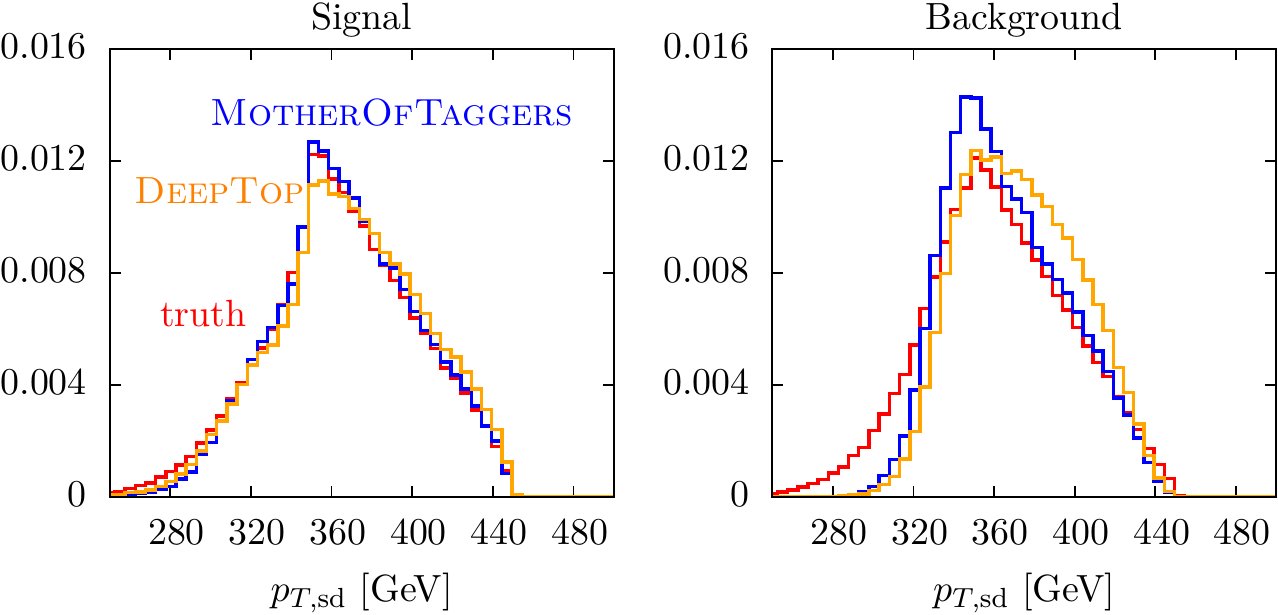}
  \caption{Reconstructed transverse momenta for events correctly
    determined to be signal or background by the \textsc{DeepTop}
    neural network and by the \textsc{MotherOfTaggers} BDT, as well
    as Monte Carlo truth.}
\label{fig:momrec}
\end{figure}

Following up on the observation that \textsc{SoftDrop} shows excellent
performance as a hypothesis test, we show in figure~\ref{fig:momrec}
the reconstructed transverse momenta of the fat jet, or the top quark
for signal events. In the left panel we see that the transverse
momentum of the un-groomed fat jet reproduces our Monte-Carlo range
$p_{T,\text{fat}} = 350~...~450$~GeV. While the transverse momentum
distributions of signal and background are very similar, applying the
BDT or DNN induces a bias which indicates a transverse momentum
dependent tagger response. The transverse momentum dependence is larger
for the DNN. A tagger turn-on with transverse momentum is unproblematic
and can be mitigated using adverserial training
techniques~\cite{pivot} if needed. In
the right panel we see that the constituents identified by the
\textsc{SoftDrop} criterion have a significantly altered transverse
momentum spectrum. To measure the transverse momentum of the top quark
we therefore need to rely on a top identification with
\textsc{SoftDrop}, but a top reconstruction based on the (groomed) fat
jet properties.

\section{Detector effects}
\label{app:detector}

\begin{figure}[t]
  \includegraphics[width=0.45\textwidth]{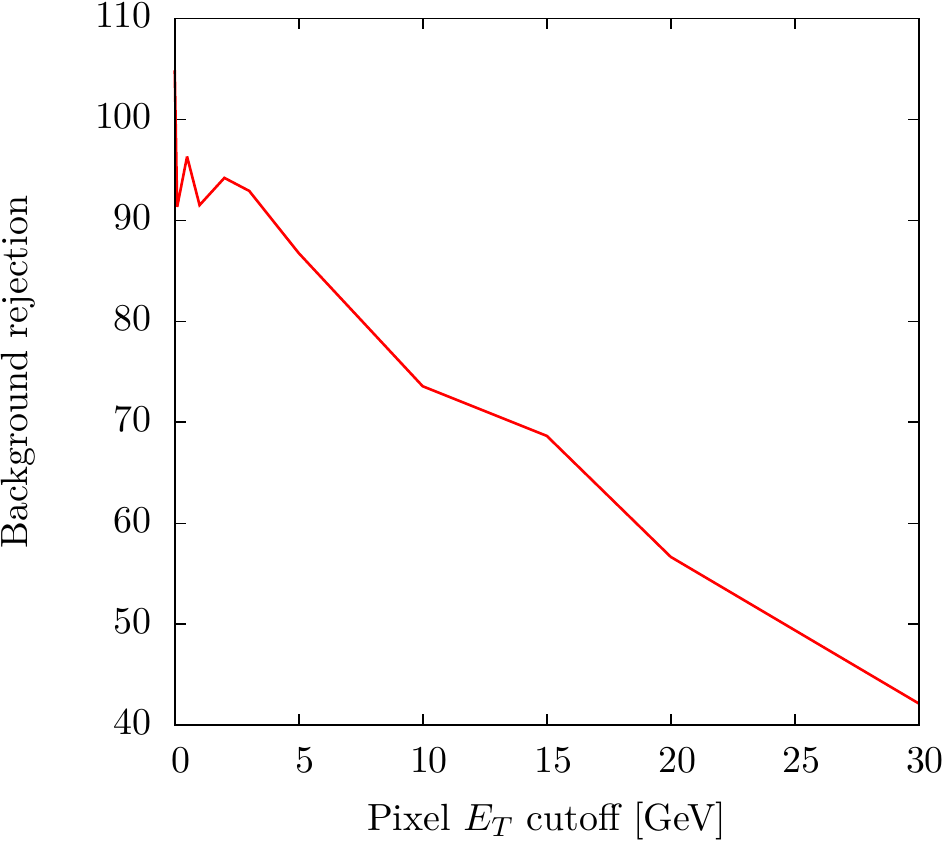}
  \hspace*{0.05\textwidth}
  \includegraphics[width=0.438\textwidth]{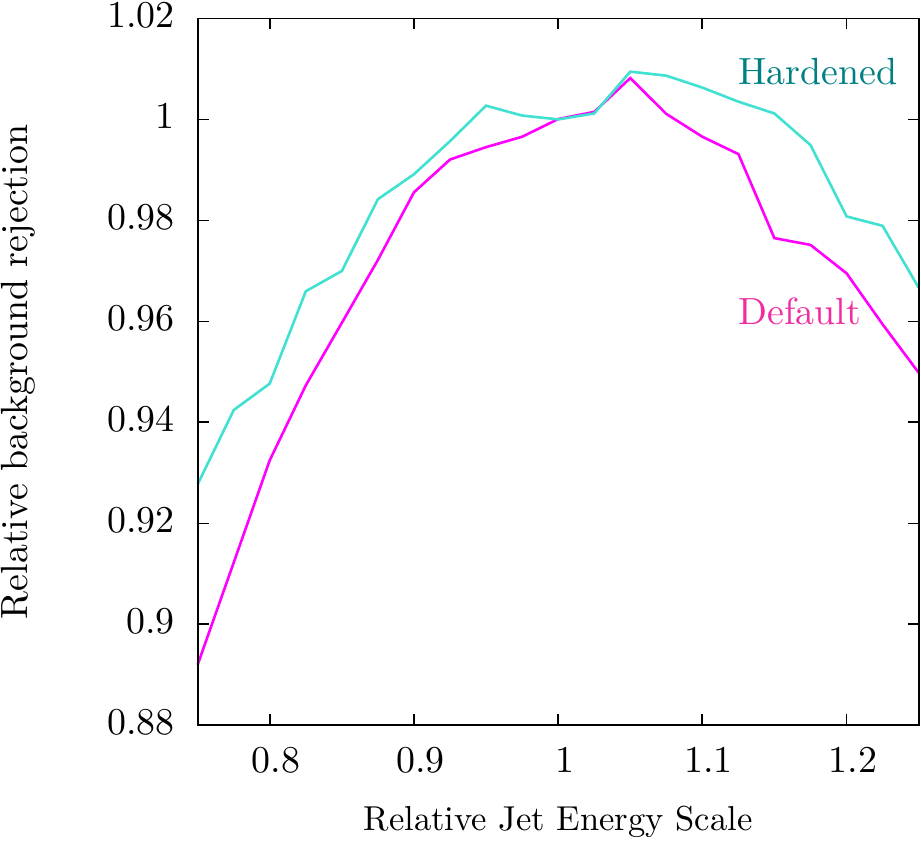}\\

  \caption{Left: Background rejection at a signal efficiency of 30~\%
    for different activation thresholds.  Right: Background rejection
    for the default and hardened training for different re-scalings of
    the jet images. The background rejection is evaluated at a signal
    efficiency of 30~\% and normalized to the rejection at nominal
    JES.}
  \label{fig:rejection_vs_X}
\end{figure}

A key question for the tagging performance is the dependence on the
activation threshold. Figure.~\ref{fig:rejection_vs_X} shows the
impact of different thresholds on the pixel activation, \ie $E_T$ used
both for training and testing the networks. Removing very soft
activity, below 3~GeV, only slightly degrades the network's
performance. Above 3~GeV the threshold leads to an approximately
linear decrease in background rejection with increasing
threshold.\medskip

An second, important experimental systematic uncertainty when working with
calorimeter images is the jet energy scale (JES). We assess the
stability of our network by evaluating the performance on jet images
where the $E_T$ pixels are globally rescaled by $\pm 25\%$.
As shown in the right panel of figure~\ref{fig:rejection_vs_X}
this leads to a decline in the tagging performance of
approximately $10\%$ when reducing the JES
and $5\%$ when increasing the JES.

Next, we train a \textit{hardened} version of the network. It uses the
same architecture as our default, but during the training procedure
each image is randomly rescaled using a Gaussian distribution with a
mean of 1.0 and a width of 0.1. New random numbers are used from epoch
to epoch.  The resulting network has a similar performance as the
default and exhibits a further reduced sensitivity to changes in the
global JES.

While other distortions of the image, such as non-uniform rescaling,
will need to be considered, the resilience of the network and our
ability to further harden it are very encouraging for experimental usage
where the mitigation and understanding of systematic uncertainties is
critical.


\end{document}